\documentclass[preprint,11pt,twocolumn,times,5p]{elsarticle}
\usepackage[utf8]{inputenc}
\usepackage{graphicx}
\usepackage{dcolumn}
\usepackage{bm}
\usepackage{color}
\usepackage{xcolor}
\usepackage{verbatim}
\usepackage{spverbatim}
\usepackage{subcaption}
\usepackage{soul}
\usepackage{amsmath}
\usepackage{amssymb}
\usepackage{listings}
\usepackage{ulem}
\usepackage{subcaption}
\usepackage[hyphens]{url}
\usepackage{hyperref} 
\hypersetup{colorlinks=true,breaklinks=true}
\graphicspath{ {./images/} }
\usepackage{todonotes}
\usepackage[export]{adjustbox}

\journal{Computer Physics Communications}

\begin{document}
\begin{frontmatter}

\title{DMFTwDFT: An open-source code combining Dynamical Mean Field Theory with various Density Functional Theory packages}

\author[a,b]{Vijay Singh\corref{author}}
\author[b]{Uthpala Herath}
\author[a]{Benny Wah}
\author[a]{Xingyu Liao}
\author[b]{Aldo H. Romero}
\author[a]{Hyowon Park}

\cortext[author]{Corresponding author.\\\textit{E-mail address:}vsingh83@uic.edu}
\address[a]{Department of Physics, University of Illinois at Chicago, Chicago, IL, 60607}
\address[b]{Department of Physics and Astronomy, West Virginia University, Morgantown, WV 26506}

\begin{abstract}
Dynamical Mean Field Theory (DMFT) is a successful method to compute the electronic structure of strongly correlated materials, especially when it is combined with density functional theory (DFT). Here, we present an open-source computational package (and a library) combining DMFT with various DFT codes interfaced through the Wannier90 package. The correlated subspace is expanded as a linear combination of Wannier functions introduced in the DMFT approach as local orbitals. In particular, we provide a library mode for computing the DMFT density matrix. This library can be linked and then internally called from any DFT package, assuming that a set of localized orbitals can be generated in the correlated subspace. The existence of this library allows developers of other DFT codes to interface with our package and achieve the charge-self-consistency within DFT+DMFT loops. To test and check our implementation, we computed the density of states and the band structure of well-known correlated materials, namely LaNiO$_3$, SrVO$_3$, and NiO. The obtained results are compared to those obtained from other DFT+DMFT implementations.\\

\noindent
{\bf PROGRAM SUMMARY} \\
\begin{small}
\noindent
{\em Program title}: DMFTwDFT\\
{\em Licensing provisions}: GNU General Public License 3\\
{\em Programming language}: Python2/3, C++, and FORTRAN\\
{\em No. of lines in distributed program, including test data.}: 9091\\
{\em No. of bytes in distributed program, including test data.}: 109,051,904\\
{\em Distributed format}: zip\\
{\em Computer}: Non-specific\\
{\em Operating system}: Unix/Linux\\
{\em RAM}: Up to several GB\\
{\em External routines}: MPI, FFTW, BLAS, LAPACK, Numpy, Scipy, mpi4py, Glib, gsl, weave, PyProcar, and PyChemia\\
{\em Has the code been vectorised or parallelized?}: Yes, parallelized using MPI\\
{\em Nature of problem}: Need for a simple, efficient, higher-level, and open-source package to study strongly correlated materials interfacing to various DFT codes regardless of basis sets used in DFT.\\
{\em Solution method}:
We present an open-source Python code which can be easily interfaced with Wannier90 and different DFT packages and perform a full charge-self-consistent DFT+DMFT calculation using a modern continuous-time quantum Monte Carlo (CTQMC) impurity solver.\\
{\em Subprograms used:} Wannier90, Siesta, VASP, CTQMC\\
\end{small}
\end{abstract}

\begin{keyword}
DFT; DMFT; strongly correlated materials; Python; condensed matter physics; many-body physics  
\end{keyword}
\end{frontmatter}

\section{Introduction}
 
One of the challenging tasks in modern material science is the theoretical design of novel materials with exceptional properties established only from their atomic species and positions based on first-principle methodologies. While DFT, a workhorse of the electronic structure calculation, is the most-used methodology to describe material properties, it has several drawbacks in the description of strongly correlated materials (SCMs). This problem is mainly due to the lack of accuracy in describing SCMs because DFT relies on a crudely approximated exchange-correlation functional neglecting significant many-body fluctuations. Therefore, the existing approximations cannot capture correctly strong correlations present in localized orbitals or dispersionless bands.
Examples of significant failures in DFT are the incorrect prediction of metallic state in strongly correlated Mott insulators observed in many transition metal oxides, and severe underestimation of the electronic effective mass in heavy fermions~\cite{mott-insulator,Heavy_fermions,Fermiliquid}.

DMFT has been one of the most successful methods treating many-body fluctuations, by including dynamical but local correlations beyond the static DFT exchange-correlation functional~\cite{Savrasov:04,Kotliar:06}.
The heart of DMFT is to solve a single-site Anderson impurity problem embedded in an electronic bath determined self-consistently as the original lattice is approximated to a quantum impurity problem.
Examples of the impurity solvers include exact-diagonalization~\cite{Exact_diag}, numerical renormalization group~\cite{RGgroup}, density-matrix renormalization group~\cite{DMRG}, and Quantum Monte Carlo (QMC)~\cite{Jarrell_MonteCarlo_1,Monte_carlo_2} methods.
While the QMC calculation can be more expensive than other solvers, it can provide a numerically exact and non-perturbative solution of the impurity problem. 
Among QMC methods, continuous-time QMC (CTQMC)~\cite{CTQMC} has been frequently used in that respect.  Several free-licensed CTQMC packages~\cite{haule_ctqmc,ALPS,ComDMFT,IQIST} are currently available and our DMFTwDFT code is interfaced to one of the CTQMC codes~\cite{haule_ctqmc} included in the EDMFTF package~\cite{EDMFT}.

In the last a few decades, great progress has been made in developing computational algorithms or packages for solving the DMFT equations~\cite{ALPS,TRIQS,w2dynamics} and combining DMFT with other electronic structure methods such as DFT or GW. Examples of available DFT+DMFT or GW+DMFT packages include EDMFTF~\cite{EDMFT}, TRIQS/DFTTools~\cite{TRIQSDFTTools}, D-core~\cite{Dcore}, AMULET~\cite{Amulet}, LMTO+DMFT~\cite{RsPt_DMFT}, Questaal~\cite{Questaal}, and ComDMFT~\cite{ComDMFT}. 
These implementations of DMFT in combination with DFT usually require the construction of local orbitals to define a correlation subspace for solving the DMFT equations.
One choice is to use local atomic orbitals which are exactly centered at the ion sites and highly localized, so called projectors. They are constructed from the atomic solution and projected to the wide energy window of Kohn-Sham (KS) wavefunctions to ensure the locality of orbitals.
These orbitals are frequently used in all-electron DFT codes where different flavors of approximations are introduced as in LMTO, LAPW, and so on~\cite{LMTOLAPW}.  
The other popular choice of localized orbitals is the Maximally Localized Wannier functions(MLWFs)~\cite{Wannier90_2012,wannier90new,Updated_wannier90_2014}. 
Wannier functions can be used to construct both the hybridization and correlation subspace from the given energy window of the DFT band structure.
Currently, the interface to Wannier90 package has been implemented in various DFT codes including VASP~\cite{VASP,PhysRevB.47.558, kresse_efficiency_1996}, Quantum espresso~\cite{QuantumExpresso}, Siesta~\cite{Siesta}, Abinit~\cite{ABINIT, gonze2002first,gonze2009abinit}, ELK~\cite{ELK}, Wien2k~\cite{WIEN2k}, and so on. 
The required overlap matrices are obtained from the DFT code and the localization of the Wannier function is performed by Wannier90.
  
While DMFT has been a powerful method for studying the electronic structure of SCMs, the full implementation of DFT+DMFT sometimes requires the combination of a DMFT implementation with licensed DFT codes. This has been a bottleneck of the wide-applicability of the DFT+DMFT methodology. In this paper, we provide a DMFT package interfaced to the Wannier90 code for its efficient extension to various free-licensed DFT codes.
Our DMFTwDFT package can 1) use the Wannier orbitals for constructing the hybridization and correlation subspaces to perform DMFT loops by taking advantage of the Wannier90 interface between various DFT codes, 2) provide the \textit{library mode} to link the module for computing a DMFT density matrix and updating a charge density within the DFT loops without modifying any DFT source codes significantly, and 3) provide a flexible Python-based interface that do not rely much on extensive user experience or specific parameters to perform DFT+DMFT calculations of SCMs. The outputs of our DMFTwDFT package currently include band structures, density of states, and total energies. Forces and the Fermi surface calculations will be available in the future release of the code. Source codes are currently located at the GitHub repository, \url{https://github.com/DMFTwDFT-project/DMFTwDFT} which also includes a documentation with examples.

Our paper is organized as follows. In section$\:$\ref{sec:method}, we describe the theoretical background used to perform DFT+DMFT loops. 
Section$\:$\ref{sec:feature} describes the essential features of our DMFTwDFT package including the library mode, the interface to different DFT codes, and automated scripts for post-processing. Section$\:$\ref{sec:example} provides some run examples of well-known SCMs including SrVO$_3$, LaNiO$_3$, and NiO and compare our results to other available DFT+DMFT codes. 
Finally, we conclude the paper in the conclusion section.

\section{Methodology}
\label{sec:method}
In this section, we explain the methodology used for implementing our code.

\subsection{Implementation of DFT+DMFT}

The formal derivation of electronic structure methods including DFT and DMFT can be achieved by constructing an effective Free energy functional, $\Gamma$, which depends on the choice of variables to write the energy functional~\cite{Savrasov:04,Kotliar:06}. 
For example, the variable of choice that parametrize the Free energy minimization in DFT is the electronic charge density $\rho(\mathbf{r})$~\cite{Kohn1964} and the corresponding $\Gamma^{DFT}$ is given by 
\begin{equation}
\begin{aligned}
\Gamma^{DFT}\left[\hat{\rho}, \hat{V}^{Hxc}\right]=&-\operatorname{Tr}\left(\ln \left[\left(i \omega_{n}+\mu\right) \hat{1}-\hat{H}^{KS}\right]\right) \\
&+\Phi^{DFT}[\hat{\rho}]-\operatorname{Tr}\left(\hat{V}^{Hxc} \hat{\rho}\right),
\end{aligned} 
\label{eq:DFT}
\end{equation}
where $\omega_n$ is the Matsubara frequency for fermions, $\mu$ is the chemical potential, $\hat{1}$ is the unit matrix, $\Phi^{DFT}[\rho]$ is the DFT interaction energy, and $\hat{H}^{KS}=-\frac{\hbar^2}{2m}\hat{\nabla}^2+\hat{V}^{ext}+\hat{V}^{Hxc}$ is the KS Hamiltonian operator, where
$\hat{V}^{ext}$ is the ionic potential operator and $\hat{V}^{Hxc}$ is the Hartree-exchange-correlation potential operator. 
Since the exact form of the functional $\Phi^{DFT}[\rho]$ is not known, it is usually approximated by using the local density approximation\cite{LDA1,LDA2} or the general gradient approximation\cite{PBE_functional}.
The stationary value of $\Gamma$ with respect to the selected variable can provide the Free energy within the electronic structure methods.
In DFT,  minimizing the functional $\Gamma^{DFT}$ with respect to $\rho(\mathbf{r})$ and $V^{Hxc}(\mathbf{r})$ leads to the self-consistent equation, also known as the KS equation.

In DMFT, the variable of choice is the dynamical Green's function $G^{cor}(i\omega_n)$. The effective many-body potential conjugate to $G^{cor}(i\omega_n)$ is the dynamical self-energy $\Sigma(i\omega_n)$. 
The main idea of DFT+DMFT is to treat dynamical correlations of localized orbitals using the DMFT functional in terms of $G^{cor}(i\omega_n)$ and $\Sigma(i\omega_n)$ within a ``correlated subspace" defined from the DFT band structure, and then to subtract a double-counting term of correlations for which both DFT and DMFT functionals are accounted.
As a result, the DFT+DMFT functional $\Gamma$ can be constructed using four operators ($\hat{\rho}$, $\hat{V}^{Hxc}$, $\hat{G}^{cor}$, and $\hat{\Sigma}$): 
\begin{equation}
\begin{aligned} 
&\Gamma\left[\hat{\rho}, \hat{V}^{Hxc}, \hat{G}^{cor}, \hat{\Sigma}\right]=-\operatorname{Tr}\left(\ln \left[\left(i \omega_{n}+\mu\right) \hat{1}-\hat{H}^{KS}\right.\right.\\
&\left.\left.-\hat{P}_{cor}^{\dagger}\left(\hat{\Sigma}-\hat{V}^{D C}\right) \hat{P}_{cor}\right]\right)+\Phi^{DFT}[\hat{\rho}] -\operatorname{Tr}\left(\hat{V}^{Hxc} \hat{\rho}\right) \\
&+\Phi\left[\hat{G}^{cor}\right] -\operatorname{Tr}\left(\hat{\Sigma} \hat{G}^{cor}\right)-E^{DC}\left[\hat{G}^{cor}\right] +\operatorname{Tr}\left(\hat{V}^{DC} \hat{G}^{cor}\right),
\end{aligned} 
\label{eq:DFTDMFT}
\end{equation}
\noindent
where 
$\hat{V}^{DC}$ is the double-counting (DC) potential operator, $E^{DC}$ is the DC energy, and $\hat{P}_{cor}$ ($\hat{P}^\dagger_{cor}$) is a projection operator defined to downfold (upfold) between the correlated subspace and the hybridization subspace.
The DMFT interaction energy, $\Phi[G^{cor}]$ is the Luttinger-Ward functional summing all vacuum-to-vacuum Feynman diagrams which are local~\cite{LuttingerI,LuttingerII,LuttingerIII}. 

The stationary solution of the Free energy functional $\Gamma$ within DFT+DMFT can be obtained by extremizing $\Gamma$ with respect to $G^{cor}(i\omega_n)$ and ${\Sigma}(i\omega_n)$, which lead to: 
\begin{eqnarray}\label{eq:SCE}
\hat{\Sigma}&=&\frac{\delta \Phi[\hat{G}^{cor}] }{\delta \hat{G}^{cor}},\\
\hat{G}^{cor}&=&\hat{P}_{cor}\:\hat{G}^{hyb}\:\hat{P}^\dagger_{cor},
\label{eq:SCE2}
\end{eqnarray}
where $\hat{G}^{hyb}=[(i\omega_n+\mu)\hat{1}-\hat{H}^{KS}-\hat{P}_{cor}^{\dagger}\hat{(\Sigma}-\hat{V}^{DC})\hat{P}_{cor}]^{-1}$ is the Green's function operator defined within the energy window where $\Sigma(i\omega_n)$ is hybridized (upfolded). 
 
Although the many-body functional $\Phi[\hat{G}^{cor}]$ in Eq.$\:$\ref{eq:DFTDMFT} needs to be evaluated only within the correlated subspace, computing the exact and non-perturbative $\Phi[\hat{G}^{cor}]$ is still a formidable task. Nevertheless, it can be approximated to a solution of an effective impurity problem within DMFT, i.e., $\Phi[\hat{G}^{cor}]\simeq\Phi[\hat{G}^{imp}]$ by assuming that the correlated Green's function of a lattice is approximated to the impurity one, i.e., $\hat{G}^{cor}\simeq\hat{G}^{imp}$.
Therefore, the numerically exact solution can be obtained by solving an impurity problem hybridized to an effective electronic bath $\hat{\Delta}(i\omega_n)$ using the QMC method.
As a result, $\Sigma(i\omega_n)=\Sigma^{imp}(i\omega_n)$ from Eq.$\:$\ref{eq:SCE} and the DMFT self-consistent condition ensures that the hybridization function operator $\hat{\Delta}(i\omega_n)=(i\omega_n+\mu)\hat{1}-\hat{\epsilon}_{imp}-\hat{\Sigma}(i\omega_n)-[\hat{G}^{cor}(i\omega_n)]^{-1}$ where $\hat{\epsilon}_{imp}$ is the matrix representing the impurity levels of correlated orbitals.

The DMFT self-consistent condition is completed by computing new $G^{cor}$ from the obtained $\Sigma(i\omega_n)$ using Eq.$\:$\ref{eq:SCE2} and by iterating the calculation until both $G^{cor}$ and $\Sigma(i\omega_n)$ are converged.
Solving Eq.$\:$\ref{eq:SCE2} requires the construction of a projection operator, $\hat{P}_{cor}$ to define the correlated subspace. To achieve this, one needs to adopt the localized
orbital $\phi_m^{\mathbf{\tau}}$ having the orbital character $m$ of the correlated atom centered at $\mathbf{\tau}$ in an unit cell, namely $\hat{P}_{cor}=\sum_{m\mathbf{\tau}}|\phi_m^{\mathbf{\tau}} \rangle\langle \phi_m^{\mathbf{\tau}}|$ and choose the energy window where these correlated orbitals will be hybridized. In this way, orbitals within the hybridization window do not mix with states outside this energy window.
Eq.$\:$\ref{eq:SCE2} can be represented as a matrix equation using the KS wavefunction basis. $\hat{H}^{KS}$ can be diagonalized within these KS basis while $\hat{\Sigma}$ is in general a non-diagonal matrix with complex numbers. 
Therefore, computing $\hat{G}^{hyb}$ requires the inversion of a non-Hermitian matrix with complex numbers and can be achieved by solving the generalized eigenvalue problem of the Hamiltonian in the KS basis at each momentum $\mathbf{k}$ and frequency $\omega_n$:
\begin{eqnarray}
\sum_j\left[\epsilon_{i}^{\mathbf{k}}\delta_{ij}+\Sigma_{ij}^{\mathbf{k}\omega_{n}}\right]C^{R}_{jl,\mathbf{k}\omega_n}=C^{R}_{il,\mathbf{k}\omega_n}\epsilon_{l}^{\mathbf{k}\omega_{n}},
\label{eq:Eig}
\end{eqnarray}
where $\epsilon_{i}^{\mathbf{k}}$ is the KS eigenvalue at the band index $i$ and the momentum $\mathbf{k}$. $\epsilon_{l}^{\mathbf{k}\omega_{n}}$ is the complex eigenvalue and $C^{R(L)}_{\mathbf{k}\omega_n}$ is the right (left) eigenfunction of the above matrix equation.
$\Sigma_{ij}^{\mathbf{k}\omega_{n}}$ is the DMFT self-energy upfolded to the KS space ($|\psi^{\mathbf{k}}\rangle$):
\begin{eqnarray}
\Sigma_{ij}^{\mathbf{k}\omega_{n}}=\sum_{mn\mathbf{\tau}}\langle\psi_{i}^{\mathbf{k}}|\phi^{\mathbf{\tau}}_m\rangle(\Sigma^{\mathbf{\tau}}_{mn}(i\omega_n)-V^{DC})\langle\phi^{\mathbf{\tau}}_n|\psi_{j}^{\mathbf{k}}\rangle.
\label{eq:sigij}
\end{eqnarray}
One can note that the self-energy matrix element can be $\mathbf{k}-$dependent in the KS basis although they are purely local in the correlated orbital basis. 
Finally, $\hat{G}^{cor}$ (Eq.$\:$\ref{eq:SCE2}) in the local orbital basis can be represented using these obtained eigenvalues and eigenfunctions: 
\begin{equation}
G^{cor}_{mn}(i\omega_n) = \frac{1}{N_{\mathbf{k}}}\sum_{\mathbf{k}ijl}
\frac{\langle\phi^{\mathbf{\tau}}_m|\psi_{i}^{\mathbf{k}}\rangle C^{R}_{il,\mathbf{k}\omega_n} \left(C^{L}_{jl,\mathbf{k}\omega_n}\right)^*\langle \psi_{j}^{\mathbf{k}}| \phi^{\mathbf{\tau}}_n\rangle
}{i\omega_{n}+\mu-\epsilon_{l}^{\mathbf{k}\omega_{n}}}.
\label{eq:Ghyb}
\end{equation}

\subsection{Construction of the hybridization and correlation subspace: Wannier orbitals}

Since the implementation of DFT+DMFT requires the construction of the correlated subspace where the self-energy $\Sigma$ is defined, one needs to construct the localized orbital $\phi_m^{\mathbf{R}}$ centered at each correlated atom. 
Here, the locality of the correlated orbital matters since $\Sigma$ is approximated as a local quantity within DMFT, i.e., $\Sigma(\mathbf{k},\omega)\simeq\Sigma(\omega)$ and the non-locality of the Coulomb interaction should be minimized. 
One choice of such orbitals is so called ``projectors", namely the orbital represents atomic character within a chosen atomic sphere.
Those projectors are exactly centered at correlated atoms and highly localized by definition. 
Usually projectors require the construction of a quite large hybridization window as these highly localized orbitals are hybridized with KS wavefunctions in a rather wide energy window. 

Wannier functions have been also frequently used as the choice of correlated orbitals. They are constructed from the unitary transform of the KS wavefunctions and can represent the isolated DFT band structure within the hybridization window exactly. 
In this way, not only the correlated orbitals but also other orbitals strongly hybridized with those correlated orbitals are also constructed.
However, the choice of the Wannier function is not unique and there have been several methods to achieve the locality of Wannier functions. They include MLWFs, selectively localized Wannier functions, symmetry-adapted Wannier functions, and so on~\cite{Wannier90_2012,wannier90new,Updated_wannier90_2014}.
Here, we show some examples of DFT+DMFT using MLWFs to construct localized orbitals within the hybridization window.
 
The MLWF $|\tilde{\phi}_n^{\mathbf{R}}\rangle$ can be constructed from the KS orbital $|\psi_i^{\mathbf{k}}\rangle$ by performing the Unitary transform $U^{\mathbf{k}}$ which minimize the sum of all Wannier orbital spreads:
\begin{equation}
|\tilde{\phi}_n^{\mathbf{R}}\rangle=\frac{1}{\sqrt{N_{\mathbf{k}}}}\sum_{i\mathbf{k}}e^{-i\mathbf{k}\cdot\mathbf{R}}|\psi_i^{\mathbf{k}}\rangle \cdot U_{in}^{\mathbf{k}}.
\end{equation}
And the KS Hamiltonian can be represented using the basis of the MLWF $|\tilde{\phi}_n^{\mathbf{R}}\rangle$ as:

\begin{equation}
\tilde{\epsilon}_{mn}(\mathbf{R_i}-\mathbf{R_j}) = \langle \phi_m^{\mathbf{R_i}}|\hat{H}^{KS}|\phi_n^{\mathbf{R_j}} \rangle = \frac{1}{N_{\mathbf{k}}}\sum_{i\mathbf{k}}e^{i\mathbf{k}\cdot(\mathbf{R_i}-\mathbf{R_j})} \tilde{\epsilon}_{mn}^{\mathbf{k}}
\label{eq:KS}
\end{equation}

\begin{equation}
\tilde{\epsilon}_{mn}^{\mathbf{k}}=\sum_i (U_{im}^{\mathbf{k}})^*\epsilon_{i}^{\mathbf{k}} U_{in}^{\mathbf{k}}
\end{equation}
where $\tilde{\epsilon}_{mn}^{\mathbf{k}}$ is the Wannier Hamiltonian matrix elements at the momentum $\mathbf{k}$ 
and $m$, $n$ are  dual indices $(\tau,\alpha)$ in which $\tau$ labels correspond to an atomic site in the unit cell and $\alpha$ labels to the orbital character of the corresponding site.
One can note that the $\mathbf{k}-$point mesh representing the Wannier Hamiltonian can be much denser than the DFT $\mathbf{k}-$point mesh by adopting the band-structure interpolation scheme~\cite{Wannier90_2012}.

Since Wannier orbitals can represent not only the correlated orbitals but also all other orbitals in the energy window where the correlated orbitals are hybridized, Eq.$\:$\ref{eq:Eig} can be solved using the Wannier orbital basis at each momentum $\mathbf{k}$ and frequency $\omega_n$:
\begin{equation}
\sum_n\left[\tilde{\epsilon}_{mn}^{\mathbf{k}}+\Sigma_{mn}(i\omega_{n})-V^{DC}\right]C^{R}_{nl,\mathbf{k}\omega_n}=C^{R}_{ml,\mathbf{k}\omega_n}\epsilon_{l}^{\mathbf{k}\omega_{n}}.
\label{eq:dmft_wan}
\end{equation}
Here, the size of the matrix for the eigenvalue problem becomes exactly the number of Wannier orbitals specified in an unit cell. Moreover, $\Sigma$ can be also a diagonal matrix ($\Sigma_{mn}\simeq \Sigma_{m}\delta_{mn}$) as the local axis for the Wannier orbital can be rotated to minimize the off-diagonal term of the hybridization function $\Delta(i\omega_n)$ (see Appendix 6.2)
and CTQMC can solve the impurity problem for this diagonal $\Delta(i\omega_n)$ matrix. 
Finally, the correlated Green's function in the Wannier basis is given by 
\begin{eqnarray}
G^{cor}_{mn}(i\omega_n) = \frac{1}{N_{\mathbf{k}}}\sum_{\mathbf{k}l}
\frac{C^{R}_{ml,\mathbf{k}\omega_n} \left(C^{L}_{nl,\mathbf{k}\omega_n}\right)^*
}{i\omega_{n}+\mu-\epsilon_{l}^{\mathbf{k}\omega_{n}}}.
\label{eq:Gwan}
\end{eqnarray}

\subsection{Charge-self-consistency in DFT+DMFT}

The charge density $\rho(\mathbf{r})$ in DFT+DMFT can be obtained when the Free energy functional $\Gamma$ in Eq.$\:$\ref{eq:DFTDMFT} is minimized by extremizing this functional with respect to the density functional potential $V^{Hxc}(\mathbf{r})$. 
As a result, the equation for the charge density $\rho(\mathbf{r})$ is obtained to be:
\begin{eqnarray}
\rho(\mathbf{r})&=&T\sum_{\omega_n}\left\langle \mathbf{r} \middle|\hat{G}\middle| \mathbf{r} \right\rangle e^{i\omega_n0^+}
\label{eq:rho}
\end{eqnarray}
where $T$ is temperature and 
the Green's function operator $\hat{G}=[(i\omega_n+\mu)\hat{1}-\hat{H}^{KS}-\hat{P}_{cor}^{\dagger}\hat{(\Sigma}-\hat{V}^{DC})\hat{P}_{cor}]^{-1}$.  
The full charge-self-consistency is achieved when both $\rho$ and $G$ are converged after DFT+DMFT loops.
$\rho(\mathbf{r})$ in Eq.$\:$\ref{eq:rho} can be computed by representing $\hat{G}$ using KS orbitals:
\begin{eqnarray}
\rho(\mathbf{r}) &=&\frac{T}{N_{\mathbf{k}}} \sum_{ij \mathbf{k}, \omega_{n}}\left\langle\mathbf{r} \middle| \psi_{i}^{\mathbf{k}}\right\rangle\left\langle\psi_{i}^{\mathbf{k}}\middle|\hat{G}\middle| \psi_{j}^{\mathbf{k}}\right\rangle\left\langle\psi_{j}^{\mathbf{k}} \middle| \mathbf{r}\right\rangle e^{i \omega_{n} 0^{+}} \nonumber \\ 
&=&\frac{1}{N_{\mathbf{k}}}\sum_{ij\mathbf{k}}\psi_{i}^{\mathbf{k}}(\mathbf{r}) \left(\psi_{j}^{\mathbf{k}}(\mathbf{r})\right)^* n_{ij}^{\mathbf{k}} 
\label{eq:rho1}
\end{eqnarray}
where $n^{\mathbf{k}}_{ij}$ is the DMFT occupancy matrix element in the KS orbital basis:
\begin{eqnarray}
\label{eq:nijk0}
n_{ij}^{\mathbf{k}}&=&T\sum_{\omega_{n}}
\left\langle\psi_{i}^{\mathbf{k}}\middle|\hat{G}\middle| \psi_{j}^{\mathbf{k}}\right\rangle e^{i \omega_{n} 0^{+}},\\
\left\langle\psi_{i}^{\mathbf{k}}\middle|\hat{G} \middle| \psi_{j}^{\mathbf{k}}\right\rangle&=&\sum_{mnl}
\frac{U_{im}^{\mathbf{k}} C^{R}_{ml,\mathbf{k}\omega_n} \left(U_{jn}^{\mathbf{k}} C^{L}_{nl,\mathbf{k}\omega_n}\right)^*
}{i\omega_{n}+\mu-\epsilon_{l}^{\mathbf{k}\omega_{n}}}.
\end{eqnarray}
One can note that the DMFT occupancy matrix, $n^{\mathbf{k}}$ contains non-diagonal matrix elements in the Kohn-Sham basis and it becomes a diagonal matrix whose elements are DFT Fermi functions (DFT occupation matrix) when dynamical self-energies are zero .  
Since DMFT self-energies are hybridized with DFT bands only within the hybridization window $W$, the DMFT occupation matrix can be given by:
\begin{eqnarray}
\label{eq:nijk}
n_{ij}^{\mathbf{k}} &=& \overline{n}_{ij}^{\mathbf{k}} \:\:\:\:\:\:\:\:\:\:\:\: if\: (\epsilon_{i\mathbf{k}},\epsilon_{j\mathbf{k}}) \in W, \nonumber \\
&=& f_i^{\mathbf{k}} \delta_{ij} \:\:\:\:\:\:\:\: otherwise.
\end{eqnarray}
Namely, $n^{\mathbf{k}}$ is a non-diagonal matrix $\overline{n}^{\mathbf{k}}$ when both $\epsilon_{i\mathbf{k}}$ and  $\epsilon_{j\mathbf{k}}$ are located inside the energy window $W$ while $n^{\mathbf{k}}$ is a DFT Fermi function outside the window $W$. 


Since $\overline{n}^{\mathbf{k}}$ is a non-diagonal but also Hermitian matrix in the KS basis, it can be also decomposed in terms of eigenvalues $w^{\mathbf{k}}_{\lambda}$ and eigenfunctions $v_\lambda^{\mathbf{k}}$ as:
\begin{equation}
\overline{n}^{\mathbf{k}}_{ij} = \sum_{\lambda}v^{\mathbf{k}}_{i\lambda}\cdot 
w^{\mathbf{k}}_{\lambda}\cdot \left(v^{\mathbf{k}}_{j\lambda}\right)^*
\end{equation}
where the eigenvalue index $\lambda$ runs over the number of bands in the window $W$.
Therefore, the DMFT occupation matrix can be diagonalized by rotating a KS wavefunction $|\psi_i^{\mathbf{k}}\rangle$ 
to a new DMFT wavefunction $|\overline{\psi}_{\lambda}^{\mathbf{k}}\rangle$
using the unitary transform whose matrix row is the eigenfunction $v_\lambda^{\mathbf{k}}$:
\begin{equation}
\langle\mathbf{r} | \overline{\psi}_{\lambda}^{\mathbf{k}}\rangle=\sum_{i}\left\langle\mathbf{r} | \psi_{i}^{\mathbf{k}}\right\rangle \cdot v_{i \lambda}^{\mathbf{k}}
\end{equation}
Now, $w^{\mathbf{k}}_{\lambda}$ will be the diagonal elements of the DMFT occupation matrix in this rotated KS basis and the sum over band indices $i,j$ in Eq.$\:$\ref{eq:rho1} can be simplified to the sum over a new index $\lambda$. 
As a result, the DFT+DMFT charge density $\rho(\mathbf{r})$ 
can be constructed as:
\begin{eqnarray}
\rho(\mathbf{r})  &=&  \frac{1}{N_{\mathbf{k}}}\sum_{\lambda,\mathbf{k}}|\overline{\psi}_{\lambda}^{\mathbf{k}}(\mathbf{r})|^2 w^{\mathbf{k}}_{\lambda} \:\:\:\:\:\:\:\: if\: (\epsilon_{i\mathbf{k}},\epsilon_{j\mathbf{k}}) \in W, \nonumber \\ 
&=& \frac{1}{N_{\mathbf{k}}}\sum_{i,\mathbf{k}}
|\psi_{i}^{\mathbf{k}}(\mathbf{r})|^2 f_i^{\mathbf{k}}
\:\:\:\:\:\:\:\: otherwise.
\label{eq:rho2}
\end{eqnarray}

Eq.$\:$\ref{eq:rho2} implies that $\rho(\mathbf{r})$ in DFT+DMFT can be computed using the existing modules for computing $\rho(\mathbf{r})$ in a DFT package without much modifications.
The major modifications within the hybridization window include 1) the change of DFT Fermi function $f_i^{\mathbf{k}}$ to the DMFT occupation function $w^{\mathbf{k}}_{\lambda}$ and 2) the unitary transform of $|\psi_i^{\mathbf{k}}\rangle$
to $|\overline{\psi}_{\lambda}^{\mathbf{k}}\rangle$
.
To facilitate the implementation of the charge calculation in a DFT package, our DMFTwDFT package provides a library mode 
such that any DFT codes can call the Fortran subroutine to obtain the necessary information to update charge density within DFT+DMFT. Specifically, one can pass the $\mathbf{k}-$points information within DFT to the subroutine $Compute\_DMFT$ and can obtain the DMFT weight $w^{\mathbf{k}}$ and the Unitary matrix $v^{\mathbf{k}}$ at each $\mathbf{k}-$point for computing the charge density $\rho(\mathbf{r})$. 
Details about the structure of this subroutine are provided in the following section.

Total number of valence electrons, $N_{tot}$ can be computed by integrating $\rho(\mathbf{r})$ over the space, or equivalently from the trace of the occupation matrix $n^{\mathbf{k}}$:
\begin{eqnarray}
N_{tot} & = & \int d\mathbf{r} \rho(\mathbf{r}) =
\frac{1}{N_{\mathbf{k}}}\sum_{i\mathbf{k}} n_{ii}^{\mathbf{k}}\nonumber\\
&=&
\frac{T}{N_{\mathbf{k}}}\sum_{\mathbf{k}l\omega_n}\frac{e^{i\omega_n0^+}}{i\omega_{n}+\mu-\epsilon_{l}^{\mathbf{k}\omega_{n}}}\nonumber\\
&=&\frac{T}{N_{\mathbf{k}}}\sum_{\mathbf{k}l\omega_n}\left(\frac{1}{i\omega_n
+\mu-\epsilon_{l}^{\mathbf{k}\omega_{n}}}-\frac{1}{i\omega_n+\mu-\epsilon_{l}^{\mathbf{k}\omega_{\infty}}}\right)\nonumber \\
&+&\frac{1}{N_{\mathbf{k}}}\sum_{\mathbf{k}l}f(\epsilon_{l}^{\mathbf{k}\omega_{\infty}}-\mu)
\end{eqnarray}
where $\epsilon_{l}^{\mathbf{k}\omega_{\infty}}$
is the eigenvalue of Eq.$\:$\ref{eq:dmft_wan} evaluated at $\omega\rightarrow\infty$ and $f(\epsilon)$ is the Fermi function.
Here, the high-frequency $\omega_n$ summation can be done analytically when $\omega_n\rightarrow\omega_{\infty}$.
The chemical potential $\mu$ can be determined by imposing the condition that the total number of valence electrons ($N_{tot}$) obtained from DFT+DMFT should be fixed during the self-consistent loop and equal to the number of valence electrons in a material usually given in DFT. 


\subsection{Total energy and double counting correction}

Once the charge-self-consistent DFT+DMFT loop is converged, the functional $\Gamma$ in Eq.$\:$\ref{eq:DFTDMFT} evaluated at the stationary point (self-consistently determined DFT+DMFT solution) delivers the electronic Free energy of a given material within DFT+DMFT.
The total energy $E$ within DFT+DMFT can be obtained from the Free energy functional in the zero temperature limit as follows:
\begin{equation}
E=E^{DFT}[\rho]+\frac{1}{N_{\mathbf{k}}} \sum_{i\mathbf{k}} \epsilon_{i}^{\mathbf{k}} \cdot\left(n_{ii}^{\mathbf{k}}-f_{i}^{ \mathbf{k}}\right)+E^{POT}
-E^{DC}
\label{eq:EDMFT}
\end{equation}
where $E^{DFT}[\rho]$ is the DFT energy evaluated using the charge density $\rho$ obtained within DFT+DMFT, $\epsilon_{i}^{\mathbf{k}}$ is the DFT KS eigenvalue, $n_{ii}^{\mathbf{k}}$ is the diagonal element of the DMFT occupancy matrix $n^{\mathbf{k}}$ (Eq.$\:$\ref{eq:nijk}), and $f_{i}^{\mathbf{k}}$ is the Fermi function (DFT occupancy matrix) with the KS band $i$ and the momentum $\mathbf{k}$.

The potential energy $E^{POT}$ is the Luttinger-Ward functional $\Phi$ evaluated using the DMFT Green's function $G^{cor}$ and can be given from the Migdal-Galistkii formula\cite{Migdal}:
\begin{equation}
E^{pot}=\frac{1}{2}\mathrm{Tr}\left[\Sigma\cdot G^{cor}\right]=\frac{1}{2}\sum_{\omega_n}\left[\Sigma(i\omega_n)\cdot G^{cor}(i\omega_n)\right].
\end{equation}

The double counting energy, $E^{DC}$ needs to subtracted from the DFT energy functional since the part of the DFT correlation energy has been already accounted in the DMFT potential energy.
A frequently used expression of $E^{DC}$ is the fully localized limit (FLL) form which has been adopted mostly in DFT+U calculations ~\cite{FLL_V}.
\begin{eqnarray}
\label{eq:Edc}
E^{DC}=\frac{U}{2}\cdot N_d\cdot(N_d-1)-\frac{J}{4}\cdot N_d\cdot(N_d-2)\\
V^{DC}=\frac{\partial E^{DC}}{\partial N_d}=U\cdot(N_d-\frac{1}{2})-\frac{J}{2}\cdot(N_d-1)
\label{eq:Vdc}
\end{eqnarray}
where $V^{DC}$ is the DC potential, $U$ is the on-site Hubbard interaction, $J$ is the Hund's coupling, and $N_d$ is the occupancy of correlated orbitals within the correlation subspace which is obtained from the result of self-consistent DFT+DMFT calculations.

While the calculation of the exact $E^{DC}$ and $V^{DC}$ values can be difficult, at the same time, some DFT+DMFT calculations also indicate that smaller $E^{DC}$ and $V^{DC}$ values than the FLL form gives better agreement of electronic structure compared to experiments~\cite{Park_dc_2014,LaNiO3_press,PhysRevB.89.161113,Projector2}.
Here, we provide three different types of $E^{DC}$ functions.
A modified $E^{DC}$ form (DC\_type=1) is given by
\begin{equation}
E^{DC}=\frac{(U-\alpha)}{2}\cdot N_d\cdot(N_d-1)-\frac{J}{4}\cdot N_d\cdot(N_d-2)
\label{eq:Edc1}
\end{equation}
where the Hubbard $U$ used in Eq.$\:$\ref{eq:Edc} becomes smaller by $\alpha$ so that the $E^{DC}$ value is reduced. Here, the value for the parameter $\alpha$ can be chosen by users. $\alpha=0$ (default setting) recovers the FLL form in Eq.$\:$\ref{eq:Edc}.
Another modified form (DC\_type=2) is given by
\begin{equation}
E^{DC}=\frac{U}{2}\cdot (N_d-\alpha)\cdot(N_d-\alpha-1)-\frac{J}{4}\cdot (N_d-\alpha)\cdot(N_d-\alpha-2)
\label{eq:Edc2}
\end{equation}
where $N_d$ used in Eq.$\:$\ref{eq:Edc} gets smaller by $\alpha$ so that the $E^{DC}$ value is reduced. Also, $\alpha=0$ setting recovers the FLL form.
The other modified $V^{DC}$ form (DC\_type=3) is given by
\begin{equation}
V^{DC}=\frac{U}{2}\cdot N_d^0\cdot(N_d^0-1)-\frac{J}{4}\cdot N_d^0\cdot(N_d^0-2)
\label{eq:Edc3}
\end{equation}
where $N_d^0$ is the nominal occupancy of the correlated orbital. Also, this nominal $V^{DC}$ potential is known to be close to an exact $V^{DC}$ form~\cite{PhysRevLett.115.196403}.

An atomic force calculation within DFT+DMFT can be performed by taking an explicit derivative of the total energy in Eq.$\:$\ref{eq:EDMFT} or the Free energy with respect to the atomic position. Some implementations of atomic force calculations in DFT+DMFT are already present. \cite{PhysRevLett.112.146401,DMFT_forces_haule} 
We are currently incorporating the force calculation within our package and the details will be given in another paper.

\section{Features of DMFTwDFT}
\label{sec:feature}
In this section, we provide the most important features of our DMFTwDFT code including the overall structure of the code, the parallelized nature, the library mode, the interface to different DFT codes, and the automated scripts.

\subsection{Overall structure}

\begin{figure}[!h]
\centering{
\includegraphics[width=0.45\textwidth]{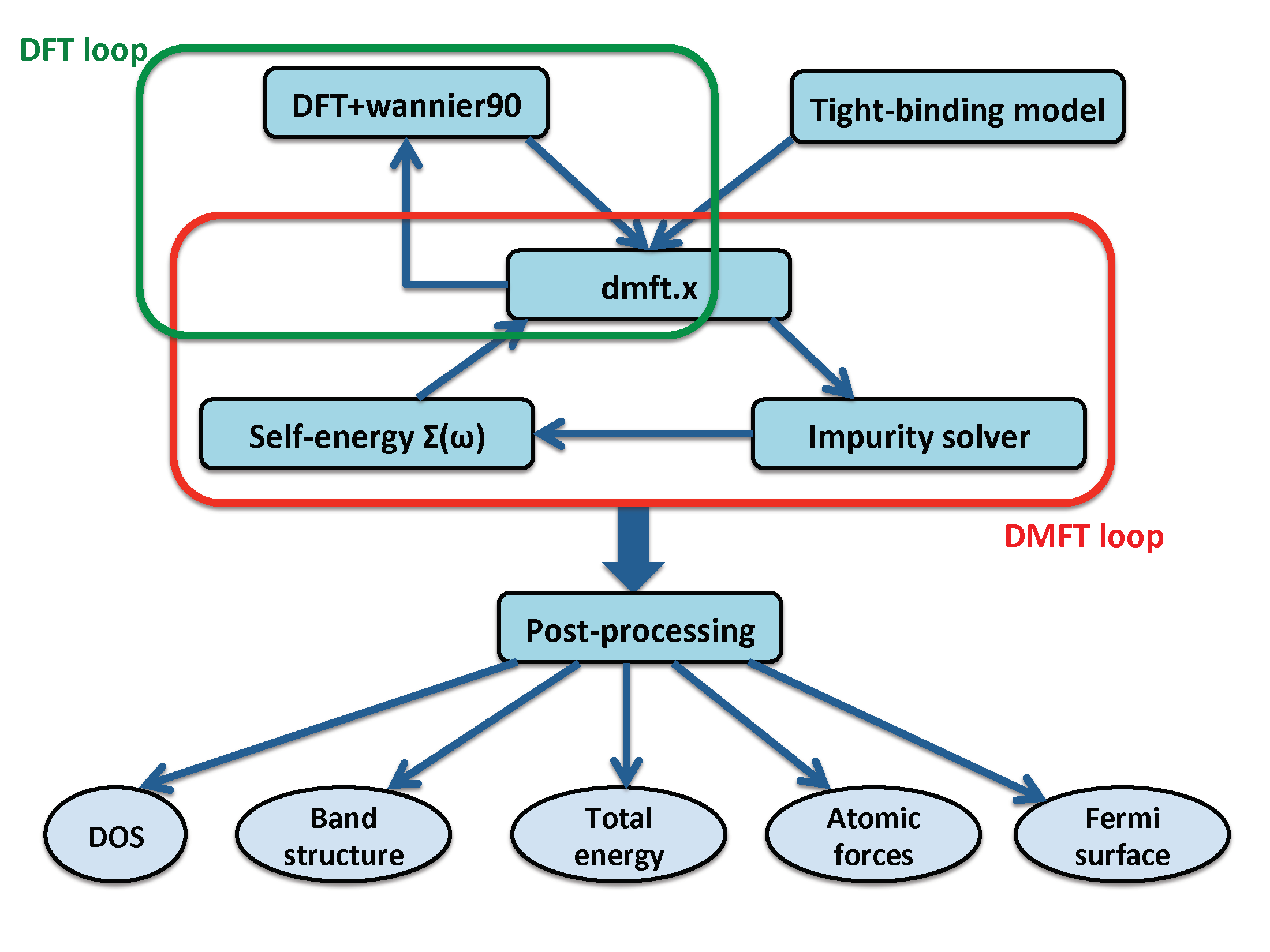}
}
\caption{The overall structure of the DMFTwDFT code.
} 
\label{fig:Figure1}
\end{figure}

Here, we describe the overall structure of our DMFTwDFT code in Fig.$\:$\ref{fig:Figure1}.
The overall DFT+DMFT loop is performed by a Python script (RUNDMFT.py). 
The DMFT loop in Fig.$\:$\ref{fig:Figure1} is performed by the main executable of the DMFTwDFT program (dmft.x). First, the local Green's function $G^{cor}(i\omega_n)$ (G\_loc.out) and the hybridization function $\Delta(i\omega_n)$ (Delta.out) are computed using inputs of a DMFT self-energy $\Sigma(i\omega_n)$ (sig.inp) and a Wannier-based Hamiltonian (see Eq.$\:$\ref{eq:dmft_wan} and Eq.$\:$\ref{eq:Gwan}).
The Wannier Hamiltonian can be obtained from DFT interfaced with the wannier90 code \cite{wannier90new} or from tight-binding parameters provided by users. 
The outputs of dmft.x including $\Delta(i\omega_n)$ (Delta.out), impurity energy levels (Ed.out), and the chemical potential (DMFT\_mu.out) are used as inputs of a DMFT impurity solver. Our code is currently interfaced with the CTQMC impurity solver.
The DMFT self-energy obtained from CTQMC is used as the input of dmft.x for the next DMFT loop.

The charge-self-consistency in DFT+DMFT is achieved by updating the charge density from the DMFT Green's function within the DFT loop in Figure$\:$\ref{fig:Figure1}. Our code provides the library mode for passing the necessary information from the DMFT calculation to a DFT code, where the DMFT occupation matrix (Eq.$\:$\ref{eq:rho2}) is included, and a new charge density and Wannier functions are obtained within the DFT loop. It is important to note that the main component of our code is interfaced to the MLWF, which is an independent basis set from a DFT-specific basis set used in obtaining the Bloch states. Thus, our code can also be interfaced straightforwardly to any electronic structure code, as long as the DFT implementation is able to obtain MLWFs. Currently, the VASP (with the full DFT+DMFT loop) and Siesta codes (the DMFT loop only) are interfaced with our program. The detailed procedure of the full charge-self-consistent DFT+DMFT calculation is as follows.

\begin{enumerate}
    \item First, a complete DFT self-consistent calculation is performed from the given atomic structure without any spin-polarization and the solution of the DFT KS equation is obtained. 
    \item The Wannier functions are constructed to represent the localized orbitals within the hybridization energy window. The KS Hamiltonian ($\hat{H}^{KS}$) in the basis of the Wannier function is obtained. In order to find the appropriate hybridization window (Wannier energy window), one may employ a projected band structure or density of states plot to identify the energy range of the hybridization subspace. One such method to achieve this is through PyProcar \cite{HERATH2019107080}, a code developed in the group of one of the authors. 
    \item Next, the DMFT loop in Figure$\:$\ref{fig:Figure1} is performed using the dmft.x executable. The inputs of dmft.x are the Wannier Hamiltonian ($\hat{H}^{KS}$) obtained from the Wannier90 outputs, the self-energy $\Sigma(i\omega_n)$, and the DC potential $V^{DC}$. Both $\Sigma(i\omega_n)$ and $V^{DC}$ can be given as an initial guess or obtained from the previous DFT+DMFT loop. The outputs of dmft.x are the chemical potential $\mu$, the impurity energy level $\epsilon_{imp}$, the hybridization function $\Delta(i\omega_n)$, and the Green's function $G^{cor}(i\omega_n)$. 
    \item A quantum impurity problem coupled to $\Delta(i\omega_n)$ is solved using a CTQMC impurity solver to obtain $\Sigma(i\omega_n)$. The new $\Sigma$ is mixed with the old $\Sigma$ and used as input of dmft.x for the next DMFT step.
    The $V^{DC}$ potential is also updated.
    \item While the DMFT loop is converging, one can achieve the full charge-self-consistent DFT+DMFT result by updating $\rho(\mathbf{r})$ from the DMFT occupancy matrix $n^{\mathbf{k}}$ (Eq.$\:$\ref{eq:rho1}). For the new charge update, a DFT code should be modified by linking our library mode to the DFT package and implement Eq.$\:$\ref{eq:rho2}.
    \item Once the new $\rho(\mathbf{r})$ is obtained, one can go back to Step 1 and a new KS equation can be solved. The Wannier functions are computed again to generate the new $\hat{H}^{KS}$. For a better convergence of $\rho(\mathbf{r})$, one can iterate $\rho(\mathbf{r})$ using the new $\hat{H}^{KS}$ while the DMFT self energy is fixed until the DFT loop (see Figure$\:$\ref{fig:Figure1}) is converged. 
    \item The full charge-self-consistent DFT+DMFT solution is achieved when both DFT and DMFT loops are converged. While the DFT+DMFT loops are converging, the information about the occupancy of correlated orbitals, the total energy, and both the Green's function and the self energy at each iteration are stored. The convergence can be checked by monitoring the change of these variables.
    \item After the DFT+DMFT loop is converged, one can perform the post-processing to obtain the band-structure and the density of states (see Section \ref{sec:automation} and Appendix).
\end{enumerate}

The source files can be found in the /src directory of our package in the github repository. After the compilation of the source codes, executable files (dmft.x, ctqmc, wannier90.x, modifed DFT code) and Python scripts (RUNDMFT.py) can be copied to the /bin directory and the path to this bin directory should be added to the path to environmental variables (\$PATH and \$PYTHONPATH). As an example, in the run\_example directory of our package we have also kept the result of LaNiO$_{3}$ for both non-charge self consistent and charge self consistent DMFT calculation. By comparing these results, we did not find any significant change in the band structure of LaNiO$_{3}$.

\subsection{Parallelization}

Our DMFTwDFT code has been implemented by adopting efficient parallelization using message passing interface (MPI). A bottleneck in running the dmft.x executable is solving the eigenvalue problem of the general complex matrix given in Eq.$\:$\ref{eq:dmft_wan} for a dense $\mathbf{k}-$point mesh and large Matsubara $\omega_n$ points. Our code adopts the $\mathbf{k}-$point paralellization so that calculations with different $\mathbf{k}-$points can be distributed to different cores. Moreover, our code is also compatible with the VASP $\mathbf{k}-$point parallel scheme (INCAR tag:KPAR) and the charge update calculation can be also performed using the $\mathbf{k}-$point parallelization.

\subsection{Library mode}

\begin{figure}[!h]
\centering{
\includegraphics[width=0.4\textwidth]{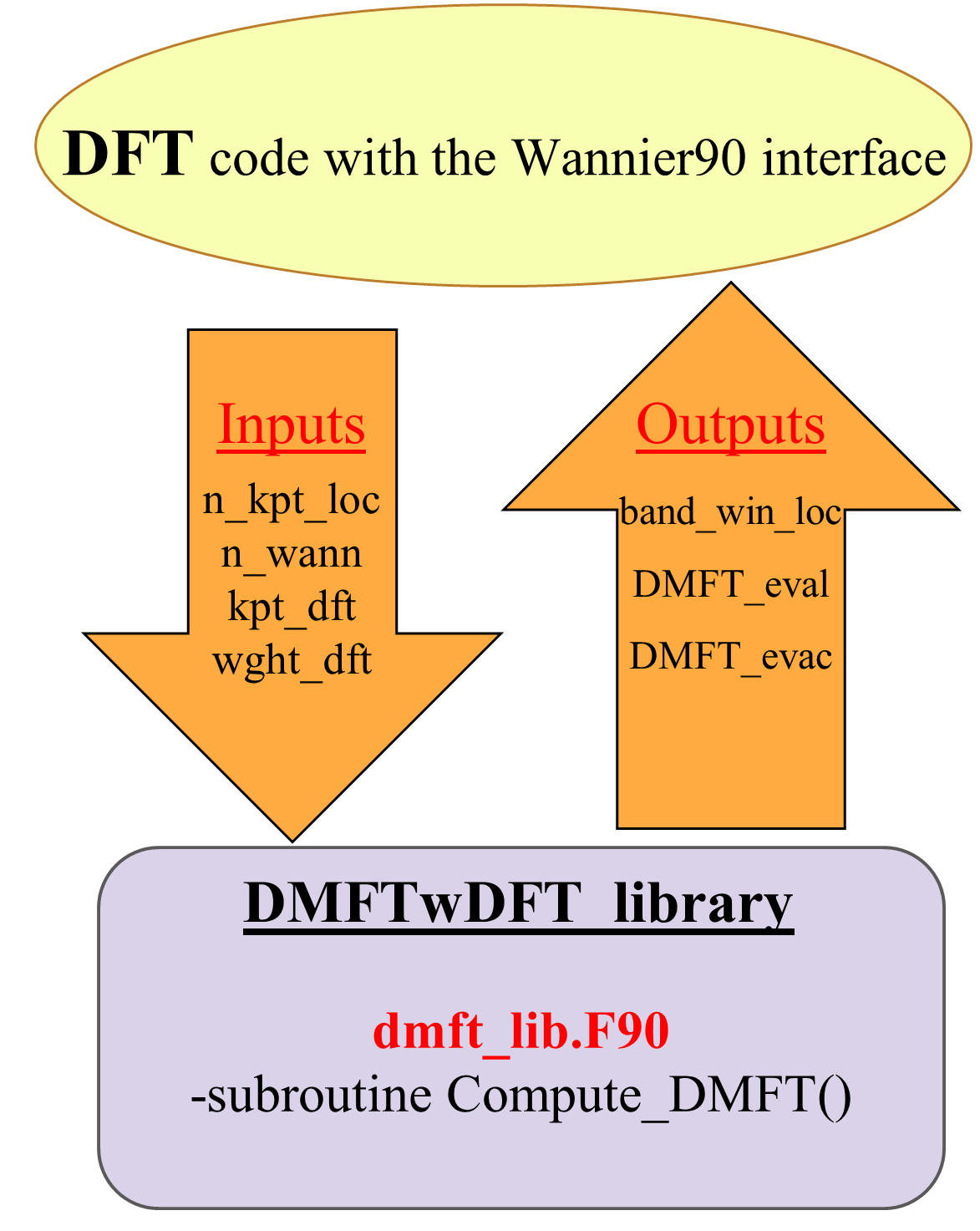}
}
\caption{Schematic of the DMFTwDFT library mode.
} 
\label{fig:Figure2}
\end{figure}

As we explained in the Method section, implementing a fully charge self-consistent solution of DFT+DMFT requires the modification of the DFT package so that $\rho(\mathbf{r})$ can be updated using Eq.$\:$\ref{eq:rho2} from eigenvalues $w^{\mathbf{k}}$ and eigenfunctions $v^{\mathbf{k}}$ of the DMFT occupation matrix $\overline{n}^{\mathbf{k}}$ obtained within our DMFTwDFT code. This can be easily achieved by employing our library mode. A schematic of our library mode is given in Figure ~\ref{fig:Figure2}. 
It is clear from the schematic that any DFT code can be linked to our $library$ mode and call the Fortran subroutine $Compute\_DMFT$ from dmft\_lib.F90 to obtain the outputs of $w^{\mathbf{k}}$ (DMFT\_eval) and $v^{\mathbf{k}}$ (DMFT\_evec) of $\overline{n}^{\mathbf{k}}$.  These outputs are used for modifying the DFT occupation (the Fermi function) and the KS wavefunction to the DMFT occupation $w^{\mathbf{k}}$ and the DMFT wavefunction $\overline{\psi}^{\mathbf{k}}$ to compute new charge density.
The structure and details of input and output parameters used in this subroutine are given as follows:

\noindent subroutine Compute\_DMFT(n\_kpts\_loc, n\_wann, kpt\_dft,\\wght\_ dft, band\_win\_loc, DMFT\_eval, DMFT\_evec)\\

  integer, intent(in) :: n\_kpts\_loc, n\_wann \\
 \indent  real(kind=dp), intent(in) :: kpt\_dft(3, n\_kpts\_loc) \\
 \indent  real(kind=dp), intent(in) :: wght\_dft(n\_kpts loc)\\
 \indent  integer, intent(out) :: band\_win\_loc(2, n\_kpts\_loc) \\
 \indent  real(kind=dp), intent(out) :: DMFT\_eval(n\_wann, n\_kpts\_loc) \\
 \indent  complex(kind=dp), intent(out) :: DMFT\_evec(n\_wann, n\_kpts\_loc ) \\
  
 Here, n\_kpts\_loc is a variable to represent the number of $\mathbf{k}-$points in DFT (It can be either $\mathbf{k}-$points in irreducible Brillouin zone (IBZ) or full BZ). n\_wann is the number of wannier orbitals in an unit cell (the size of the Wannier Hamiltonian). kpt\_dft is the list of $\mathbf{k}-$points with fractional coordinates. wght\_dft is the weight of each $\mathbf{k}-$point in BZ. The sum of weights should be one. band\_win\_loc is the range of the band index (minimum and maximum values) for the Wannier subspace $W$ at each $\mathbf{k}-$point. This will be needed for computing charge density within the subspace. DMFT\_eval is the eigenvalue ($w^{\mathbf{k}}_{\lambda}$) of the DMFT occupancy matrix $n^{\mathbf{k}}$. And DMFT\_evec is the eigenvector ($v^{\mathbf{k}}_{i\lambda}$) of $n^{\mathbf{k}}$. \\

\subsection{Interfacing with various DFT codes} 
\label{sec:interface}

Due to the object oriented nature of the DMFTwDFT code it is possible to interface our library to a variety of DFT codes. Modern DFT codes are interfaced with the Wannier90 package. An initial DFT+wannier90 calculation is all it takes to feed inputs to the DMFT loop. However, for full charge self-consistent DFT+DMFT calculations the DFT codes must be modified to sum the DFT and DMFT charge densities, as the total charge density.  Our DMFTwDFT library mode mentioned in the features section renders this possibility.  Currently we have the full charge DFT+DMFT self-consistent calculation interfaced to VASP~\cite{Park_dc_2014} and the self-consistent DMFT calculation interfaced to Siesta.

\subsubsection{VASP}

The Vienna Ab initio Simulation Package (VASP)\cite{VASP_hafner}, is a package for performing first principles electronic structure calculations using either Vanderbilt pseudopotentials\cite{Pseudo_van}, or the projector augmented wave (PAW) method\cite{PAW}. This code uses a plane wave basis set for the KS orbitals, which has several advantages while performing the electronic structure calculations. The basic methodology employed in VASP is DFT, but it also allows use of post-DFT corrections such as hybrid functionals mixing DFT and Hartree–Fock exchange, many-body perturbation theory (the GW method) and dynamical electronic correlations within the random phase approximation (RPA).
VASP uses fast iterative techniques for the diagonalization of the DFT Hamiltonian and allows to perform total‐energy calculations and structural optimizations for systems with thousands of atoms. Also, ab-initio molecular dynamics simulations for ensembles with a few hundred atoms extending over several tens of picosecond is possible using this code. It also has an interface with Wannier90 code. 
More details on the DMFT implementation and how was interfaced with VASP, using the projector augmented wave method, can be found in Ref.~\cite{Park_dc_2014}.

\subsubsection{Siesta}
Siesta~\cite{Siesta} is a DFT code is a Spanish initiative to perform electronic structure calculations and Ab initio molecular dynamics based on localized basis sets and with a large global community. Unlike VASP, this code uses numerical atomic orbitals as the basis set for the KS orbitals. It also supports interfacing with wannier90 which enables the implementation of our DMFT code to it. As VASP is a commercial code, it was decided to pursue developments with free license codes and Siesta was our first choice. Currently, we have interfaced our DMFTwDFT code with latest Siesta code (SIESTA version 4 or greater) to perform self-consistent DMFT calculations. We have checked our implementation by performing DMFT calculation on SrVO$_{3}$. For SrVO$_{3}$, we have used  Troullier-Martins norm-conserving pseudopotentials scheme as implemented in the SIESTA code.\cite{Troullier19911993} Exchange and correlation functional was approximated using generalized gradient approximation (GGA) of Perdew-Burke-Ernzerhof (PBE)\cite{PBE_functional}, with a plane wave energy cutoff of 600 Ry and a 8$\times$8$\times$8  k-point mesh. We selected the multiple-zeta basis set, split and fixed the orbital confining cut-off to 0.02 Ry. The split norm used was 0.15. Geometry optimizations were performed using the conjugate gradient algorithm until all residual forces were smaller than 0.001 eV/Å. In the following, we discuss the procedure to run DMFT with Siesta. 

\begin{enumerate}
    \item Initially, run siesta to find the Fermi energy and the Total energy. We need this for the DMFT calculation.
For this initial run, the users can comment out the wannier blocks in the .fdf file.

\begin{verbatim}
    siesta<SrVO3.fdf>SrVO3.out
\end{verbatim}

\item Now, the Fermi energy and Total energy can be extracted from the .out file and save it in the files \textbf{DFT\_mu.out} and \textbf{siesta\_ETOT}, respectively. 

\item Run \textbf{RUNDMFT\_siesta.py}.
\begin{verbatim}
    python RUNDMFT_siesta.py
\end{verbatim}
\end{enumerate}

\begin{figure}[ht!]
    \centering
    \includegraphics[width=\linewidth]{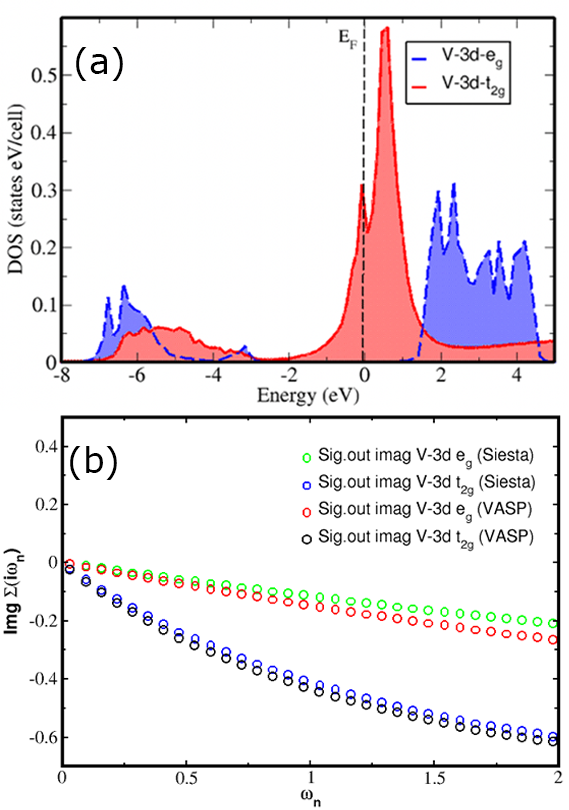}
    \caption{(a)SrVO$_3$ DMFT DOS, and (b) Compared the imaginary part of the self-energy ($\Sigma$) as a function of Matsubara frequencies which was obtained by the Siesta and VASP DMFT calculations.}
    \label{fig:Figure3}
\end{figure}

The convergence procedure is very similar to the VASP case.
Once the calculation is complete, one can perform the post-processing similarly to VASP+DMFT. Obtained DMFT projected density of states  using siesta interface for SrVO$_3$ is shown in Figure ~\ref{fig:Figure3}a. In addition, we have compared the imaginary part of the self energy for V-3d e$_g$ and t$_{2g}$ states as a function of Matsubara frequency obtained by both VASP and siesta DMFT interface. A comparison is shown in Figure ~\ref{fig:Figure3}b. Our results clearly revealed a similar trend for the self-energy on both electronic structure codes. Thus, the obtained correlation is very similar in both siesta and VASP. As of now, siesta+DMFT only performs charge self consistent calculations within DMFT. We are currently working on achieving full charge self consistency within the complete DFT+DMFT loop which will be available in the next code release.

\subsection{Automated scripts}
\label{sec:automation}
We have created a set of Python scripts to automate the complete DFT+DMFT calculation and the post-processing procedure. The Python scripts are in the /scripts directory of our package and should be copied to the /bin directory. The functionality of the scripts are described below.

\subsubsection{DMFT.py}

This script performs the DFT+DMFT calculation. Running DMFT.py -h displays a help message providing instructions. The calculation has the following options:

\begin{itemize}
    \item -dft:\\
    The choice of DFT code. Currently, VASP and Siesta are supported.
    
    \item -relax:\\
    This flag turns on DFT convergence testing. If the forces are not converged a convergence calculation is attempted and if it fails the user is asked to modify convergence parameters. This is useful for vacancy and defect calculations where a force convergence is required after the vacancy or defect is created in order to obtain a relaxed structure to perform DFT+DMFT calculation. Currently supported for VASP. This uses PyChemia~\cite{PyChemia} to check for convergence. The relaxation occurs inside a  ``DFT\_relax'' directory. 
    
    \item -structurename:\\
    DFT codes such as Siesta uses input files that contain the name of the system e.g. SrVO$_3$.fdf. Therefore when performing DFT+DMFT calculations with Siesta this flag is required.
    
    \item -dmft: \\
    This flag performs the DMFT calculation using the results from the DFT calculation if a previous DMFT calculation in the same directory is incomplete. 
    
    \item -hf:\\
    This flag performs the Hartree-Fock (HF) calculation to the correlated orbitals specified in INPUT.py if a previous HF calculation in the same directory is incomplete. 
    
    \item -force:\\
    This flag forces a DMFT or HF calculation even if a previous calculation has been completed. The option to check for completeness is helpful when running many DMFT/HF jobs on a cluster.
    
    \item -kmeshtol:\\
    This controls the tolerance of two k-points belonging to the the same shell in the wannier90 calculation. 
    
The calculations are performed in an automatically generated ``DMFT'' or ``HF'' directory where the script was run from. 
E.g.:
\begin{verbatim}
    $DMFT.py -dft vasp -relax -dmft
    $DMFT.py -dft siesta -structurename 
    SrVO3 -dmft
\end{verbatim}

\end{itemize}

\subsubsection{postDMFT.py}
This script performs analytical continuation, density of states and band structure calculations on the DMFT/HF data. Once the DMFT/HF calculations are complete, this script should be initiated within the ``DMFT'' or ``HF'' directories. Running postDMFT.py -h displays a help message providing instructions. The calculations has the following options:

\begin{itemize}
    \item ac:\\
    This function performs the Analytic Continuation to obtain the Self Energies on the real axis. For detail refer the Appendix. It has the option \texttt{-siglistindx} to specify the last number of Self Energy files to average for the calculation. 
    
    \item dos:\\
    This function performs the partial density of states of the correlated orbitals. It has the following options:
        \begin{itemize}
            \item emin : Minimum energy value for interpolation
            \item emax : Maximum energy value for interpolation
            \item rom : Number of Matsubara Frequency ($\omega$) points
            \item broaden : Broadening of the dos
            \item show : Display the density of states 
            \item elim : The energy range to plot
        \end{itemize}
        
    \item bands:\\
    This function performs the DMFT band structure calculations. It has the following options:
        \begin{itemize}
            \item emin : Minimum energy value for interpolation
            \item emax : Maximum energy value for interpolation
            \item rom : Number of Matsubara Frequency ($\omega$) points
            \item kpband : Number of k-points for band structure calculation
            \item kn : A list of labels for k-points
            \item kp : A list of k-points corresponding to the the k-point labels
            \item plotplain : Flag to plot a plain band structure
            \item plotpartial : Flag to plot a projected band structure
            \item wo : List of Wannier orbitals to project onto the band structure
            \item vlim : Spectral intensity range
            \item show : Display the bands
        \end{itemize}
The projected bands are especially helpful in determining the contribution to bands from different orbitals.     
\end{itemize}

The calculations are stored in directories ac, dos and bands, respectively. 
The following are some example commands to perform post-processing.
e.g.:
\begin{verbatim}
    $postDMFT.py ac -siglistindx 4
    $postDMFT.py dos -show
    $postDMFT.py bands -plotplain
    $postDMFT.py bands -plotpartial -wo 4 5 6
\end{verbatim}


\section{Examples}
\label{sec:example}
 
 In this section, we illustrate the capabilities of our DMFTwDFT code by describing electronic structure of three different correlated systems: (1) SrVO$_{3}$, a paramagnetic $d-$orbital system; (ii) LaNiO$_{3}$, a paramagnetic system with a $p-d$ covalent bonding; (iii) NiO, a charge-transfer insulator. 
 
\subsection{SrVO$_3$}

\begin{figure}[ht!]
    \centering
    \includegraphics[width=0.8\linewidth]{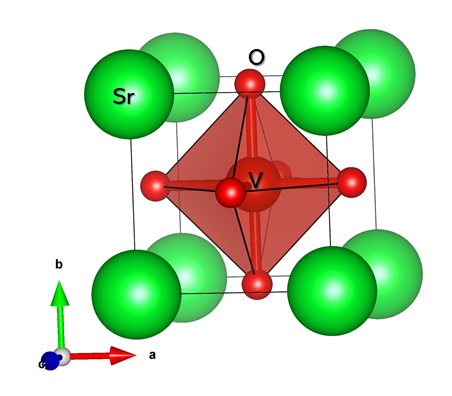}
    \caption{Crystal structure of SrVO$_{3}$.}
    \label{fig:Figure4}
\end{figure}

SrVO$_{3}$ forms a perovksite crystal structure with an ideal cubic Pm$\bar{3}$m symmetry, containing one V ion in the unit cell~\cite{SrVO3_exp_crystal}. In the cell, V-ion is coordinated by 6 oxygens and forms an undistorted VO$_{6}$ octahedra (see Figure~\ref{fig:Figure4}). Due to the cubic symmetry, the $d$ orbitals split into two sets of three t$_{2g}$ and two e$_{g}$ orbitals. The expected electronic configuration for V-ion is 3$d^1$ following from the formal oxidation V$^{4+}$.
One $d$ electron partially occupied in the t$_{2g}$ shell can exhibit the correlation effect. 
Therefore, SrVO$_{3}$ has been the subject of many experimental and theoretical investigations using DFT+DMFT~\cite{SrVO3_e1, SrVO3_e2,SrVO3_e3} as a benchmark material for strong correlation physics.

Previous electronic-structure studies show that SrVO$_{3}$ exhibits pronounced lower and upper Hubbard bands, which cannot be explained by conventional DFT~\cite{SrVO3_LDA1, SrVO3_LDA2, SrVO3_LDA}. 
Here, we perform the DFT+DMFT calculation of SrVO$_3$ using our DMFTwDFT package interfaced with VASP.
To prepare the necessary input for our DMFT run, we begin by performing DFT calculations of SrVO$_{3}$ using the VASP code~\cite{VASP,PAW} with the PBE exchange and correlation functional~\cite{PBE_functional}. The plane-wave energy cutoff was chosen as 600 eV and a 8 $\times$ 8 $\times$ 8 Monkhorst-pack grid~\cite{MPgrid} was used for defining a $k-$space. 

\begin{figure}[ht!]
    \centering
    \includegraphics[width=\linewidth]{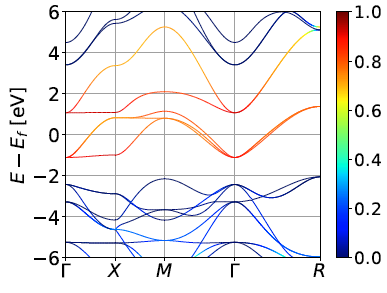}
    \caption{The V 3$d$ orbital projected band structure of SrVO$_3$ (red). The energy range that encloses these projected orbitals is used to construct the Wannier window for the DMFT calculation.}
    \label{fig:Figure5}
\end{figure}

The DFT band-structure plotted using our recently implemented DFT pre/post-processing software PyProcar \cite{HERATH2019107080}, is shown in Figure \ref{fig:Figure5}. 
V-3d(t$_{2g}$) states are located near the Fermi energy between -1.0 eV and 1.0 eV and V-3d(e$_{g}$) bands are between 1.0 eV and 5.0 eV. 
Oxygen 2p states are below -2.0 eV
and mixed with some of V-e$_g$ bands.
While many DFT+DMFT calculations of SrVO$_3$ used the V-t$_{2g}$ orbitals as correlated orbitals and chose the Wannier energy window between -1.0 eV and 1.0 eV \cite{SrVO3_t2g_m}, we use the wider energy window of [-8.0:5.0] eV from the Fermi energy to ensure the highly localized nature of V-$d$ orbitals. 
Therefore, we construct the MLWFs of V-3d and O-2p orbitals using the Wannier90 code~\cite{Wannier90_2012,Updated_wannier90_2014} interfaced to VASP and converge the gauge-dependent spread of Wannier orbitals in 1000 steps.
We compared the original band structure obtained from VASP with our Wannier-interpolated band structure and a good agreement was obtained. 

\begin{figure}[ht!]
    \centering
    \includegraphics[width=\linewidth]{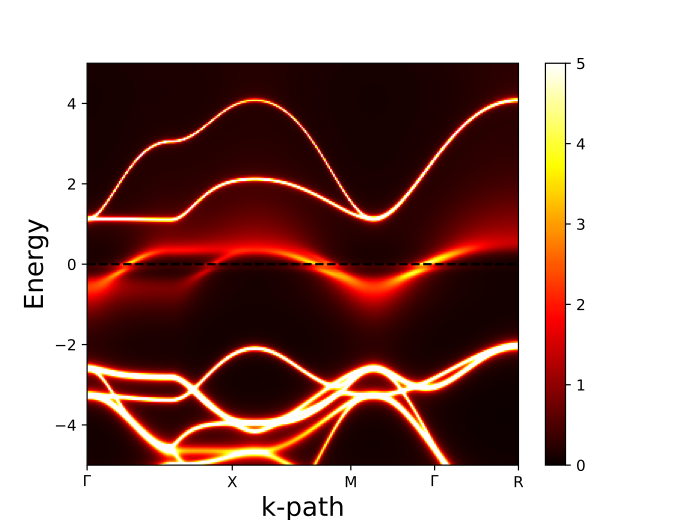}
    \caption{ SrVO$_3$ DMFT band structure along the high symmetry direction of BZ, $\Gamma$ - X - M - $\Gamma$ - R.}
    \label{fig:Figure6}
\end{figure}

\begin{figure}[ht!]
    \centering
    \includegraphics[width=\linewidth]{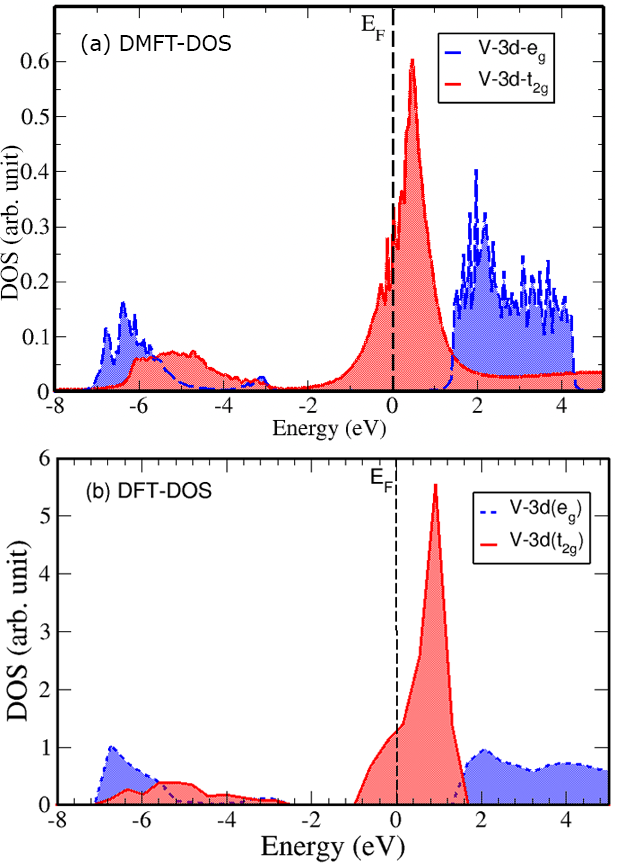}
    \caption{ The projected density of states of SrVO$_3$ for the correlated V 3d-e$_{g}$ and 3d-t$_{2g}$ orbitals obtained using (a) DMFT and (b) DFT calculations. }
    \label{fig:Figure7}
\end{figure}

Our DMFTwDFT code can perform the DMFT calculations of SrVO$_{3}$ by copying required DFT and Wannier90 output files from the DFT run directory using the python script {$"Copy\_input.py"$}.
In the case of SrVO$_{3}$, we treat V as a correlated site and V-3d (t$_{2g}$ and e$_{g}$) orbitals as the correlated orbitals. 
We used CTQMC~\cite{CTQMC} as the DMFT impurity solver using the local Coulomb repulsion U = 5.0 eV, a Hund's exchange coupling J=1.0 eV~\cite{SrVO3_Uvalue}, and temperature as low as 0.01 eV $\approx$ 110K.
To avoid the double counting of the Coulomb interaction, we also used the modified DC correction, DC\_type = 1 (Eq.$\:$\ref{eq:Edc1}) with $\alpha$ = 0.2 for SrVO$_{3}$ DMFT calculation~\cite{Park_dc_2014}. Our results are not sensitive to the choice of different DC corrections as the t$_{2g}$ orbitals are rather separated from other orbitals. We have used  20$\times$20$\times$20 k-points while doing the DMFT calculations.
After DFT+DMFT calculations are converged, we used our 
post-processing script as we discussed earlier and calculated the $k-$resolved spectral function A(k, $\omega$) (Fig.$\:$\ref{fig:Figure6}) and density of states A($\omega$) (Fig.$\:$ \ref{fig:Figure7}a) for SrVO$_{3}$. 

In Figure \ref{fig:Figure6}, we present the DMFT band structure A($k$, $\omega$) plotted following a $k-$path in the BZ for the energy $\omega$ between -5.0 and 5.0 eV. Comparison of our DMFT bands to DFT bands shows that V-t$_{2g}$ bands near the Fermi energy are renormalized and slightly incoherent due to the broadening of the self-energy while e$_g$ and oxygen p bands are very similar to DFT bands. This is also consistent with the fact that the imaginary part of $\Sigma(i\omega_n)$ for both t$_{2g}$ and e$_{g}$ orbitals are very small at $\omega_n=0$ but the t$_{2g}$ orbital has the larger $\Sigma(i\omega_n)$ than the e$_{g}$ orbital as $\omega_n$ increases. 
Our calculated mass renormalization factor for t$_{2g}$ states is 1.7, which is  slightly smaller than the experimental mass renormalization factor that ranges from 1.8-2\cite{Zfactor_srvo3_1_exp,Zfactor_srvo3_2_exp,Zfactor_srvo3_3_exp}. This is a measure of the reduction in quasiparticle weight that can be easily inferred from the slope of the Matsubara-axis self-energy at $\omega$ = 0 [ Z $\approx$ 1/(1-Im$\Sigma(i\omega_n)/\omega_{n}$)]. The reciprocal of Z can be considered as a mass renormalization factor.
The overall DMFT spectra A($\omega$) for SrVO$_{3}$ (Fig.\ref{fig:Figure7}a) exhibits noticeable changes of the local spectral function compared to the DFT DOS (Fig.\ref{fig:Figure7}b). Namely, a narrowing of the t$_{2g}$ quasi-particle (QP) bands close to the Fermi level occurs and the QP spectral weights move to lower and upper Hubbard bands, whose positions are dependent on the choice of the Hubbard U~\cite{Vollhardt_SrVO3}.



\subsection{LaNiO$_3$}

\begin{figure}[ht!]
    \centering
    \includegraphics[width=0.8\linewidth]{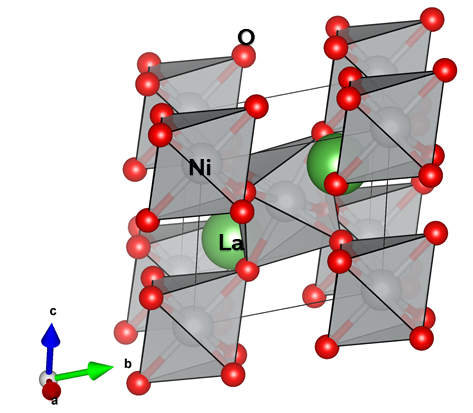}
    \caption{Crystal structure of LaNiO$_3$ }
    \label{fig:Figure8}
\end{figure}

LaNiO$_3$ is the only known paramagnetic and metallic compound among the rare-earth nickelate series down to very lowest temperatures~\cite{LaNiO3_Metal_para1,LaNiO3_Metal_para2}. In spite of the metallic state, LaNiO$_3$ resides on very close to the Mott insulator phase boundary, and moreover various experimental probes including Angle-resolved Photoemission Spectroscopy (ARPES)~\cite{PRB.79.115122}, optical conductivity~\cite{PRB.83.075125,PRB.82.165112}, and thermo-dynamical measurements~\cite{PRB.48.1112} show that LaNiO$_3$ is still correlated. 
Moreover, recent discovery for superconductivity in the infinite-layer rare-earth nickelate has also resurged the correlation effects in nickelates \cite{nature_NdNiO2_sup,Hepting_2020}.
LaNiO$_3$ has a rhombodedral symmetry, described by R$\bar{3}$c space group. The crystal structure of LaNiO$_3$ is shown in Figure~\ref{fig:Figure8}. 
 
Several DFT calculations have been performed to study the electronic and lattice properties of LaNiO$_3$. 
Guo et al~\cite{Structure_Rhombo_cubic_laNiO3} found
the A1g Raman mode, whose frequency is sensitive to the electronic band structure method, is a useful signature to characterize the octahedral rotations in rhombohedral LaNiO$_3$.
The authors also found that DFT with local spin density approximation (LSDA) accurately reproduces the delocalized nature of the valence states in LaNiO$_3$ and gives the best agreement with the available experimental data~\cite{LaNiO3_PES} for the electronic structure. Surprisingly, they have found that NiO$_6$ rotation angle $\theta$, the order parameter characterizing the structural phase transition in LaNiO$_3$, is highly sensitive to the exchange correlation (XC) functional. Even calculations with the same functional but different pseudopotentials (e.g., the LSDA calculations performed with the VASP and Quantum espresso codes) yield $\theta$ values with obvious differences. Therefore, the authors suggest that an accurate and comprehensive study of various theoretical approximations for the description of octahedra rotation angles in rhombohedral perovskite oxides is needed which remains a mystery until now. This implies an accurate calculation of the forces is indeed necessary to obtained the accurate structure of LaNiO$_3$, which remain the next target of the current project. Calculation of the forces using DMFT will be available with the next release of DMFTwDFT code.

\begin{figure}[ht!]
    \centering
    \includegraphics[width=\linewidth]{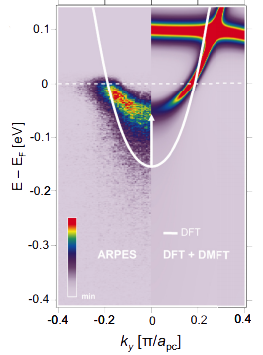}
    \caption{Comparison of ARPES spectrum (left side) and DFT + DMFT spectral function (right side), both along the momentum space cut ($\pi$/2a$_{pc}$, k$_{y}$ , 0.7$\pi$/a$_{pc}$) (a$_{pc}$ is the lattice constant of the primitive cell) to the DFT band structure, calculated in the bulk R$\bar{3}$c structure (white line).Adapted from Ref.\cite{Nowadnik_ARPES_LaNiO3}}
    \label{fig:Figure9}
\end{figure}

Nevertheless, recently, Nowadnik, et al\cite{Nowadnik_ARPES_LaNiO3} performed DFT+DMFT calculations in LaNiO$_3$ using the early version of our DMFTwDFT code and quantified the electronic correlation strength by comparing with ARPES measurements. Their results established that the LaNiO$_3$ is indeed a moderately correlated Fermi liquid. Obtained DFT+DMFT spectral-function of LaNiO$_3$ along the momentum space cut ($\pi$/2a$_{pc}$, k$_{y}$, 0.7$\pi$/a$_{pc}$) which is in good agreement with existing ARPES data as shown in Figure \ref{fig:Figure9} (COPY WRITE for the Figure ~\ref{fig:Figure9} is provided by the American Physical Society and  Scientific Publishing and Remittance Integration services (SciPris)). 
In both experiment (left side) and DFT + DMFT (right side), there is a shallow band crossing the Fermi level with a band bottom at ~50 meV and a Fermi level crossing at ky = -0.2$\pi$/a$_{pc}$. This band is substantially renormalized by electron correlations relative to the rhombohedral DFT band structure. By considering the frequency derivative of the electron self-energy obtained by the DMFT, authors have also calculated the theoretical mass renormalization for LaNiO$_{3}$, which is 3.5. This implies, $m^{*}_{DMFT}$ = 3.5 $\times m^{*}_{band}$, where $m^{*}_{band}$ is the electron effective mass calculated from DFT and $m^{*}_{DMFT}$ is the mass approximated from the DMFT spectral function. This is in a good agreement with the soft X-ray ARPES mass renormalization value which is 3. 
In the following, for completeness, we discuss the full DFT+DMFT spectral function along the high-symmetry points of the BZ and density of states of LaNiO$_3$ obtained using our DMFTwDFT code. 

\begin{figure}[ht!]
    \centering
    \includegraphics[width=\linewidth]{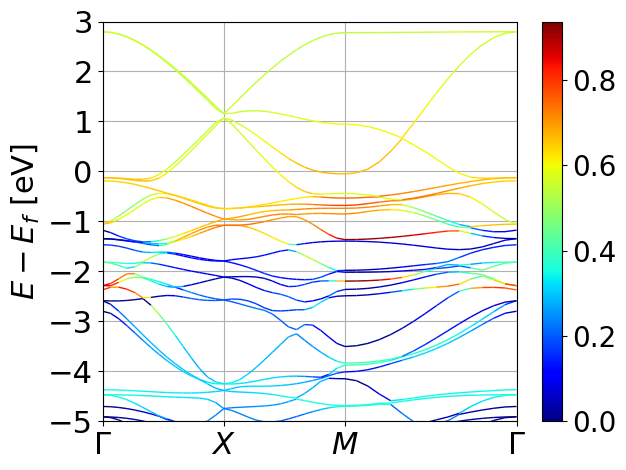}
    \caption{Ni-3d orbital projected DFT band structure of LaNiO$_3$ (orange). The energy range that encloses these projected orbitals is used to construct the Wannier window for the DMFT calculation.}
    \label{fig:Figure10}
\end{figure}

Similar to the SrVO$_3$ case, 
we first perform DFT calculations using the VASP code~\cite{VASP} by consider the bulk LaNiO$_{3}$ in a rhombohedral crystal structure (space group R$\bar{3}$c, a$^{-}$a$^{-}$a$^{-}$ in Glazer notation)~\cite{Structure_Rhombo_cubic_laNiO3}. As we mentioned earlier, by comparing different DFT functionals Guo et al. reach to a conclusion that the LDA functional is the best functional for LaNiO$_{3}$~\cite{Structure_Rhombo_cubic_laNiO3}. Thus, to treat exchange-correlation in LaNiO$_3$ we have also employed LDA functional and core electrons were defined within the PAW methodology~\cite{PAW} as implemented in the VASP code. We have used a 600 eV plane-wave cutoff and, for structural relaxations, a force convergence tolerance of 2 meV/A.  We used 8 $\times$ 8 $\times$ 8 $k-$points meshes. The obtained DFT band-structure along the high symmetry points in BZ is shown in Figure~\ref{fig:Figure10}. 

\begin{figure}[ht!]
    \centering
    \includegraphics[width=\linewidth]{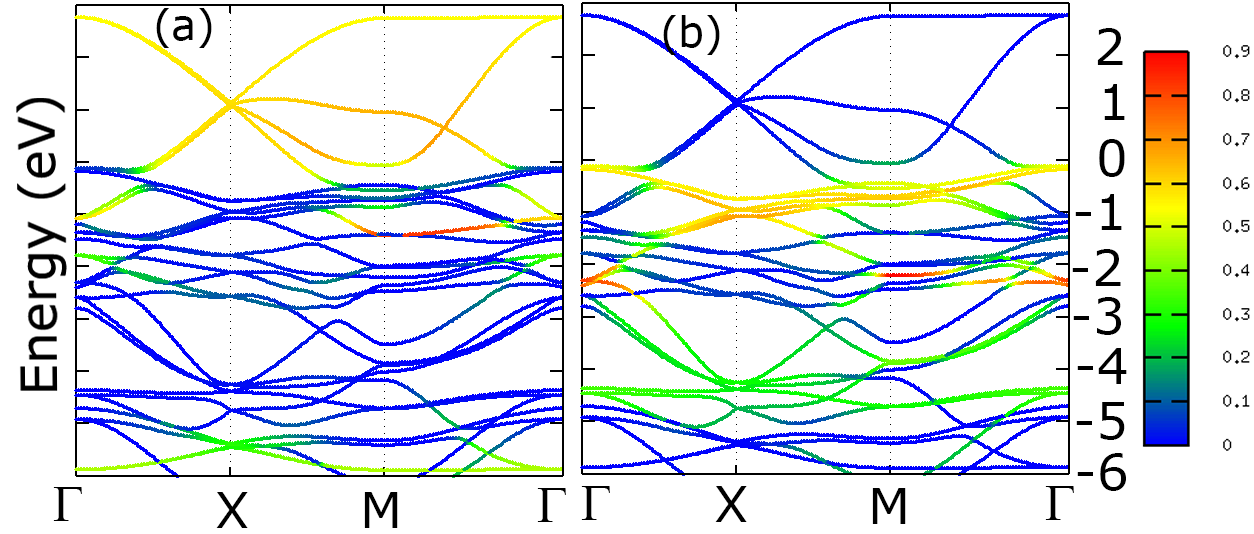}
    \caption{Wannier-interpolated band structure with (a) Ni-3d(e$_{g}$) and (b) Ni-3d(t$_{2g}$) states of LaNiO$_3$. The zero energy is the Fermi level. }
    \label{fig:Figure11}
\end{figure}

From the DFT band structure, we construct the Ni 3d and O 2p MLWFs using the Wannier90~\cite{Wannier90_2012, Updated_wannier90_2014} code over the $\sim$11 eV range ([-8:3.2]eV from the Fermi energy) spanned by the p-d manifold as the hybridization window and treated Ni d orbitals as the correlated orbitals. The obtained Wannier interpolated band-structure is shown in Figure ~\ref{fig:Figure11} which is in good agreement with the DFT band structure as shown in Figure ~\ref{fig:Figure10} . In LaNiO$_3$, the nominal configuration is Ni d$^7$ with fully filled t$_{2g}$ (Figure ~\ref{fig:Figure11}a) band and quarter filled e$_{g}$ band (Figure ~\ref{fig:Figure11}b). 
Our results clearly revealed that Ni 3d t$_{2g}$ bands are completely filled at energy between -2.0 eV and 0.0 eV, and e$_g$ bands are partially filled in the range of -1.0 eV and 3.0 eV. Different from the SrVO$_3$ band structure, oxygen 2p states are much closer to the Fermi energy and e$_g$ bands are covalently mixed with O p states. Therefore, including all Ni $d$ and O $p$ orbitals in the hybridization window will be important for a better description of LaNiO$_3$. 

\begin{figure}[ht!]
    \centering
    \includegraphics[width=\linewidth]{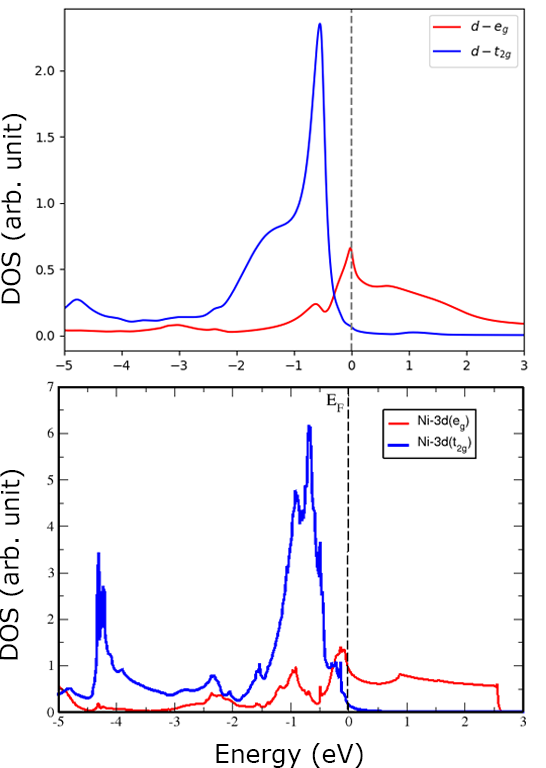}
    \caption{The projected density of states of LaNiO$_3$ for the correlated Ni 3d-$e_{g}$ and 3d-$t_{2g}$ orbitals obtained by (a) DMFT and (b) DFT calculations. }
    \label{fig:Figure12}
\end{figure}

\begin{figure}[ht!]
    \centering
    \includegraphics[width=\linewidth]{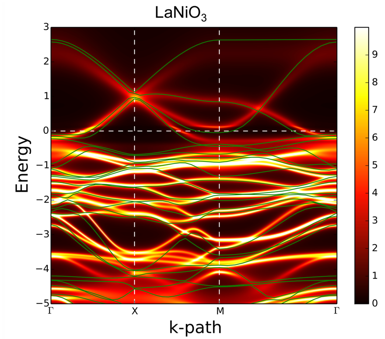}
    \caption{The DFT (green) and DFT+DMFT (red) bandstructures of LaNiO$_3$}
    \label{fig:Figure13}
\end{figure}

DMFT calculations were performed such that the
correlated subspace is treated using the Hubbard interaction strength U=5eV and the Hund’s interaction J=1eV. For the double-counting correction required in DFT+DMFT, we use the parametrization of U as U $-$ $\alpha$ (DC\_type = 1. $\alpha$ = 0.2), which was found to correctly reproduce the pressure phase diagram of the RNiO$_3$~\cite{LaNiO3_press}. 
The DMFT impurity problem is solved using the CTQMC method~\cite{CTQMC} with the temperature set to 0.01 eV $\approx$ 110 K. 
Using our post processing scripts, we obtained the Ni-3d projected DOS (see Figure ~\ref{fig:Figure12}a) and $\mathbf{k}-$resolve spectral function along the high-symmetry direction of the BZ (see Figure ~\ref{fig:Figure13}) for LaNiO$_{3}$. 
Both results clearly reveal that moderate correlation is associated with the renormalized Ni-e$_g$ manifold near the Fermi energy while t$_{2g}$ state is almost filled and broader than the DFT DOS, Figure ~\ref{fig:Figure12}b. 


\subsection{NiO}

  \begin{figure}[ht!]
    \centering
    \includegraphics[width=\linewidth]{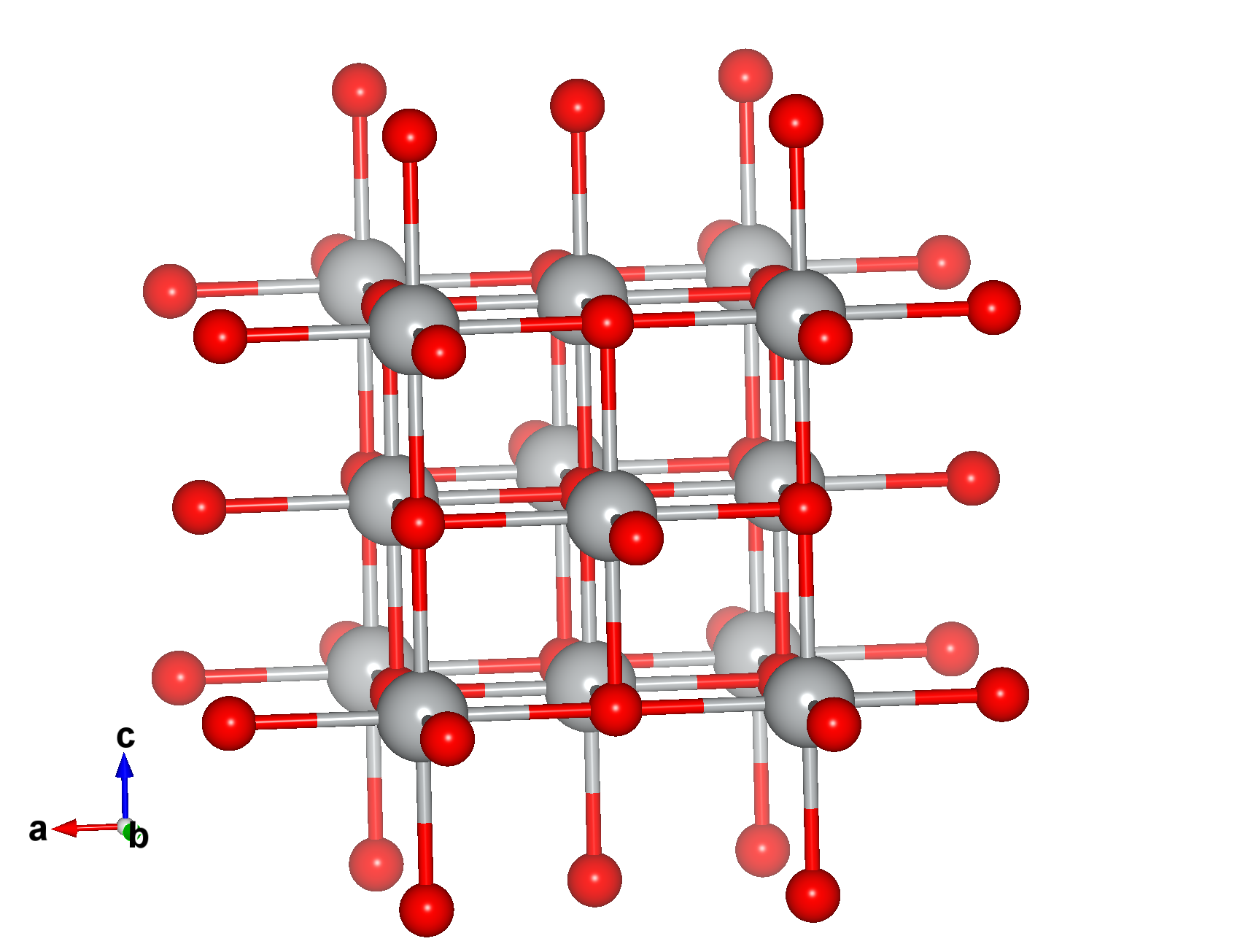}
    \caption{Crystal structure of NiO. Ni and O atoms are in gray and red color, respectively. Rocksalt crystal of NiO, visualized as NiO$_{6}$ octahedral networks.}
    \label{fig:Figure14}
\end{figure}
Crystal structure of NiO adopts a cubic rock-salt (B1) structure with octahedral Ni$^{+2}$ and O$^{-2}$ sites as shown in Figure ~\ref{fig:Figure14}.
NiO has been extensively studied experimentally and theoretically. It is a strongly correlated charge-transfer insulator with a large insulating gap of 4.3 eV and antiferromagnetic (AFM) ordering temperature (T$_N$) of  = 523K~\cite{nio_band_gap2,Allen_nio_band_gap,cox2010_nio_band_gap1}. Conventional band theories cannot explain this large gap and predicted wrongly NiO to be metallic~\cite{nio_con_dft}. Spin-polarized DFT calculations using local spin density approximation (LSDA) found the AFM insulating state but obtained local magnetic moment at Ni sites are considerably smaller than the experimental values~\cite{sp_dft_nio}. There has been several studies which made an effort to go beyond DFT including self-interaction-corrected density functional theory (SIC-DFT)~\cite{SIC_nio}, the LDA+U method~\cite{LAD+U_nio}, and the GW approximation~\cite{Gw_nio,GW1_niO}. These methods represent some corrections of the single-particle Kohn-Sham potential and provide the improvements over the L(S)DA results for the values of the energy gap and local moments. It is important to note that in these methods the self-energy is static and hence does not take dynamical correlation effects into account adequately. Also, different GW schemes give quite different results regarding the value of the insulating gap and the relative position of the energy bands~\cite{mark_ScGW_niO,GW_nio_Stefen,GW1_niO,Gw_nio}.

  \begin{figure}[ht!]
    \centering
    \includegraphics[width=\linewidth]{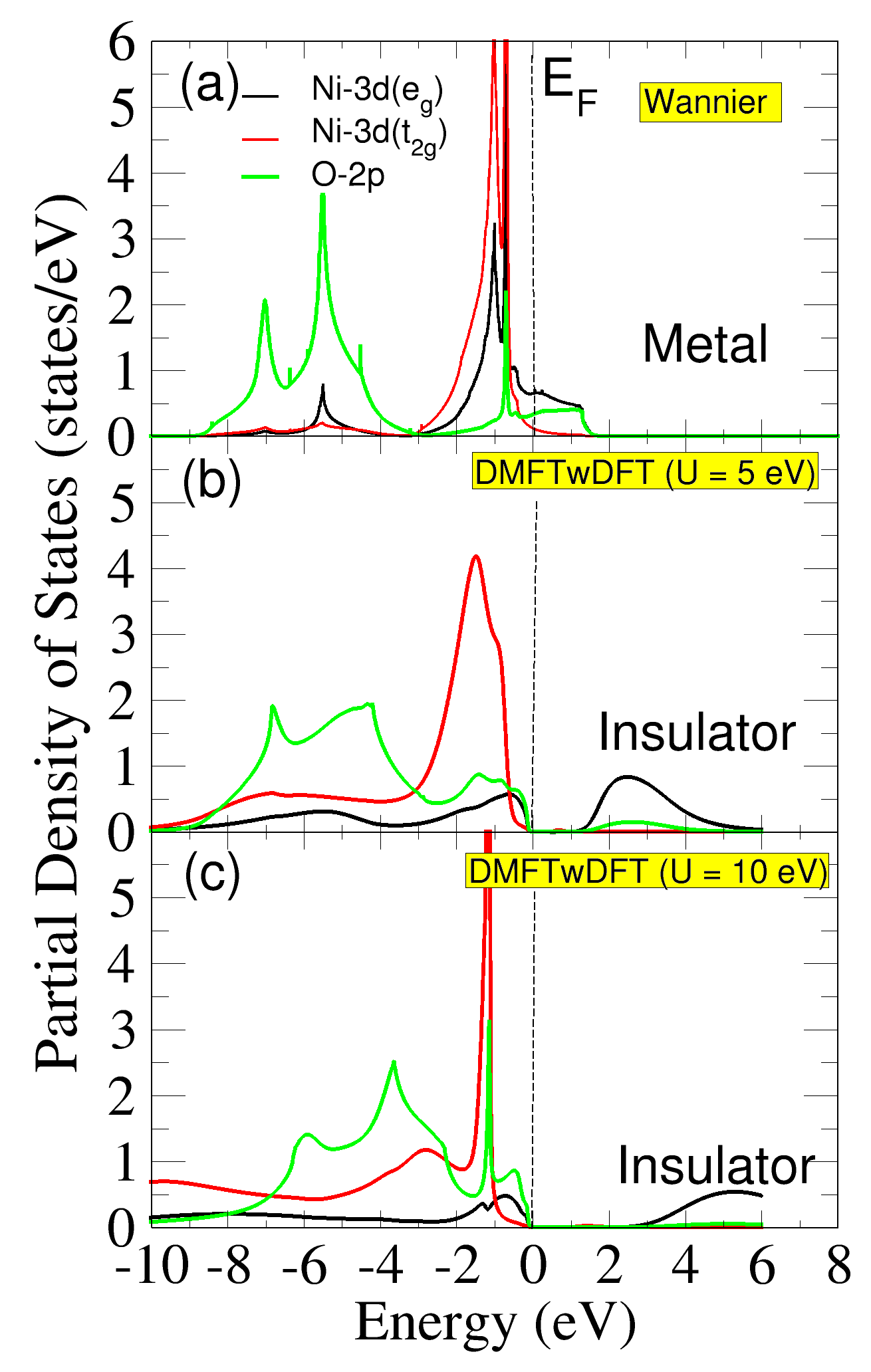}
    \caption{Compared the projected density of states of NiO obtained by Wannier (a) and DMFT calculation for U = 5 eV (b) and 10 eV (c). }
    \label{fig:Figure15}
\end{figure}

Experimentally, it has been found that both the local magnetic moment and the energy band gap for NiO are essentially unchanged even above the Neel temperature~\cite{nio_para_nochange}. Also, in other experiments, it has been found that long range magnetic order do not has significance influence on the valence band photo emission spectra~\cite{nio_magnetic_influ} and the electron density distributions~\cite{nio_charge_density}. Therefore, the role of magnetism and correlation is still not clear in NiO. To resolve the above controversy, Ren \textit{et al}~\cite{Vollhardt_nio} has employed the LDA+DMFT approach and concluded that a large insulating gap in NiO is due to the strong electronic correlations in the paramagnetic state. They also suggest that AFM long-range order has no significant influence on the electronic structure of NiO. Recently, using ab-initio LQSGW + DMFT, Kang \textit{et al}~\cite{kang2019nature} claimed that they have resolved the long standing controversy of two-peak structure in the valence band photoemission spectra of NiO~\cite{nio_EXP2, Allen_nio_band_gap}. They suggest that, the two peak structure is driven by the concerted effect of AFM ordering and inter-site electron hopping. Surprisingly, the two peak structure has also been obtained by Ren \textit{et al}~\cite{Vollhardt_nio} where authors used LDA+DMFT approach for T = 1160 K and U = 8 eV, J = 1 eV. Thus, although considerable progress was made in the theoretical understanding of NiO from first principles, several important issues are still open. This is certainly not the goal of the present manuscript. However, in the following we will discuss our DFT+DMFT results for NiO and compare with other existing DMFT results and experiments. 
 
  \begin{figure}[ht!]
    \centering
    \includegraphics[width=\linewidth]{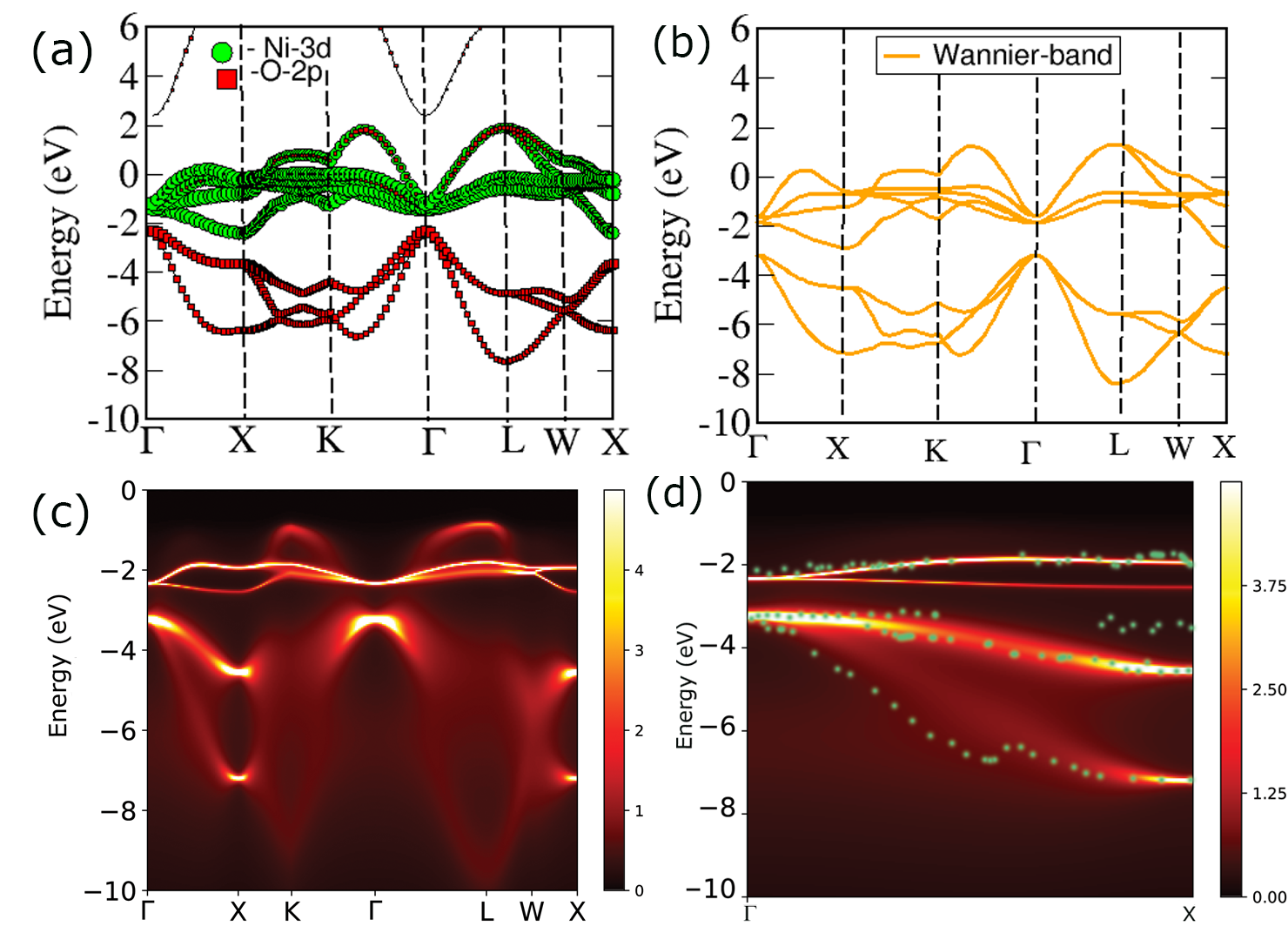}
    \caption{(a) NiO DFT fatbands, (b) Ni-3d and O-2p projected Wannier interpolated band structure, (c) a DMFT momentum-resolve spectral function along the high symmetry direction of the BZ obtained using U=10eV and J=1eV, and (d) the spectral function of NiO compared with the ARPES data (green dots) obtained by Shen et al\cite{Shen}.  }
    \label{fig:Figure16}
\end{figure}
 
 To prepare the required input for our DMFT run, we first performed the first principles DFT calculations followed by DMFT calculations. First principles DFT calculations of NiO were performed using VASP code~\cite{VASP}. We have used PAW pseudopotantials\cite{PAW}, and PBEsol\cite{PBEsol} exchange and correlation functional for NiO, which as per our knowledge is not tested before. In practice, while PBEsol provides better crystal cell parameters than PBE\cite{NiO_PBESol}, it is not clear that all properties are improved overall. Here we use PBEsol to demonstrate that DMFT is less sensible to the exchange correlation details and provide quite similar results than those obtained from LDA and PBE. The plane wave energy cutoff was chosen 600 eV for NiO. 6 $\times$ 6 $\times$ 6 Monkhorst-pack k-point grids\cite{MPgrid} were used for reciprocal space integration.  After obtaining the self-consistent ground state, we perform a self-consistent calculation on uniform grid of k-points without changing the potential.

  \begin{figure}[ht!]
    \centering
    \includegraphics[width=\linewidth]{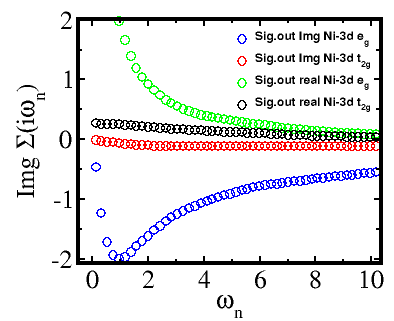}
    \caption{Imaginary part of the Ni self energy of NiO for Ni-3d e$_{g}$ and t$_{2g}$ states as a function of Matsubara frequencies obtained using U=10eV and J=1eV.  }
    \label{fig:Figure17}
\end{figure}

 Obtained DFT density of states clearly revealed that NiO is metal with Ni-3d(eg) state and O-2p states are strongly mixed due to the nature of the charge-transfer insulator (refer Figure ~\ref{fig:Figure15}a).
 Ni-3d(t$_{2g}$) states are fully filled, and Ni-3d(eg) and O-2p states are widely formed between -8 eV and 2 eV (refer Figure ~\ref{fig:Figure15}a). We also construct MLWFs as we did for other examples. Projections onto atom centred Ni-3d and O-2p function are used to construct the initial guess, and further Wannier90 is used to obtained the MLWFs\cite{Wannier90_2012,Updated_wannier90_2014}. To obtained the correct energy window of [-8.0 - 3.0]eV for wannier-function, we compared the original band structure obtained from DFT calculation with our Wannier-interpolated band structure as shown in Figure ~\ref{fig:Figure16}a and b, respectively.
 In the case of NiO, we treat Ni as a correlated site and Ni-3d (t$_{2g}$ and e$_{g}$) orbitals as the correlated orbitals. The Coulomb interaction U=5.0 eV as well as 10.0 eV and a Hund's exchange coupling J=1.0 eV are used. Previously, suggested value of Hubbard U on Ni-3d orbital vary in the range of 4-10 eV.\cite{NiO_U1,NiO_U2,kang2019nature} We have used the FLL double counting correction (DC\_type =1 and $\alpha$ = 0.0).
 Temperature as low as 0.03 eV $\approx$ 300K are used and set of 24$\times$24$\times$24 k-points have been used for the DMFT calculations.

\begin{figure}[ht!]
    \centering
    \includegraphics[width=\linewidth]{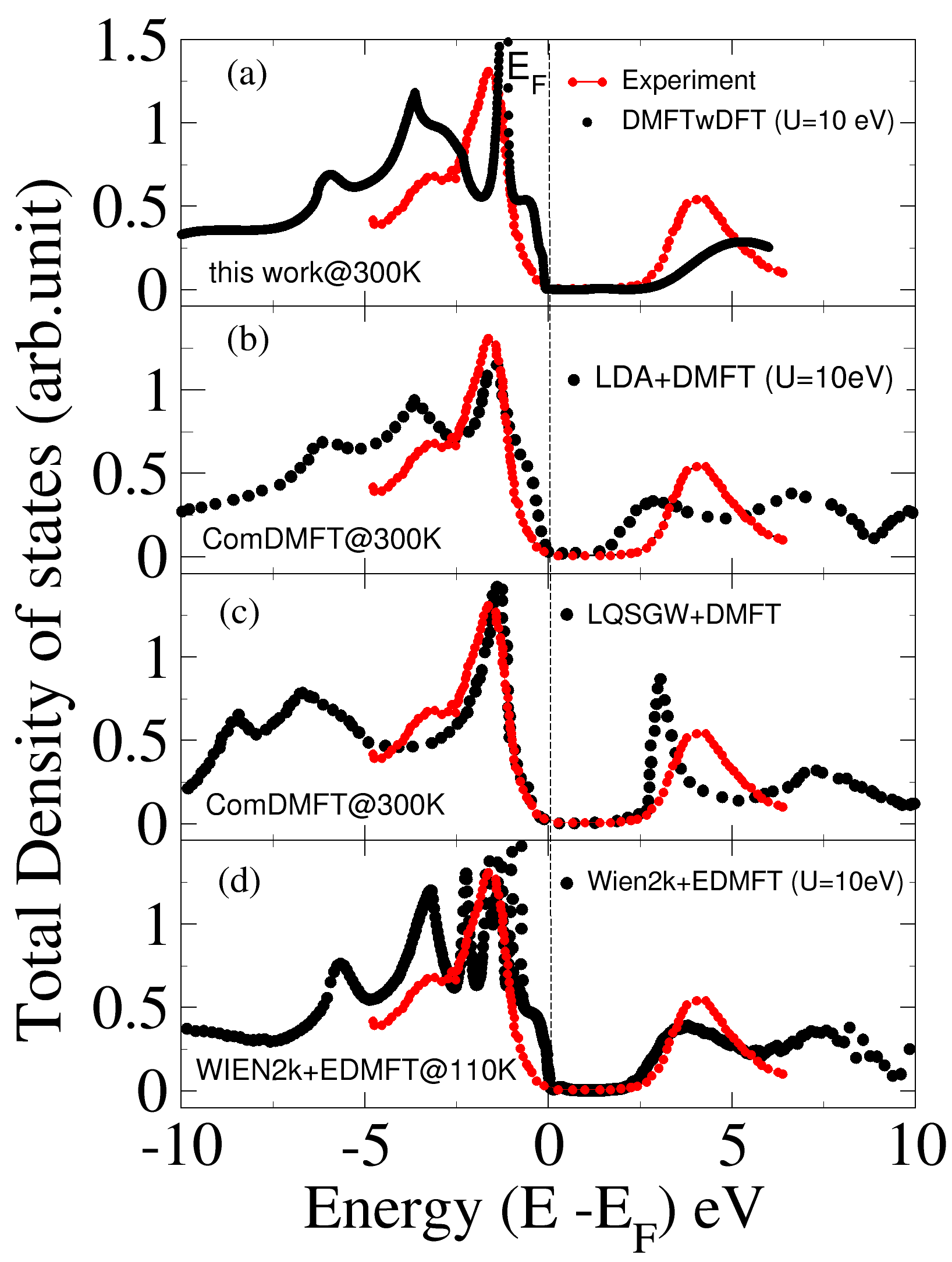}
    \caption{Compared the total density of states of NiO obtained by DMFTwDFT code (this work) with existing experimental PES as well as with other DFT+DMFT codes such as WIEN2k+EDMFT and ComDMFT\cite{ComDMFT} (LQSCGW + DMFT and LDA+DMFT).}
    \label{fig:Figure18}
\end{figure}

 Using the post-processing tools, we have calculated the DMFT density of states as well as k-resolved spectral function A(k, $\omega$) for NiO. Interestingly, our DMFT results clearly revealed that including the dynamical correlation leads to an insulating state in NiO as shown in Figure ~\ref{fig:Figure15}b and c. 
 As U increases, the insulating gap gets larger and Ni 3d states below the Fermi energy are strongly hybridized with O-2p orbitals.
 As a result, both e$_g$ and t$_{2g}$ orbitals exhibit longer tails below the Fermi energy and the separation between O-2p and Ni-3d states has been reduced at U=10eV (see Figure~\ref{fig:Figure15}c). 
 Compared to the experimental spectra (see Figure~\ref{fig:Figure18}), U=10eV produces better agreements with experiments than U=5eV.
 Moreover, a divergent nature of the self-energy ($\Sigma(i\omega_n)\sim 1/(i\omega_n-\omega_0)$ for Ni 3d-e$_{g}$ states for low Matsubara frequency clearly indicate that the e$_{g}$ state develops a Mott gap while t$_{2g}$ states behaves as a band insulator due to much smaller $\Sigma(i\omega_n)$ (refer to Figure~\ref{fig:Figure17}). 
 
 In Figure~\ref{fig:Figure16}c and d, we also present the $\mathbf{k}-$resolved DMFT spectra for NiO along the high symmetry direction of BZ. In Figure~\ref{fig:Figure16}c, dispersionless Ni t$_{2g}$ bands are located near -2eV and the mixture of Ni e$_g$ and O bands are dispersing slightly above -2eV. Below -4eV, the bands are mostly O-2p characters while they are also strongly mixed with t$_{2g}$ and e$_g$ bands, therefore those mixed bands are strongly incoherent. We have also compared our obtained DMFT band structure with the experimental band structure along the $\Gamma$ - X direction in Figure ~\ref{fig:Figure16}d. The obtained result is in very good agreement with the experimental result~\cite{Shen}.

We also compared our DMFT density of the states with existing experimental data~\cite{Allen_nio_band_gap,nio_EXP2} as well as theoretical results obtained using other DMFT codes in Figure ~\ref{fig:Figure16}. We compared our U=10eV result, which is in good agreement with experimental results. In Figure~\ref{fig:Figure18}(b-d), we also presented total density of states of NiO obtained by other equally important DMFT tools including WIEN2k+EDMFT~\cite{EDMFT}, LQSCGW + DMFT~\cite{ComDMFT}, and LDA+DMFT~\cite{ComDMFT}. Note that the presented DMFT spectra of LQSCGW + DMFT and LDA +DMFT was taken from the example directory of ComDMFT~\cite{ComDMFT}. Surprisingly, we observed different DMFT tools at the same U and J values give slight variations of energy-band gaps for NiO. However, the overall features of peak positions are in good agreement with experiment. Namely, a small bump with the Ni e$_g$ and O-2p mixture below the Fermi energy, the t$_{2g}$ peak at around -2eV, and the O-2p peak below the t$_{2g}$ state are all consistent in different codes. The peak positions are also comparable to experimental data. 

\section{Conclusions} 
We have implemented a computational package (DMFTwDFT) combining the DMFT methodology to different DFT codes to improve our theoretical description of SCMs. Our package can perform a charge-self-consistent
DFT+DMFT calculation adopting Wannier functions as localized orbitals, which are constructed from the Wannier90 package interfaced to many DFT codes. 
Our current implementation has been interfaced to two different DFT codes, VASP (a commercial package) and SIESTA (a non-commercial package). 
We also provide a library mode to link our package to different DFT codes without much modifications. 

We applied our package to compute the band structure and the density of states of different SCMs, namely SrVO$_{3}$, LaNiO$_{3}$, and NiO. Results of SrVO$_{3}$ obtained from both VASP+DMFT and SIESTA+DMFT are in good agreement showing the moderate mass enhancement of t${2g}$ orbitals near the Fermi energy. Both SrVO$_{3}$ and LaNiO$_{3}$ are correlated metallic systems and the quasi-particle band renormalizations near the Fermi energy are captured properly by DMFT, consistently with experiments. 
Moreover, our NiO calculation shows that Ni e$_g$ orbital develops a Mott gap near the Fermi energy (the divergence of the self-energy).
and band structures below the Fermi energy are consistent with ARPES measurements.
Calculations of NiO with different DFT+DMFT codes with the same $U$ and $J$ parameters also exhibit the similar density of states compared to our results.



\textbf{Acknowledgements}
The authors thank Javier Junquera from Cantabria University
for insightful discussions and help with the interface with Siesta. This work is supported by the NSF SI2-SSE Grant 1740112. 
Uthpala Herath and Aldo H. Romero are also supported by DMREF-NSF 1434897 and DOE DE-SC0016176 grants.
Xingyu Liao is supported by ACS-PRF grant 60617.
This work used the XSEDE which is supported by National Science Foundation grant number ACI-1053575 and allocation number TG-PHY190035. The authors also acknowledge the support from the Texas Advances Computer Center (with the Stampede2 and Bridges supercomputers). We acknowledge the West Virginia University supercomputing clusters (Spruce Knob and Thorny Flat) and the Advanced Cyberinfrastructure for Education and Research (ACER) group at the University of Illinois at Chicago for providing HPC resources which were used for the development of the library. 

\bigskip

\section{Appendix}

\subsection{Input file}

Here, the input parameters for DMFT+DFT calculations are described in INPUT.py file. 

\begin{itemize}
    \item {\bf Niter}: The value of Niter defines the maximum number of iterations used in the DFT+DMFT loop.
    \item {\bf Nit}: The value of Nit defines the maximum number of iterations of the DMFT self-consistent calculations.
    \item {\bf Ndft}: The value of Ndft defines the maximum number of iterations of the DFT self-consistent calculations.
    \item {\bf n-tot}: The value of n-tot defines the total number of electrons in the Wannier subspace. For example, in LaNiO$_{3}$, we have 2 Ni$^{3+}$ ions with 7 d-electrons/Ni and 6 O$^{2-}$ ions with 6 p-electrons/O, therefore totally 50 electrons.
    \item {\bf nf}:  The value of nf defines the nominal occupancy of d- or f- electrons in a correlated atom. This is used for the initial guess of self-energy. One can initialize it as the DFT occupancy of the correlated atom or the nominal electron number.
    \item {\bf nspin}:\\
    Default value: 1\\
    The value of nspin defines the number of spins in DMFT calculations. nspin=2 means the spin-polarized calculation.  It is important to note that to start spin-polarized DMFT calculations we still need non-spin-polarized DFT calculations. 
    \item {\bf atomnames}: The value of atomnames defines the name of atoms where the Wannier projection will be taken.
    \item {\bf orbs}:  The value of orbs defines the name of Wannier orbitals in each atom in atomnames.
    \item {\bf L-rot}: L-rot defines whether the wannier projection axis will be rotated along the local axis. 1: rotated, 0: non-rotated.
    \item {\bf cor-at}: cor-at is the list of all correlated atoms in the material. For LaNiO$_{3}$, it is Ni1 and Ni2. 
    \item {\bf cor-orb}: cor-orb is the list of all correlated orbitals in each correlated atom.
    \item {\bf U}: U is the value of local Hubbard interaction on correlated atoms.
    \item {\bf J}: J is the value of the Hund's coupling.
    \item  {\bf alpha}: \\
    Default value: 0\\
    $\alpha$ is the double counting correction parameter. alpha=0 means the conventional double counting in the fully localized limit (FLL). 
    \item{\bf mix-sig}:
    Default value: 0.2
    mix-sig is the mixing parameter between the previous and the current self energies.
    \item {\bf q}: q is the number of k-points while doing DMFT self-consistent calculations. We are using the Wannier interpolation technique, therefore large numbers of q-points will be possible. Usually 2-3 times larger than DFT k-points will be necessary for better convergence.
    \item {\bf ewin}: ewin is the energy window for Wannier projection with respect to the DFT fermi energy.
\end{itemize}

\subsection{Wannier90 calculation}

During the DFT+DMFT calculation, one should generate the hybridization subspace of the MLWFs by using the "Projection" technique adopted in the wannier90 package (www.wannier.org) within a certain energy window. 
For example, in the LaNiO$_3$ case with the rhombohedral structure, the wannier90.win file can be constructed as follows.\\

\noindent dis\_win\_ min  = -0.3014 \\
dis\_win\_max  = 10.6986 \\
num\_wann   = 28 \\
num\_iter   = 100 \\
\newline
begin projections \\
f=0.00000000,0.00000000,0.00000000:l=2:z=-0.29712138,0.77011009,0.56449032:x=-0.52491861,-0.62540726,0.57692379 \\
f=0.50000000,0.50000000,0.50000000:l=2:,\\
z=0.29712138,0.77011009,-0.56449032:x=0.52491861,-0.62540726,-0.57692379 \\
O:p \\
end projections \\

Here, the energy window is specified by [dis\_win\_min : dis\_win\_max], and it is usually determined from the band structure as explained in the DMFTwDFT workflow section. In this case, Ni $3d$ and O $2p$ bands are entangled in a range of -8.0eV to 3.0eV. Since the Fermi energy from a DFT calculation is 7.6986eV, the energy window is chosen as above.
The number of wannier functions is 28 since we have 10 $d-$orbital and 18 $p-$orbital Wannier functions.

The projection orbitals are needed for the initial guess of Wannier orbitals and, for Ni $d-$orbitals, one needs to choose the projection axis to be aligned to the local Ni-O octahedron axis. This is because we want to minimize the off-diagonal terms of the Hamiltonian in the $d-$orbital basis, as we are using the ctqmc impurity solver and the calculation of off-diagonal terms will be very inefficient. One can use "generate\_win.py" file to obtain this projection axis. This file is accessible in the source bin directory.  Using this win file along with other input data obtained by the DFT calculation, one should converge the maximal localization of wannier obitals and obtain wannier90.chk and wannier90.eig files.

\subsection{DFT+DMFT calculations}

After the DFT and wannier90 output files are obtained, the next step is to perform DFT+DMFT calculations. Copy\_input.py file will copy necessary output files and rename them according the the defined notation used in the inputs of the DFT+DMFT.:\\

{python Copy-input.py path-to-DFT-folder}\\

 For non-charge-self-consistent (NCSC) calculations (i.e. Niter=1 tag in INPUT.py), DFT-mu.out, DMFT-mu.out,  INPUT.py,  OSZICAR,  RUNDMFT.py, sig.inp, wannier90.chk, and wannier90.eig files are needed.
 
 For charge-self-consistent (CSC) runs (i.e. Niter$>$1 tag in INPUT.py), INCAR,  KPOINTS, OUTCAR, POSCAR, POTCAR, wannier90.win, wannier90.amn, and WAVECAR files are additionally needed.
 
 The input parameters for the DFT+DMFT calculations will be stored in INPUT.py file. After all input files from the DFT calculations are created, the self-energy file, sig.inp, can be generated by using sigzero.py file. Note that if this file has already been created in a previous calculation, it can be reuse to accelerate the calculation.

 Once we have all input files, the program RUNDMFT.py is executed. During the run, dmft.x, ctqmc, and dft excutables will be run using mpi. Therefore one should put the mpi commands in para-com.dat. For example, one can put the following line in submit script.\\
 echo mpirun -machinefile PBS-NODEFILE -n XX $>$ para-com.dat \\

\begin{table*}[ht]
\caption{Format of an Output file, INFO-ITER obtained by non-charge-self-consistent DMFT calculation for LaNiO$_3$ system.}
\centering
\begin{tabular}{p{0.09\linewidth}p{0.09\linewidth}p{0.09\linewidth}p{0.09\linewidth}p{0.09\linewidth}p{0.09\linewidth}p{0.09\linewidth}p{0.09\linewidth}p{0.09\linewidth}}
\hline
total interaction steps & DMFT iteration steps & lattice occupancy & impurity occupancy &  $$\Sigma^{lattice}_{(\omega=\infty)}- Vdc$$  & $$\Sigma^{impurity}_{(\omega=\infty)}-Vdc$$ & total energy Migdal-Galitskii method & total energy ctqmc & charge difference \\
\hline
 1 &  10  &   7.798655 &     7.797084 &      1.669025 &      1.644641  &   $-68.440528$    &   $-68.086689$ & 0.000000\\
  1  & 11 &    7.798642 &    7.797683 &       1.668703    &   1.644087    &   $-68.444524$ &       $-68.088970$      &    0.000000\\
  1 & 12    &  7.798855 &    7.796977   &    1.669338  &     1.644704 &      $-68.446219$     &  $-68.090861$  &       0.000000\\
  1  & 13 &    7.798939 &    7.797447 &      1.669546 &       1.644125 &       $-68.446738$ &       $-68.096304$ &         0.000000\\
  1 &  14 &     7.798821    & 7.797382 &       1.669375 &       1.644365 &       $-68.443074$ &       $-68.092242$ &         0.000000\\
  1  & 15  &   7.798739 &     7.797317 &       1.669131 &       1.644268     &  $-68.442809$  &     $-68.091764$ &         0.000000\\
\hline
\end{tabular}
\end{table*}

\subsection{Output files}

The main output files of DFT+DMFT calculation are INFO-ITER, INFO-ENERGY, INFO-KSUM, INFO-DM, and INFO-TIME. The examples of LaNiO$_{3}$ runs for both NCSC (i.e. Niter=1) and CSC (i.e. Niter$>$1) can be found in run-exmaples directory within the source directory.

\subsubsection{INFO-ITER}

INFO-ITER records all iteration information necessary for monitoring convergence. For example, INFO-ITER file for LaNiO$_{3}$ DMFT calculation (i.e. Niter=1) will show something like Table 1.\\

  To check the convergence of DFT+DMFT calculation, one must check if the local lattice quantity and the impurity quantity are converging (getting equal). Here, the first number is total interaction steps and the second number is DMFT iteration steps. The third and fourth number compares the d-occupancy from local lattice calculation and the impurity calculations of ctqmc. The fifth and sixth numbers compare the $\Sigma_{(\omega=\infty)}$- Vdc  for lattice and impurity where $\Sigma_{(\omega=\infty)}$ is the self-energy at $\omega\rightarrow\infty$ and Vdc is double counting potential. The seventh and eighth numbers are the total energy computed using the Migdal-Galitskii method and the ctqmc sampling. The last number is the charge difference between two consecutive steps.
  
  If you perform charge-self-consistent calculations for LaNiO$_{3}$, INFO\_ITER file will see something like Table 2.\\

\begin{table*}[ht]
\caption{Format of an Output file, INFO-ITER obtained by charge-self-consistent DMFT calculation for LaNiO$_3$ system.}
\centering
\begin{tabular}{p{0.09\linewidth}p{0.09\linewidth}p{0.09\linewidth}p{0.09\linewidth}p{0.09\linewidth}p{0.09\linewidth}p{0.09\linewidth}p{0.09\linewidth}p{0.09\linewidth}}
\hline
total interaction steps & DMFT iteration steps & lattice occupancy & impurity occupancy &  $$\Sigma^{lattice}_{(\omega=\infty)}-Vdc$$  & $$\Sigma^{impurity}_{(\omega=\infty)}-Vdc$$ & total energy Migdal-Galitskii method & total energy ctqmc & charge difference \\
\hline
22   & 1    &  8.013450 &     8.008304 &       1.497393 &       1.475141 &       $-68.094131$      &  $-67.725218$ &         0.004445 \\
 23  &  1  &    8.013099 &     8.008603 &       1.497078 &       1.474757 &      $-68.081961$      & $-67.715811$ &         0.006589\\
 24 &  1 &     8.012837 &     8.009002 &       1.496692  &      1.474657 &       $-68.072933$      & $-67.715758$ &         0.011620\\
 25  & 1    &  8.012751 &     8.009157 &       1.497266 &       1.474415 &       $-68.067251$      & $-67.713353$ &         0.011801\\
 26 &  1    &  8.012565 &     8.008943 &       1.497044 &       1.474825    &   $-68.064955$  &     $-67.714023$ &         0.006227\\
 27  &  1 &    8.012501 &     8.009220        & 1.497517  &      1.474323  &      $-68.065745$ &      $-67.714654$ &         0.007282\\
 28  &  1  &    8.012436 &     8.009272 &       1.497843 &       1.474266 &       $-68.067957$ &       $-67.714704$ &         0.010110\\
 29  &  1 &     8.012442 &     8.009222 &       1.498471 &       1.474300  &      $-68.067595$ &       $-67.716171$ &         0.012198\\
 30 &  1 &     8.012360 &     8.008898 &       1.498468 &       1.474642    &    $-68.068845$     &  $-67.713375$ &        0.009332\\
\hline
\end{tabular}
\end{table*}
 
 Now you can see that we performed total iteration of 30 steps with 1 DMFT step. The d-occupancy is increased to near 8.0 and the last number is updated for the charge difference.
 
 Nevertheless, users are strongly encouraged to read the manual, which contains additional information about these files.

\subsection{Post-processing tools}

\subsubsection{Analytic continuation - maximum entropy method}

Since the ctqmc impurity solver samples the self energy (sig.inp) on the imaginary axis, one should perform the analytic continuation to obtain the self energy on the real axis. i.e.
\begin{equation*}
\Sigma_{(i\omega)} \rightarrow \Sigma_{(\omega)}
\end{equation*}

Here, we use the Maximum Entropy method \cite{Jarell_ana_cont} developed by Jarrell et al. to perform this analytic continuation. The source file can be compiled in post-tools/maxent-source directory.

The procedure of performing analytic continuation with the Maximum Entropy method, \textbf{max-ent}, is as follows.

\begin{enumerate}
    \item Compile the maxent-source codes and copy *.so files and *.py files to the \textbf{/bin} directory. An empty directory to perform max-ent should be created inside the DMFT run directory. 
    \item Copy few of the last self energy data sig.inp.XXX to the directory. This ensures that only the converged self energies are used for the calculations, provided enough DMFT iterations have been performed to reach convergence. 
    \item Run "sigaver.py" to take the average of the self energies.
    \begin{verbatim}
    sigaver.py sig.inp.*
    \end{verbatim}
    This results in an averaged self energy file, \textbf{sig.inpx} which is used in max-ent.
    \item Copy maxent-params.dat file from the source directory into the current working directory.
    \item Perform the analytic continuation with max-ent.
    \begin{verbatim}
    maxent_run.py sig.inpx
    \end{verbatim}
    The analytically continued self-energy will be stored in the file \textbf{Sig.out}.
\end{enumerate}

Once the \textbf{Sig.out} file is obtained it could be used for further post processing such as plotting band structures and density of states. 

\subsubsection{Density of states}

The DMFT density of states could be calculated from the imaginary part of the local Green's function on the real axis obtained from the self energy on the real axis retrieved in the previous section. 

\begin{equation*}
    A(\omega)=-\frac{1}{\pi} \operatorname{lm} G(\omega)
\end{equation*}

\begin{enumerate}
   \item Create a new directory and copy necessary files by executing the \textbf{Copy\_input.py} program, found in the \textbf{/bin} directory.
   \begin{verbatim}
   Copy_input.py <path-to-DMFT-results> -dos    
   \end{verbatim}
   \item Generate the real axis self energies on a denser mesh by interpolating self energies from \textbf{Sig.out} obtained previously. Run:
   \begin{verbatim}
       Interpol_sig_real.py 
   \end{verbatim}
   This provides the interpolated self energy file, \textbf{sig.inp\_real}.
   \item Run \textbf{dmft\_dos.x} to obtain the local Green's function, \textbf{G\_loc.out} on the real axis.
   \begin{verbatim}
       mpirun -n X dmft-dos.x
   \end{verbatim}
   where, X is the number of cores
   
   \item Finally, run \textbf{plotDMFTDOS.py} located in the \textbf{/scripts} directory to obtain a projected density of states plot.
   \begin{verbatim}
       plotDMFTDOS.py
   \end{verbatim}
   
   Modifying \textbf{plotDMFTDOS.py} allows changing projections that would be plotted. 
 
\end{enumerate} 

\subsubsection{Band Structure}

\noindent
Once the analytic continuation has been completed as explained in the previous section, one may use the self energies to plot the DMFT band structure. DMFTwDFT is capable of plotting a variety of different bandstructures. Unlike in the density of states case, the Spectral Function for band structures is a function of both the Matsubara Frequency, $\omega$ and $k$-vectors as seen below.

\begin{equation*}
    A(k, \omega)=-\frac{1}{\pi}\frac{I m \Sigma} {\left(\omega-\epsilon_{k}-Re \Sigma\right)^{2}+(I m \Sigma)^{2}}
\end{equation*}

In the following sections we explain how to obtain the different types of band structure. 

\begin{itemize}
    \item Plain DMFT bandstructure: \\
    This is the most basic type of DMFT band structure obtainable. The following steps are pursued to obtain it.
    \begin{enumerate}
    
        \item Create a new directory and copy necessary files by executing \textbf{Copy\_input.py}.
        \begin{verbatim}
Copy_input.py <path-to-DMFT-results> 
-bands   
         \end{verbatim}
         
        \item Interpolate the real axis self energies.
        \begin{verbatim}
Interp_Sig.py  
         \end{verbatim}
         
          \item Generate a $k$-path for the band structure. One may define the number of $k$-points and the $k$-path in \textbf{kgen.py}. Then run:
        \begin{verbatim}
kgen.py  
         \end{verbatim}
        This results in the file \textbf{klist.dat} which contains the $k$-path data.
        
        \item Now run \textbf{dmft\_ksum\_band} to obtain the local Green's function data, \textbf{Gk.out}. This could be run in parallel as follows:
        \begin{verbatim}
mpirun -np 16 dmft_ksum_band  
         \end{verbatim}
        
         \item Finally, one may run \textbf{plot\_Gk.py} to obtain the band structure output in the .eps format. 

    \end{enumerate}
   
    \item Spin polarized DMFT bandstructure:\\
    This is useful to study strongly correlated magnetic systems. The steps are similar to the above. Instead of \textbf{Interp\_Sig.py}, \textbf{dmft\_ksum\_band} and \textbf{plot\_Gk.py} for spin polarized band structure calculations, \textbf{Interp\_Sig\_sp.py}, \textbf{dmft\_ksum\_band\_sp} and \textbf{plot\_Gk\_sp.py} are used, respectively. 
    
    \item Orbital projected bandstructure:\\
    This type of band structure comes in handy to study the material properties based on their individual orbital contributions.  
    This is performed similar to the plain band structure but by using \textbf{dmft\_ksum\_band\_partial} and \textbf{plot\_Gk\_partial.py}. The orbitals to be projected are specified in \textbf{plot\_Gk\_partial.py} and follows the Wannier orbital ordering. 
    
    \item DFT and DMFT band structure comparison:\\
    This helps to clearly visualize the effects of correlations on the band structure as seen in Figure ~\ref{fig:Figure13}. Once the initial steps for the plain band structure is performed, \textbf{plot\_Gk\_compare.py} is used to obtain this band structure. This plots the DFT and DMFT band structures in a single plot making it more convenient for comparison.
\end{itemize}

\bibliographystyle{elsarticle-num}
\bibliography{references}

\begin{thebibliography}{100}
\expandafter\ifx\csname url\endcsname\relax
  \def\url#1{\texttt{#1}}\fi
\expandafter\ifx\csname urlprefix\endcsname\relax\def\urlprefix{URL }\fi
\expandafter\ifx\csname href\endcsname\relax
  \def\href#1#2{#2} \def\path#1{#1}\fi

\bibitem{mott-insulator}
S.~L. Dudarev, G.~A. Botton, S.~Y. Savrasov, C.~J. Humphreys, A.~P. Sutton,
  \href{https://link.aps.org/doi/10.1103/PhysRevB.57.1505}{Electron-energy-loss
  spectra and the structural stability of nickel oxide: An lsda+u study}, Phys.
  Rev. B 57 (1998) 1505--1509 (Jan 1998).
\newblock \href {https://doi.org/10.1103/PhysRevB.57.1505}
  {\path{doi:10.1103/PhysRevB.57.1505}}.
\newline\urlprefix\url{https://link.aps.org/doi/10.1103/PhysRevB.57.1505}

\bibitem{Heavy_fermions}
S.~Wirth, F.~Steglich,
  \href{https://doi.org/10.1038/natrevmats.2016.51}{Exploring heavy fermions
  from macroscopic to microscopic length scales}, Nature Reviews Materials 1
  (2016) 16051 (Aug 2016).
\newblock \href {https://doi.org/10.1038/natrevmats.2016.51}
  {\path{doi:10.1038/natrevmats.2016.51}}.
\newline\urlprefix\url{https://doi.org/10.1038/natrevmats.2016.51}

\bibitem{Fermiliquid}
H.~v. L\"ohneysen, A.~Rosch, M.~Vojta, P.~W\"olfle,
  \href{https://link.aps.org/doi/10.1103/RevModPhys.79.1015}{Fermi-liquid
  instabilities at magnetic quantum phase transitions}, Rev. Mod. Phys. 79
  (2007) 1015--1075 (Aug 2007).
\newblock \href {https://doi.org/10.1103/RevModPhys.79.1015}
  {\path{doi:10.1103/RevModPhys.79.1015}}.
\newline\urlprefix\url{https://link.aps.org/doi/10.1103/RevModPhys.79.1015}

\bibitem{Savrasov:04}
S.~Y. Savrasov, G.~Kotliar, Spectral density functionals for electronic
  structure calculations, Phys. Rev. B 69 (2004) 245101 (Jun 2004).
\newblock \href {https://doi.org/10.1103/PhysRevB.69.245101}
  {\path{doi:10.1103/PhysRevB.69.245101}}.

\bibitem{Kotliar:06}
G.~Kotliar, S.~Y. Savrasov, K.~Haule, V.~S. Oudovenko, O.~Parcollet, C.~A.
  Marianetti, Electronic structure calculations with dynamical mean-field
  theory, Rev. Mod. Phys. 78 (2006) 865--951 (2006).
\newblock \href {https://doi.org/10.1103/RevModPhys.78.865}
  {\path{doi:10.1103/RevModPhys.78.865}}.

\bibitem{Exact_diag}
Y.~Lu, M.~W. Haverkort,
  \href{https://doi.org/10.1140/epjst/e2017-70042-4}{Exact diagonalization as
  an impurity solver in dynamical mean field theory}, The European Physical
  Journal Special Topics 226~(11) (2017) 2549--2564 (Jul 2017).
\newblock \href {https://doi.org/10.1140/epjst/e2017-70042-4}
  {\path{doi:10.1140/epjst/e2017-70042-4}}.
\newline\urlprefix\url{https://doi.org/10.1140/epjst/e2017-70042-4}

\bibitem{RGgroup}
R.~Bulla, T.~A. Costi, T.~Pruschke,
  \href{https://link.aps.org/doi/10.1103/RevModPhys.80.395}{Numerical
  renormalization group method for quantum impurity systems}, Rev. Mod. Phys.
  80 (2008) 395--450 (Apr 2008).
\newblock \href {https://doi.org/10.1103/RevModPhys.80.395}
  {\path{doi:10.1103/RevModPhys.80.395}}.
\newline\urlprefix\url{https://link.aps.org/doi/10.1103/RevModPhys.80.395}

\bibitem{DMRG}
K.~Hallberg, D.~J. Garc{\'{\i}}a, P.~S. Cornaglia, J.~I. Facio,
  Y.~N{\'{u}}{\~{n}}ez-Fern{\'{a}}ndez,
  \href{https://doi.org/10.1209%2F0295-5075%2F112%2F17001}{State-of-the-art
  techniques for calculating spectral functions in models for correlated
  materials}, {EPL} (Europhysics Letters) 112~(1) (2015) 17001 (oct 2015).
\newblock \href {https://doi.org/10.1209/0295-5075/112/17001}
  {\path{doi:10.1209/0295-5075/112/17001}}.
\newline\urlprefix\url{https://doi.org/10.1209%2F0295-5075%2F112%2F17001}

\bibitem{Jarrell_MonteCarlo_1}
M.~Jarrell, \href{https://link.aps.org/doi/10.1103/PhysRevLett.69.168}{Hubbard
  model in infinite dimensions: A quantum monte carlo study}, Phys. Rev. Lett.
  69 (1992) 168--171 (Jul 1992).
\newblock \href {https://doi.org/10.1103/PhysRevLett.69.168}
  {\path{doi:10.1103/PhysRevLett.69.168}}.
\newline\urlprefix\url{https://link.aps.org/doi/10.1103/PhysRevLett.69.168}

\bibitem{Monte_carlo_2}
{Shinaoka, H.}, {Assaad, F.}, {Bl\"umer, N.}, {Werner, P.},
  \href{https://doi.org/10.1140/epjst/e2017-70050-x}{Quantum monte carlo
  impurity solvers for multi-orbital problems and frequency-dependent
  interactions}, Eur. Phys. J. Special Topics 226~(11) (2017) 2499--2523
  (2017).
\newblock \href {https://doi.org/10.1140/epjst/e2017-70050-x}
  {\path{doi:10.1140/epjst/e2017-70050-x}}.
\newline\urlprefix\url{https://doi.org/10.1140/epjst/e2017-70050-x}

\bibitem{CTQMC}
E.~Gull, A.~J. Millis, A.~I. Lichtenstein, A.~N. Rubtsov, M.~Troyer, P.~Werner,
  \href{https://link.aps.org/doi/10.1103/RevModPhys.83.349}{Continuous-time
  monte carlo methods for quantum impurity models}, Rev. Mod. Phys. 83 (2011)
  349--404 (May 2011).
\newblock \href {https://doi.org/10.1103/RevModPhys.83.349}
  {\path{doi:10.1103/RevModPhys.83.349}}.
\newline\urlprefix\url{https://link.aps.org/doi/10.1103/RevModPhys.83.349}

\bibitem{haule_ctqmc}
K.~Haule, \href{https://link.aps.org/doi/10.1103/PhysRevB.75.155113}{Quantum
  monte carlo impurity solver for cluster dynamical mean-field theory and
  electronic structure calculations with adjustable cluster base}, Phys. Rev. B
  75 (2007) 155113 (Apr 2007).
\newblock \href {https://doi.org/10.1103/PhysRevB.75.155113}
  {\path{doi:10.1103/PhysRevB.75.155113}}.
\newline\urlprefix\url{https://link.aps.org/doi/10.1103/PhysRevB.75.155113}

\bibitem{ALPS}
A.~Gaenko, A.~Antipov, G.~Carcassi, T.~Chen, X.~Chen, Q.~Dong, L.~Gamper,
  J.~Gukelberger, R.~Igarashi, S.~Iskakov, M.~KÃ¶nz, J.~LeBlanc, R.~Levy,
  P.~Ma, J.~Paki, H.~Shinaoka, S.~Todo, M.~Troyer, E.~Gull,
  \href{https://www.scopus.com/inward/record.uri?eid=2-s2.0-85009160076&doi=10.1016%2fj.cpc.2016.12.009&partnerID=40&md5=1e9d065d7250d70b1378829223826120}{Updated
  core libraries of the alps project}, Computer Physics Communications 213
  (2017) 235--251, cited By 22 (2017).
\newblock \href {https://doi.org/10.1016/j.cpc.2016.12.009}
  {\path{doi:10.1016/j.cpc.2016.12.009}}.
\newline\urlprefix\url{https://www.scopus.com/inward/record.uri?eid=2-s2.0-85009160076&doi=10.1016%2fj.cpc.2016.12.009&partnerID=40&md5=1e9d065d7250d70b1378829223826120}

\bibitem{ComDMFT}
S.~Choi, P.~Semon, B.~Kang, A.~Kutepov, G.~Kotliar,
  \href{http://www.sciencedirect.com/science/article/pii/S0010465519302140}{Comdmft:
  A massively parallel computer package for the electronic structure of
  correlated-electron systems}, Computer Physics Communications 244 (2019) 277
  -- 294 (2019).
\newblock \href {https://doi.org/https://doi.org/10.1016/j.cpc.2019.07.003}
  {\path{doi:https://doi.org/10.1016/j.cpc.2019.07.003}}.
\newline\urlprefix\url{http://www.sciencedirect.com/science/article/pii/S0010465519302140}

\bibitem{IQIST}
L.~Huang, Y.~Wang, Z.~Meng, L.~Du, P.~Werner, X.~Dai,
  \href{https://www.scopus.com/inward/record.uri?eid=2-s2.0-84932197725&doi=10.1016%2fj.cpc.2015.04.020&partnerID=40&md5=4377b9f3a6c80dea21ed8a4e252cac29}{iqist:
  An open source continuous-time quantum monte carlo impurity solver toolkit},
  Computer Physics Communications 195 (2015) 140--160, cited By 18 (2015).
\newblock \href {https://doi.org/10.1016/j.cpc.2015.04.020}
  {\path{doi:10.1016/j.cpc.2015.04.020}}.
\newline\urlprefix\url{https://www.scopus.com/inward/record.uri?eid=2-s2.0-84932197725&doi=10.1016%2fj.cpc.2015.04.020&partnerID=40&md5=4377b9f3a6c80dea21ed8a4e252cac29}

\bibitem{EDMFT}
K.~Haule, C.-H. Yee, K.~Kim,
  \href{https://link.aps.org/doi/10.1103/PhysRevB.81.195107}{Dynamical
  mean-field theory within the full-potential methods: Electronic structure of
  ${\text{ceirin}}_{5}$, ${\text{cecoin}}_{5}$, and ${\text{cerhin}}_{5}$},
  Phys. Rev. B 81 (2010) 195107 (May 2010).
\newblock \href {https://doi.org/10.1103/PhysRevB.81.195107}
  {\path{doi:10.1103/PhysRevB.81.195107}}.
\newline\urlprefix\url{https://link.aps.org/doi/10.1103/PhysRevB.81.195107}

\bibitem{TRIQS}
O.~Parcollet, M.~Ferrero, T.~Ayral, H.~Hafermann, I.~Krivenko, L.~Messio,
  P.~Seth,
  \href{https://www.scopus.com/inward/record.uri?eid=2-s2.0-84942199229&doi=10.1016%2fj.cpc.2015.04.023&partnerID=40&md5=732bb6841cd04bec02acd927199440ab}{Triqs:
  A toolbox for research on interacting quantum systems}, Computer Physics
  Communications 196 (2015) 398--415, cited By 110 (2015).
\newblock \href {https://doi.org/10.1016/j.cpc.2015.04.023}
  {\path{doi:10.1016/j.cpc.2015.04.023}}.
\newline\urlprefix\url{https://www.scopus.com/inward/record.uri?eid=2-s2.0-84942199229&doi=10.1016%2fj.cpc.2015.04.023&partnerID=40&md5=732bb6841cd04bec02acd927199440ab}

\bibitem{w2dynamics}
M.~Wallerberger, A.~Hausoel, P.~Gunacker, A.~Kowalski, N.~Parragh, F.~Goth,
  K.~Held, G.~Sangiovanni,
  \href{http://www.sciencedirect.com/science/article/pii/S0010465518303217}{w2dynamics:
  Local one- and two-particle quantities from dynamical mean field theory},
  Computer Physics Communications 235 (2019) 388 -- 399 (2019).
\newblock \href {https://doi.org/https://doi.org/10.1016/j.cpc.2018.09.007}
  {\path{doi:https://doi.org/10.1016/j.cpc.2018.09.007}}.
\newline\urlprefix\url{http://www.sciencedirect.com/science/article/pii/S0010465518303217}

\bibitem{TRIQSDFTTools}
M.~Aichhorn, L.~Pourovskii, P.~Seth, V.~Vildosola, M.~Zingl, O.~Peil, X.~Deng,
  J.~Mravlje, G.~Kraberger, C.~Martins, M.~Ferrero, O.~Parcollet,
  \href{https://www.scopus.com/inward/record.uri?eid=2-s2.0-84992295854&doi=10.1016%2fj.cpc.2016.03.014&partnerID=40&md5=20adaa27ac1610695ce05ccd1e56bdc0}{Triqs/dfttools:
  A triqs application for ab initio calculations of correlated materials},
  Computer Physics Communications 204 (2016) 200--208, cited By 35 (2016).
\newblock \href {https://doi.org/10.1016/j.cpc.2016.03.014}
  {\path{doi:10.1016/j.cpc.2016.03.014}}.
\newline\urlprefix\url{https://www.scopus.com/inward/record.uri?eid=2-s2.0-84992295854&doi=10.1016%2fj.cpc.2016.03.014&partnerID=40&md5=20adaa27ac1610695ce05ccd1e56bdc0}

\bibitem{Dcore}
\href{https://issp-center-dev.github.io/DCore/master/index.htm}{Dcore - dcore
  1.0.0 documentation}.
\newline\urlprefix\url{https://issp-center-dev.github.io/DCore/master/index.htm}

\bibitem{Amulet}
\href{http://www.amulet-code.org/publication-list}{Amulet}.
\newline\urlprefix\url{http://www.amulet-code.org/publication-list}

\bibitem{RsPt_DMFT}
O.~Grånäs, I.~D. Marco, P.~Thunström, L.~Nordström, O.~Eriksson,
  T.~Björkman, J.~Wills,
  \href{http://www.sciencedirect.com/science/article/pii/S092702561100646X}{Charge
  self-consistent dynamical mean-field theory based on the full-potential
  linear muffin-tin orbital method: Methodology and applications},
  Computational Materials Science 55 (2012) 295 -- 302 (2012).
\newblock \href
  {https://doi.org/https://doi.org/10.1016/j.commatsci.2011.11.032}
  {\path{doi:https://doi.org/10.1016/j.commatsci.2011.11.032}}.
\newline\urlprefix\url{http://www.sciencedirect.com/science/article/pii/S092702561100646X}

\bibitem{Questaal}
D.~Pashov, S.~Acharya, W.~R. Lambrecht, J.~Jackson, K.~D. Belashchenko,
  A.~Chantis, F.~Jamet, M.~van Schilfgaarde,
  \href{http://www.sciencedirect.com/science/article/pii/S0010465519303868}{Questaal:
  A package of electronic structure methods based on the linear muffin-tin
  orbital technique}, Computer Physics Communications (2019) 107065 (2019).
\newblock \href {https://doi.org/https://doi.org/10.1016/j.cpc.2019.107065}
  {\path{doi:https://doi.org/10.1016/j.cpc.2019.107065}}.
\newline\urlprefix\url{http://www.sciencedirect.com/science/article/pii/S0010465519303868}

\bibitem{LMTOLAPW}
O.~K. Andersen, \href{https://link.aps.org/doi/10.1103/PhysRevB.12.3060}{Linear
  methods in band theory}, Phys. Rev. B 12 (1975) 3060--3083 (Oct 1975).
\newblock \href {https://doi.org/10.1103/PhysRevB.12.3060}
  {\path{doi:10.1103/PhysRevB.12.3060}}.
\newline\urlprefix\url{https://link.aps.org/doi/10.1103/PhysRevB.12.3060}

\bibitem{Wannier90_2012}
N.~Marzari, A.~A. Mostofi, J.~R. Yates, I.~Souza, D.~Vanderbilt,
  \href{https://link.aps.org/doi/10.1103/RevModPhys.84.1419}{Maximally
  localized wannier functions: Theory and applications}, Rev. Mod. Phys. 84
  (2012) 1419--1475 (Oct 2012).
\newblock \href {https://doi.org/10.1103/RevModPhys.84.1419}
  {\path{doi:10.1103/RevModPhys.84.1419}}.
\newline\urlprefix\url{https://link.aps.org/doi/10.1103/RevModPhys.84.1419}

\bibitem{wannier90new}
G.~Pizzi, V.~Vitale, R.~Arita, S.~Blügel, F.~Freimuth, G.~Géranton,
  M.~Gibertini, D.~Gresch, C.~Johnson, T.~Koretsune, J.~Ibañez-Azpiroz,
  H.~Lee, J.-M. Lihm, D.~Marchand, A.~Marrazzo, Y.~Mokrousov, J.~I. Mustafa,
  Y.~Nohara, Y.~Nomura, L.~Paulatto, S.~Poncé, T.~Ponweiser, J.~Qiao,
  F.~Thöle, S.~S. Tsirkin, M.~Wierzbowska, N.~Marzari, D.~Vanderbilt,
  I.~Souza, A.~A. Mostofi, J.~R. Yates, Wannier90 as a community code: new
  features and applications (2019).
\newblock \href {http://arxiv.org/abs/1907.09788} {\path{arXiv:1907.09788}}.

\bibitem{Updated_wannier90_2014}
A.~A. Mostofi, J.~R. Yates, G.~Pizzi, Y.-S. Lee, I.~Souza, D.~Vanderbilt,
  N.~Marzari,
  \href{http://www.sciencedirect.com/science/article/pii/S001046551400157X}{An
  updated version of wannier90: A tool for obtaining maximally-localised
  wannier functions}, Computer Physics Communications 185~(8) (2014) 2309 --
  2310 (2014).
\newblock \href {https://doi.org/https://doi.org/10.1016/j.cpc.2014.05.003}
  {\path{doi:https://doi.org/10.1016/j.cpc.2014.05.003}}.
\newline\urlprefix\url{http://www.sciencedirect.com/science/article/pii/S001046551400157X}

\bibitem{VASP}
G.~Kresse, J.~Furthmüller,
  \href{http://www.sciencedirect.com/science/article/pii/0927025696000080}{Efficiency
  of ab-initio total energy calculations for metals and semiconductors using a
  plane-wave basis set}, Computational Materials Science 6~(1) (1996) 15 -- 50
  (1996).
\newblock \href {https://doi.org/https://doi.org/10.1016/0927-0256(96)00008-0}
  {\path{doi:https://doi.org/10.1016/0927-0256(96)00008-0}}.
\newline\urlprefix\url{http://www.sciencedirect.com/science/article/pii/0927025696000080}

\bibitem{PhysRevB.47.558}
G.~Kresse, J.~Hafner,
  \href{https://link.aps.org/doi/10.1103/PhysRevB.47.558}{Ab initio molecular
  dynamics for liquid metals}, Phys. Rev. B 47 (1993) 558--561 (Jan 1993).
\newblock \href {https://doi.org/10.1103/PhysRevB.47.558}
  {\path{doi:10.1103/PhysRevB.47.558}}.
\newline\urlprefix\url{https://link.aps.org/doi/10.1103/PhysRevB.47.558}

\bibitem{kresse_efficiency_1996}
G.~Kresse, J.~Furthmüller,
  \href{http://www.sciencedirect.com/science/article/pii/0927025696000080}{Efficiency
  of ab-initio total energy calculations for metals and semiconductors using a
  plane-wave basis set}, Computational Materials Science 6~(1) (1996) 15 -- 50
  (1996).
\newblock \href {https://doi.org/https://doi.org/10.1016/0927-0256(96)00008-0}
  {\path{doi:https://doi.org/10.1016/0927-0256(96)00008-0}}.
\newline\urlprefix\url{http://www.sciencedirect.com/science/article/pii/0927025696000080}

\bibitem{QuantumExpresso}
P.~Giannozzi, S.~Baroni, N.~Bonini, M.~Calandra, R.~Car, C.~Cavazzoni,
  D.~Ceresoli, G.~L. Chiarotti, M.~Cococcioni, I.~Dabo, A.~D. Corso,
  S.~de~Gironcoli, S.~Fabris, G.~Fratesi, R.~Gebauer, U.~Gerstmann,
  C.~Gougoussis, A.~Kokalj, M.~Lazzeri, L.~Martin-Samos, N.~Marzari, F.~Mauri,
  R.~Mazzarello, S.~Paolini, A.~Pasquarello, L.~Paulatto, C.~Sbraccia,
  S.~Scandolo, G.~Sclauzero, A.~P. Seitsonen, A.~Smogunov, P.~Umari, R.~M.
  Wentzcovitch,
  \href{https://doi.org/10.1088%2F0953-8984%2F21%2F39%2F395502}{{QUANTUM}
  {ESPRESSO}: a modular and open-source software project for quantum
  simulations of materials}, Journal of Physics: Condensed Matter 21~(39)
  (2009) 395502 (sep 2009).
\newblock \href {https://doi.org/10.1088/0953-8984/21/39/395502}
  {\path{doi:10.1088/0953-8984/21/39/395502}}.
\newline\urlprefix\url{https://doi.org/10.1088%2F0953-8984%2F21%2F39%2F395502}

\bibitem{Siesta}
J.~M. Soler, E.~Artacho, J.~D. Gale, A.~Garc{\'{\i}}a, J.~Junquera,
  P.~Ordej{\'{o}}n, D.~S{\'{a}}nchez-Portal,
  \href{https://doi.org/10.1088%2F0953-8984%2F14%2F11%2F302}{The {SIESTA}
  method forab initioorder-nmaterials simulation}, Journal of Physics:
  Condensed Matter 14~(11) (2002) 2745--2779 (mar 2002).
\newblock \href {https://doi.org/10.1088/0953-8984/14/11/302}
  {\path{doi:10.1088/0953-8984/14/11/302}}.
\newline\urlprefix\url{https://doi.org/10.1088%2F0953-8984%2F14%2F11%2F302}

\bibitem{ABINIT}
X.~Gonze, F.~Jollet, F.~A. Araujo, D.~Adams, B.~Amadon, T.~Applencourt,
  C.~Audouze, J.-M. Beuken, J.~Bieder, A.~Bokhanchuk, E.~Bousquet, F.~Bruneval,
  D.~Caliste, M.~Côté, F.~Dahm, F.~D. Pieve, M.~Delaveau, M.~D. Gennaro,
  B.~Dorado, C.~Espejo, G.~Geneste, L.~Genovese, A.~Gerossier, M.~Giantomassi,
  Y.~Gillet, D.~Hamann, L.~He, G.~Jomard, J.~L. Janssen, S.~L. Roux, A.~Levitt,
  A.~Lherbier, F.~Liu, I.~Lukačević, A.~Martin, C.~Martins, M.~Oliveira,
  S.~Poncé, Y.~Pouillon, T.~Rangel, G.-M. Rignanese, A.~Romero, B.~Rousseau,
  O.~Rubel, A.~Shukri, M.~Stankovski, M.~Torrent, M.~V. Setten, B.~V. Troeye,
  M.~Verstraete, D.~Waroquiers, J.~Wiktor, B.~Xu, A.~Zhou, J.~Zwanziger,
  \href{http://www.sciencedirect.com/science/article/pii/S0010465516300923}{Recent
  developments in the abinit software package}, Computer Physics Communications
  205 (2016) 106 -- 131 (2016).
\newblock \href {https://doi.org/https://doi.org/10.1016/j.cpc.2016.04.003}
  {\path{doi:https://doi.org/10.1016/j.cpc.2016.04.003}}.
\newline\urlprefix\url{http://www.sciencedirect.com/science/article/pii/S0010465516300923}

\bibitem{gonze2002first}
X.~Gonze, J.-M. Beuken, R.~Caracas, F.~Detraux, M.~Fuchs, G.-M. Rignanese,
  L.~Sindic, M.~Verstraete, G.~Zerah, F.~Jollet, et~al., First-principles
  computation of material properties: the abinit software project,
  Computational Materials Science 25~(3) (2002) 478--492 (2002).

\bibitem{gonze2009abinit}
X.~Gonze, B.~Amadon, P.-M. Anglade, J.-M. Beuken, F.~Bottin, P.~Boulanger,
  F.~Bruneval, D.~Caliste, R.~Caracas, M.~C{\^o}t{\'e}, et~al., Abinit:
  First-principles approach to material and nanosystem properties, Computer
  Physics Communications 180~(12) (2009) 2582--2615 (2009).

\bibitem{ELK}
\href{http://elk.sourceforge.net/}{The elk code}.
\newline\urlprefix\url{http://elk.sourceforge.net/}

\bibitem{WIEN2k}
K.~Schwarz, P.~Blaha,
  \href{http://www.sciencedirect.com/science/article/pii/S0927025603001125}{Solid
  state calculations using wien2k}, Computational Materials Science 28~(2)
  (2003) 259 -- 273, proceedings of the Symposium on Software Development for
  Process and Materials Design (2003).
\newblock \href {https://doi.org/https://doi.org/10.1016/S0927-0256(03)00112-5}
  {\path{doi:https://doi.org/10.1016/S0927-0256(03)00112-5}}.
\newline\urlprefix\url{http://www.sciencedirect.com/science/article/pii/S0927025603001125}

\bibitem{Kohn1964}
P.~{Hohenberg}, W.~{Kohn}, {Inhomogeneous Electron Gas}, Physical Review
  136~(3B) (1964) 864--871 (Nov 1964).
\newblock \href {https://doi.org/10.1103/PhysRev.136.B864}
  {\path{doi:10.1103/PhysRev.136.B864}}.

\bibitem{LDA1}
S.~H. Vosko, L.~Wilk, M.~Nusair,
  \href{https://doi.org/10.1139/p80-159}{Accurate spin-dependent electron
  liquid correlation energies for local spin density calculations: a critical
  analysis}, Canadian Journal of Physics 58~(8) (1980) 1200--1211 (1980).
\newblock \href {http://arxiv.org/abs/https://doi.org/10.1139/p80-159}
  {\path{arXiv:https://doi.org/10.1139/p80-159}}, \href
  {https://doi.org/10.1139/p80-159} {\path{doi:10.1139/p80-159}}.
\newline\urlprefix\url{https://doi.org/10.1139/p80-159}

\bibitem{LDA2}
D.~M. Ceperley, B.~J. Alder,
  \href{https://link.aps.org/doi/10.1103/PhysRevLett.45.566}{Ground state of
  the electron gas by a stochastic method}, Phys. Rev. Lett. 45 (1980) 566--569
  (Aug 1980).
\newblock \href {https://doi.org/10.1103/PhysRevLett.45.566}
  {\path{doi:10.1103/PhysRevLett.45.566}}.
\newline\urlprefix\url{https://link.aps.org/doi/10.1103/PhysRevLett.45.566}

\bibitem{PBE_functional}
K.~Burke, J.~P. Perdew, M.~Ernzerhof,
  \href{https://onlinelibrary.wiley.com/doi/abs/10.1002/\%28SICI\%291097-461X\%281997\%2961\%3A2\%3C287\%3A\%3AAID-QUA11\%3E3.0.CO\%3B2-9}{Why
  the generalized gradient approximation works and how to go beyond it},
  International Journal of Quantum Chemistry 61~(2) (1997) 287--293 (1997).
\newblock \href
  {http://arxiv.org/abs/https://onlinelibrary.wiley.com/doi/pdf/10.1002/\%28SICI\%291097-461X\%281997\%2961\%3A2\%3C287\%3A\%3AAID-QUA11\%3E3.0.CO\%3B2-9}
  {\path{arXiv:https://onlinelibrary.wiley.com/doi/pdf/10.1002/\%28SICI\%291097-461X\%281997\%2961\%3A2\%3C287\%3A\%3AAID-QUA11\%3E3.0.CO\%3B2-9}},
  \href
  {https://doi.org/10.1002/(SICI)1097-461X(1997)61:2<287::AID-QUA11>3.0.CO;2-9}
  {\path{doi:10.1002/(SICI)1097-461X(1997)61:2<287::AID-QUA11>3.0.CO;2-9}}.
\newline\urlprefix\url{https://onlinelibrary.wiley.com/doi/abs/10.1002/\%28SICI\%291097-461X\%281997\%2961\%3A2\%3C287\%3A\%3AAID-QUA11\%3E3.0.CO\%3B2-9}

\bibitem{LuttingerI}
W.~Kohn, J.~M. Luttinger,
  \href{https://link.aps.org/doi/10.1103/PhysRev.118.41}{Ground-state energy of
  a many-fermion system}, Phys. Rev. 118 (1960) 41--45 (Apr 1960).
\newblock \href {https://doi.org/10.1103/PhysRev.118.41}
  {\path{doi:10.1103/PhysRev.118.41}}.
\newline\urlprefix\url{https://link.aps.org/doi/10.1103/PhysRev.118.41}

\bibitem{LuttingerII}
J.~M. Luttinger, J.~C. Ward,
  \href{https://link.aps.org/doi/10.1103/PhysRev.118.1417}{Ground-state energy
  of a many-fermion system. ii}, Phys. Rev. 118 (1960) 1417--1427 (Jun 1960).
\newblock \href {https://doi.org/10.1103/PhysRev.118.1417}
  {\path{doi:10.1103/PhysRev.118.1417}}.
\newline\urlprefix\url{https://link.aps.org/doi/10.1103/PhysRev.118.1417}

\bibitem{LuttingerIII}
J.~M. Luttinger,
  \href{https://link.aps.org/doi/10.1103/PhysRev.121.942}{Analytic properties
  of single-particle propagators for many-fermion systems}, Phys. Rev. 121
  (1961) 942--949 (Feb 1961).
\newblock \href {https://doi.org/10.1103/PhysRev.121.942}
  {\path{doi:10.1103/PhysRev.121.942}}.
\newline\urlprefix\url{https://link.aps.org/doi/10.1103/PhysRev.121.942}

\bibitem{Migdal}
V.~Galitskii, A.~Migdal, Application of quantum field theory methods to the
  many body problem, Zhur. Eksptl'. me Teoret. Fiz. Vol: 34 (1 1958).

\bibitem{FLL_V}
A.~G. Petukhov, I.~I. Mazin, L.~Chioncel, A.~I. Lichtenstein,
  \href{https://link.aps.org/doi/10.1103/PhysRevB.67.153106}{Correlated metals
  and the $\mathrm{LDA}+u$ method}, Phys. Rev. B 67 (2003) 153106 (Apr 2003).
\newblock \href {https://doi.org/10.1103/PhysRevB.67.153106}
  {\path{doi:10.1103/PhysRevB.67.153106}}.
\newline\urlprefix\url{https://link.aps.org/doi/10.1103/PhysRevB.67.153106}

\bibitem{Park_dc_2014}
H.~Park, A.~J. Millis, C.~A. Marianetti,
  \href{https://link.aps.org/doi/10.1103/PhysRevB.90.235103}{Computing total
  energies in complex materials using charge self-consistent dft + dmft}, Phys.
  Rev. B 90 (2014) 235103 (Dec 2014).
\newblock \href {https://doi.org/10.1103/PhysRevB.90.235103}
  {\path{doi:10.1103/PhysRevB.90.235103}}.
\newline\urlprefix\url{https://link.aps.org/doi/10.1103/PhysRevB.90.235103}

\bibitem{LaNiO3_press}
H.~Park, A.~J. Millis, C.~A. Marianetti,
  \href{https://link.aps.org/doi/10.1103/PhysRevB.89.245133}{Total energy
  calculations using dft+dmft: Computing the pressure phase diagram of the rare
  earth nickelates}, Phys. Rev. B 89 (2014) 245133 (Jun 2014).
\newblock \href {https://doi.org/10.1103/PhysRevB.89.245133}
  {\path{doi:10.1103/PhysRevB.89.245133}}.
\newline\urlprefix\url{https://link.aps.org/doi/10.1103/PhysRevB.89.245133}

\bibitem{PhysRevB.89.161113}
H.~T. Dang, A.~J. Millis, C.~A. Marianetti,
  \href{https://link.aps.org/doi/10.1103/PhysRevB.89.161113}{Covalency and the
  metal-insulator transition in titanate and vanadate perovskites}, Phys. Rev.
  B 89 (2014) 161113(R) (Apr 2014).
\newblock \href {https://doi.org/10.1103/PhysRevB.89.161113}
  {\path{doi:10.1103/PhysRevB.89.161113}}.
\newline\urlprefix\url{https://link.aps.org/doi/10.1103/PhysRevB.89.161113}

\bibitem{Projector2}
K.~Haule, T.~Birol, G.~Kotliar,
  \href{https://link.aps.org/doi/10.1103/PhysRevB.90.075136}{Covalency in
  transition-metal oxides within all-electron dynamical mean-field theory},
  Phys. Rev. B 90 (2014) 075136 (Aug 2014).
\newblock \href {https://doi.org/10.1103/PhysRevB.90.075136}
  {\path{doi:10.1103/PhysRevB.90.075136}}.
\newline\urlprefix\url{https://link.aps.org/doi/10.1103/PhysRevB.90.075136}

\bibitem{PhysRevLett.115.196403}
K.~Haule, \href{https://link.aps.org/doi/10.1103/PhysRevLett.115.196403}{Exact
  double counting in combining the dynamical mean field theory and the density
  functional theory}, Phys. Rev. Lett. 115 (2015) 196403 (Nov 2015).
\newblock \href {https://doi.org/10.1103/PhysRevLett.115.196403}
  {\path{doi:10.1103/PhysRevLett.115.196403}}.
\newline\urlprefix\url{https://link.aps.org/doi/10.1103/PhysRevLett.115.196403}

\bibitem{PhysRevLett.112.146401}
I.~Leonov, V.~I. Anisimov, D.~Vollhardt,
  \href{https://link.aps.org/doi/10.1103/PhysRevLett.112.146401}{First-principles
  calculation of atomic forces and structural distortions in strongly
  correlated materials}, Phys. Rev. Lett. 112 (2014) 146401 (Apr 2014).
\newblock \href {https://doi.org/10.1103/PhysRevLett.112.146401}
  {\path{doi:10.1103/PhysRevLett.112.146401}}.
\newline\urlprefix\url{https://link.aps.org/doi/10.1103/PhysRevLett.112.146401}

\bibitem{DMFT_forces_haule}
K.~Haule, G.~L. Pascut, Forces for structural optimizations in correlated
  materials within a dft+embedded dmft functional approach, Physical Review B
  94~(19) (11 2016).
\newblock \href {https://doi.org/10.1103/PhysRevB.94.195146}
  {\path{doi:10.1103/PhysRevB.94.195146}}.

\bibitem{HERATH2019107080}
U.~Herath, P.~Tavadze, X.~He, E.~Bousquet, S.~Singh, F.~Muñoz, A.~H. Romero,
  \href{http://www.sciencedirect.com/science/article/pii/S0010465519303935}{Pyprocar:
  A python library for electronic structure pre/post-processing}, Computer
  Physics Communications (2019) 107080 (2019).
\newblock \href {https://doi.org/https://doi.org/10.1016/j.cpc.2019.107080}
  {\path{doi:https://doi.org/10.1016/j.cpc.2019.107080}}.
\newline\urlprefix\url{http://www.sciencedirect.com/science/article/pii/S0010465519303935}

\bibitem{VASP_hafner}
J.~Hafner,
  \href{https://onlinelibrary.wiley.com/doi/abs/10.1002/jcc.21057}{Ab-initio
  simulations of materials using vasp: Density-functional theory and beyond},
  Journal of Computational Chemistry 29~(13) (2008) 2044--2078 (2008).
\newblock \href
  {http://arxiv.org/abs/https://onlinelibrary.wiley.com/doi/pdf/10.1002/jcc.21057}
  {\path{arXiv:https://onlinelibrary.wiley.com/doi/pdf/10.1002/jcc.21057}},
  \href {https://doi.org/10.1002/jcc.21057} {\path{doi:10.1002/jcc.21057}}.
\newline\urlprefix\url{https://onlinelibrary.wiley.com/doi/abs/10.1002/jcc.21057}

\bibitem{Pseudo_van}
D.~Vanderbilt, \href{https://link.aps.org/doi/10.1103/PhysRevB.41.7892}{Soft
  self-consistent pseudopotentials in a generalized eigenvalue formalism},
  Phys. Rev. B 41 (1990) 7892--7895 (Apr 1990).
\newblock \href {https://doi.org/10.1103/PhysRevB.41.7892}
  {\path{doi:10.1103/PhysRevB.41.7892}}.
\newline\urlprefix\url{https://link.aps.org/doi/10.1103/PhysRevB.41.7892}

\bibitem{PAW}
P.~E. Bl\"ochl,
  \href{https://link.aps.org/doi/10.1103/PhysRevB.50.17953}{Projector
  augmented-wave method}, Phys. Rev. B 50 (1994) 17953--17979 (Dec 1994).
\newblock \href {https://doi.org/10.1103/PhysRevB.50.17953}
  {\path{doi:10.1103/PhysRevB.50.17953}}.
\newline\urlprefix\url{https://link.aps.org/doi/10.1103/PhysRevB.50.17953}

\bibitem{Troullier19911993}
N.~Troullier, J.~Martins,
  \href{https://www.scopus.com/inward/record.uri?eid=2-s2.0-33645426115&doi=10.1103%2fPhysRevB.43.1993&partnerID=40&md5=37f208c157cbb8eeac2646aabdeadc57}{Efficient
  pseudopotentials for plane-wave calculations}, Physical Review B 43~(3)
  (1991) 1993--2006, cited By 12670 (1991).
\newblock \href {https://doi.org/10.1103/PhysRevB.43.1993}
  {\path{doi:10.1103/PhysRevB.43.1993}}.
\newline\urlprefix\url{https://www.scopus.com/inward/record.uri?eid=2-s2.0-33645426115&doi=10.1103%2fPhysRevB.43.1993&partnerID=40&md5=37f208c157cbb8eeac2646aabdeadc57}

\bibitem{PyChemia}
\href{https://github.com/MaterialsDiscovery/PyChemia}{Pychemia,}.
\newline\urlprefix\url{https://github.com/MaterialsDiscovery/PyChemia}

\bibitem{SrVO3_exp_crystal}
B.~Chamberland, P.~Danielson,
  \href{http://www.sciencedirect.com/science/article/pii/0022459671900351}{Alkaline-earth
  vanadium (iv) oxides having the avo3 composition}, Journal of Solid State
  Chemistry 3~(2) (1971) 243 -- 247 (1971).
\newblock \href {https://doi.org/https://doi.org/10.1016/0022-4596(71)90035-1}
  {\path{doi:https://doi.org/10.1016/0022-4596(71)90035-1}}.
\newline\urlprefix\url{http://www.sciencedirect.com/science/article/pii/0022459671900351}

\bibitem{SrVO3_e1}
F.~Lechermann, A.~Georges, A.~Poteryaev, S.~Biermann, M.~Posternak,
  A.~Yamasaki, O.~K. Andersen,
  \href{https://link.aps.org/doi/10.1103/PhysRevB.74.125120}{Dynamical
  mean-field theory using wannier functions: A flexible route to electronic
  structure calculations of strongly correlated materials}, Phys. Rev. B 74
  (2006) 125120 (Sep 2006).
\newblock \href {https://doi.org/10.1103/PhysRevB.74.125120}
  {\path{doi:10.1103/PhysRevB.74.125120}}.
\newline\urlprefix\url{https://link.aps.org/doi/10.1103/PhysRevB.74.125120}

\bibitem{SrVO3_e2}
E.~Pavarini, S.~Biermann, A.~Poteryaev, A.~I. Lichtenstein, A.~Georges, O.~K.
  Andersen, \href{https://link.aps.org/doi/10.1103/PhysRevLett.92.176403}{Mott
  transition and suppression of orbital fluctuations in orthorhombic $3{d}^{1}$
  perovskites}, Phys. Rev. Lett. 92 (2004) 176403 (Apr 2004).
\newblock \href {https://doi.org/10.1103/PhysRevLett.92.176403}
  {\path{doi:10.1103/PhysRevLett.92.176403}}.
\newline\urlprefix\url{https://link.aps.org/doi/10.1103/PhysRevLett.92.176403}

\bibitem{SrVO3_e3}
M.~Imada, A.~Fujimori, Y.~Tokura,
  \href{https://link.aps.org/doi/10.1103/RevModPhys.70.1039}{Metal-insulator
  transitions}, Rev. Mod. Phys. 70 (1998) 1039--1263 (Oct 1998).
\newblock \href {https://doi.org/10.1103/RevModPhys.70.1039}
  {\path{doi:10.1103/RevModPhys.70.1039}}.
\newline\urlprefix\url{https://link.aps.org/doi/10.1103/RevModPhys.70.1039}

\bibitem{SrVO3_LDA1}
A.~Sekiyama, H.~Fujiwara, S.~Imada, S.~Suga, H.~Eisaki, S.~I. Uchida,
  K.~Takegahara, H.~Harima, Y.~Saitoh, I.~A. Nekrasov, G.~Keller, D.~E.
  Kondakov, A.~V. Kozhevnikov, T.~Pruschke, K.~Held, D.~Vollhardt, V.~I.
  Anisimov,
  \href{https://link.aps.org/doi/10.1103/PhysRevLett.93.156402}{Mutual
  experimental and theoretical validation of bulk photoemission spectra of
  ${\mathrm{s}\mathrm{r}}_{1\ensuremath{-}x}{\mathrm{c}\mathrm{a}}_{x}{\mathrm{v}\mathrm{o}}_{3}$},
  Phys. Rev. Lett. 93 (2004) 156402 (Oct 2004).
\newblock \href {https://doi.org/10.1103/PhysRevLett.93.156402}
  {\path{doi:10.1103/PhysRevLett.93.156402}}.
\newline\urlprefix\url{https://link.aps.org/doi/10.1103/PhysRevLett.93.156402}

\bibitem{SrVO3_LDA2}
I.~A. Nekrasov, G.~Keller, D.~E. Kondakov, A.~V. Kozhevnikov, T.~Pruschke,
  K.~Held, D.~Vollhardt, V.~I. Anisimov,
  \href{https://link.aps.org/doi/10.1103/PhysRevB.72.155106}{Comparative study
  of correlation effects in $\mathrm{Ca}\mathrm{V}{\mathrm{o}}_{3}$ and
  $\mathrm{Sr}\mathrm{V}{\mathrm{o}}_{3}$}, Phys. Rev. B 72 (2005) 155106 (Oct
  2005).
\newblock \href {https://doi.org/10.1103/PhysRevB.72.155106}
  {\path{doi:10.1103/PhysRevB.72.155106}}.
\newline\urlprefix\url{https://link.aps.org/doi/10.1103/PhysRevB.72.155106}

\bibitem{SrVO3_LDA}
B.~Amadon, F.~Lechermann, A.~Georges, F.~Jollet, T.~O. Wehling, A.~I.
  Lichtenstein,
  \href{https://link.aps.org/doi/10.1103/PhysRevB.77.205112}{Plane-wave based
  electronic structure calculations for correlated materials using dynamical
  mean-field theory and projected local orbitals}, Phys. Rev. B 77 (2008)
  205112 (May 2008).
\newblock \href {https://doi.org/10.1103/PhysRevB.77.205112}
  {\path{doi:10.1103/PhysRevB.77.205112}}.
\newline\urlprefix\url{https://link.aps.org/doi/10.1103/PhysRevB.77.205112}

\bibitem{MPgrid}
H.~J. Monkhorst, J.~D. Pack,
  \href{https://link.aps.org/doi/10.1103/PhysRevB.13.5188}{Special points for
  brillouin-zone integrations}, Phys. Rev. B 13 (1976) 5188--5192 (Jun 1976).
\newblock \href {https://doi.org/10.1103/PhysRevB.13.5188}
  {\path{doi:10.1103/PhysRevB.13.5188}}.
\newline\urlprefix\url{https://link.aps.org/doi/10.1103/PhysRevB.13.5188}

\bibitem{SrVO3_t2g_m}
I.~A. Nekrasov, K.~Held, G.~Keller, D.~E. Kondakov, T.~Pruschke, M.~Kollar,
  O.~K. Andersen, V.~I. Anisimov, D.~Vollhardt,
  \href{https://link.aps.org/doi/10.1103/PhysRevB.73.155112}{Momentum-resolved
  spectral functions of srvo$_{3}$ calculated by $\mathrm{LDA}+\mathrm{DMFT}$},
  Phys. Rev. B 73 (2006) 155112 (Apr 2006).
\newblock \href {https://doi.org/10.1103/PhysRevB.73.155112}
  {\path{doi:10.1103/PhysRevB.73.155112}}.
\newline\urlprefix\url{https://link.aps.org/doi/10.1103/PhysRevB.73.155112}

\bibitem{SrVO3_Uvalue}
E.~Pavarini, S.~Biermann, A.~Poteryaev, A.~I. Lichtenstein, A.~Georges, O.~K.
  Andersen, \href{https://link.aps.org/doi/10.1103/PhysRevLett.92.176403}{Mott
  transition and suppression of orbital fluctuations in orthorhombic $3d^{1}$
  perovskites}, Phys. Rev. Lett. 92 (2004) 176403 (Apr 2004).
\newblock \href {https://doi.org/10.1103/PhysRevLett.92.176403}
  {\path{doi:10.1103/PhysRevLett.92.176403}}.
\newline\urlprefix\url{https://link.aps.org/doi/10.1103/PhysRevLett.92.176403}

\bibitem{Zfactor_srvo3_1_exp}
T.~Yoshida, M.~Hashimoto, T.~Takizawa, A.~Fujimori, M.~Kubota, K.~Ono,
  H.~Eisaki, \href{https://link.aps.org/doi/10.1103/PhysRevB.82.085119}{Mass
  renormalization in the bandwidth-controlled mott-hubbard systems $srvo_{3}$
  and $cavo_{3}$ studied by angle-resolved photoemission spectroscopy}, Phys.
  Rev. B 82 (2010) 085119 (Aug 2010).
\newblock \href {https://doi.org/10.1103/PhysRevB.82.085119}
  {\path{doi:10.1103/PhysRevB.82.085119}}.
\newline\urlprefix\url{https://link.aps.org/doi/10.1103/PhysRevB.82.085119}

\bibitem{Zfactor_srvo3_2_exp}
M.~Takizawa, M.~Minohara, H.~Kumigashira, D.~Toyota, M.~Oshima, H.~Wadati,
  T.~Yoshida, A.~Fujimori, M.~Lippmaa, M.~Kawasaki, H.~Koinuma, G.~Sordi,
  M.~Rozenberg,
  \href{https://link.aps.org/doi/10.1103/PhysRevB.80.235104}{Coherent and
  incoherent $d$ band dispersions in ${\text{srvo}}_{3}$}, Phys. Rev. B 80
  (2009) 235104 (Dec 2009).
\newblock \href {https://doi.org/10.1103/PhysRevB.80.235104}
  {\path{doi:10.1103/PhysRevB.80.235104}}.
\newline\urlprefix\url{https://link.aps.org/doi/10.1103/PhysRevB.80.235104}

\bibitem{Zfactor_srvo3_3_exp}
T.~Yoshida, K.~Tanaka, H.~Yagi, A.~Ino, H.~Eisaki, A.~Fujimori, Z.-X. Shen,
  \href{https://link.aps.org/doi/10.1103/PhysRevLett.95.146404}{Direct
  observation of the mass renormalization in ${\mathrm{srvo}}_{3}$ by angle
  resolved photoemission spectroscopy}, Phys. Rev. Lett. 95 (2005) 146404 (Sep
  2005).
\newblock \href {https://doi.org/10.1103/PhysRevLett.95.146404}
  {\path{doi:10.1103/PhysRevLett.95.146404}}.
\newline\urlprefix\url{https://link.aps.org/doi/10.1103/PhysRevLett.95.146404}

\bibitem{Vollhardt_SrVO3}
I.~Leonov, V.~I. Anisimov, D.~Vollhardt,
  \href{https://link.aps.org/doi/10.1103/PhysRevLett.112.146401}{First-principles
  calculation of atomic forces and structural distortions in strongly
  correlated materials}, Phys. Rev. Lett. 112 (2014) 146401 (Apr 2014).
\newblock \href {https://doi.org/10.1103/PhysRevLett.112.146401}
  {\path{doi:10.1103/PhysRevLett.112.146401}}.
\newline\urlprefix\url{https://link.aps.org/doi/10.1103/PhysRevLett.112.146401}

\bibitem{LaNiO3_Metal_para1}
A.~Y. Dobin, K.~R. Nikolaev, I.~N. Krivorotov, R.~M. Wentzcovitch, E.~D.
  Dahlberg, A.~M. Goldman,
  \href{https://link.aps.org/doi/10.1103/PhysRevB.68.113408}{Electronic and
  crystal structure of fully strained ${\mathrm{lanio}}_{3}$ films}, Phys. Rev.
  B 68 (2003) 113408 (Sep 2003).
\newblock \href {https://doi.org/10.1103/PhysRevB.68.113408}
  {\path{doi:10.1103/PhysRevB.68.113408}}.
\newline\urlprefix\url{https://link.aps.org/doi/10.1103/PhysRevB.68.113408}

\bibitem{LaNiO3_Metal_para2}
J.~L. Garc\'{\i}a-Mu\~noz, J.~Rodr\'{\i}guez-Carvajal, P.~Lacorre, J.~B.
  Torrance,
  \href{https://link.aps.org/doi/10.1103/PhysRevB.46.4414}{Neutron-diffraction
  study of r${\mathrm{nio}}_{3}$ (r=la,pr,nd,sm): Electronically induced
  structural changes across the metal-insulator transition}, Phys. Rev. B 46
  (1992) 4414--4425 (Aug 1992).
\newblock \href {https://doi.org/10.1103/PhysRevB.46.4414}
  {\path{doi:10.1103/PhysRevB.46.4414}}.
\newline\urlprefix\url{https://link.aps.org/doi/10.1103/PhysRevB.46.4414}

\bibitem{PRB.79.115122}
R.~Eguchi, A.~Chainani, M.~Taguchi, M.~Matsunami, Y.~Ishida, K.~Horiba,
  Y.~Senba, H.~Ohashi, S.~Shin, Fermi surfaces, electron-hole asymmetry, and
  correlation kink in a three-dimensional fermi liquid {LaNiO$_{3}$}, Phys.
  Rev. B 79 (2009) 115122 (2009).
\newblock \href {https://doi.org/10.1103/PhysRevB.79.115122}
  {\path{doi:10.1103/PhysRevB.79.115122}}.

\bibitem{PRB.83.075125}
M.~K. Stewart, C.-H. Yee, J.~Liu, M.~Kareev, R.~K. Smith, B.~C. Chapler,
  M.~Varela, P.~J. Ryan, K.~Haule, J.~Chakhalian, D.~N. Basov, Optical study of
  strained ultrathin films of strongly correlated {LaNiO$_{3}$}, Phys. Rev. B
  83 (2011) 075125 (2011).
\newblock \href {https://doi.org/10.1103/PhysRevB.83.075125}
  {\path{doi:10.1103/PhysRevB.83.075125}}.

\bibitem{PRB.82.165112}
D.~G. Ouellette, S.~Lee, J.~Son, S.~Stemmer, L.~Balents, A.~J. Millis, S.~J.
  Allen, Optical conductivity of {LaNiO$_{3}$}: Coherent transport and
  correlation driven mass enhancement, Phys. Rev. B 82 (2010) 165112 (2010).
\newblock \href {https://doi.org/10.1103/PhysRevB.82.165112}
  {\path{doi:10.1103/PhysRevB.82.165112}}.

\bibitem{PRB.48.1112}
X.~Q. Xu, J.~L. Peng, Z.~Y. Li, H.~L. Ju, R.~L. Greene, Resisitivity,
  thermopower, and susceptibility of {RNiO$_{3}$(R =La,Pr)}, Phys. Rev. B 48
  (1993) 1112--1118 (1993).
\newblock \href {https://doi.org/10.1103/PhysRevB.48.1112}
  {\path{doi:10.1103/PhysRevB.48.1112}}.

\bibitem{nature_NdNiO2_sup}
D.~Li, K.~Lee, B.~Wang, M.~Osada, S.~Crossley, H.~Lee, Y.~Cui, Y.~Hikita,
  H.~Hwang,
  \href{https://www.scopus.com/inward/record.uri?eid=2-s2.0-85071446435&doi=10.1038%2fs41586-019-1496-5&partnerID=40&md5=43105fc7ff5bb8f330825757ea2cc65d}{Superconductivity
  in an infinite-layer nickelate}, Nature 572~(7771) (2019) 624--627, cited By
  6 (2019).
\newblock \href {https://doi.org/10.1038/s41586-019-1496-5}
  {\path{doi:10.1038/s41586-019-1496-5}}.
\newline\urlprefix\url{https://www.scopus.com/inward/record.uri?eid=2-s2.0-85071446435&doi=10.1038%2fs41586-019-1496-5&partnerID=40&md5=43105fc7ff5bb8f330825757ea2cc65d}

\bibitem{Hepting_2020}
M.~Hepting, D.~Li, C.~J. Jia, H.~Lu, E.~Paris, Y.~Tseng, X.~Feng, M.~Osada,
  E.~Been, Y.~Hikita, et~al.,
  \href{http://dx.doi.org/10.1038/s41563-019-0585-z}{Electronic structure of
  the parent compound of superconducting infinite-layer nickelates}, Nature
  Materials (Jan 2020).
\newblock \href {https://doi.org/10.1038/s41563-019-0585-z}
  {\path{doi:10.1038/s41563-019-0585-z}}.
\newline\urlprefix\url{http://dx.doi.org/10.1038/s41563-019-0585-z}

\bibitem{Structure_Rhombo_cubic_laNiO3}
G.~Gou, I.~Grinberg, A.~M. Rappe, J.~M. Rondinelli,
  \href{https://link.aps.org/doi/10.1103/PhysRevB.84.144101}{Lattice normal
  modes and electronic properties of the correlated metal lanio${}_{3}$}, Phys.
  Rev. B 84 (2011) 144101 (Oct 2011).
\newblock \href {https://doi.org/10.1103/PhysRevB.84.144101}
  {\path{doi:10.1103/PhysRevB.84.144101}}.
\newline\urlprefix\url{https://link.aps.org/doi/10.1103/PhysRevB.84.144101}

\bibitem{LaNiO3_PES}
K.~Horiba, R.~Eguchi, M.~Taguchi, A.~Chainani, A.~Kikkawa, Y.~Senba, H.~Ohashi,
  S.~Shin,
  \href{https://link.aps.org/doi/10.1103/PhysRevB.76.155104}{Electronic
  structure of $\mathrm{La}\mathrm{Ni}{\mathrm{o}}_{3\ensuremath{-}x}$: An in
  situ soft x-ray photoemission and absorption study}, Phys. Rev. B 76 (2007)
  155104 (Oct 2007).
\newblock \href {https://doi.org/10.1103/PhysRevB.76.155104}
  {\path{doi:10.1103/PhysRevB.76.155104}}.
\newline\urlprefix\url{https://link.aps.org/doi/10.1103/PhysRevB.76.155104}

\bibitem{Nowadnik_ARPES_LaNiO3}
E.~A. Nowadnick, J.~P. Ruf, H.~Park, P.~D.~C. King, D.~G. Schlom, K.~M. Shen,
  A.~J. Millis,
  \href{https://link.aps.org/doi/10.1103/PhysRevB.92.245109}{Quantifying
  electronic correlation strength in a complex oxide: A combined dmft and arpes
  study of ${\text{lanio}}_{3}$}, Phys. Rev. B 92 (2015) 245109 (Dec 2015).
\newblock \href {https://doi.org/10.1103/PhysRevB.92.245109}
  {\path{doi:10.1103/PhysRevB.92.245109}}.
\newline\urlprefix\url{https://link.aps.org/doi/10.1103/PhysRevB.92.245109}

\bibitem{nio_band_gap2}
T.~Bredow, A.~R. Gerson,
  \href{https://link.aps.org/doi/10.1103/PhysRevB.61.5194}{Tempeffect of
  exchange and correlation on bulk properties of $mgo$, $nio$, and $coo$},
  Phys. Rev. B 61 (2000) 5194--5201 (Feb 2000).
\newblock \href {https://doi.org/10.1103/PhysRevB.61.5194}
  {\path{doi:10.1103/PhysRevB.61.5194}}.
\newline\urlprefix\url{https://link.aps.org/doi/10.1103/PhysRevB.61.5194}

\bibitem{Allen_nio_band_gap}
G.~A. Sawatzky, J.~W. Allen,
  \href{https://link.aps.org/doi/10.1103/PhysRevLett.53.2339}{Magnitude and
  origin of the band gap in nio}, Phys. Rev. Lett. 53 (1984) 2339--2342 (Dec
  1984).
\newblock \href {https://doi.org/10.1103/PhysRevLett.53.2339}
  {\path{doi:10.1103/PhysRevLett.53.2339}}.
\newline\urlprefix\url{https://link.aps.org/doi/10.1103/PhysRevLett.53.2339}

\bibitem{cox2010_nio_band_gap1}
P.~Cox, \href{https://books.google.com/books?id=9dDYZ7XT8hoC}{Transition Metal
  Oxides: An Introduction to Their Electronic Structure and Properties}, The
  International series of monographs on chemistry, OUP Oxford, 2010 (2010).
\newline\urlprefix\url{https://books.google.com/books?id=9dDYZ7XT8hoC}

\bibitem{nio_con_dft}
L.~F. Mattheiss,
  \href{https://link.aps.org/doi/10.1103/PhysRevB.5.306}{Electronic structure
  of the $3d$ transition-metal monoxides. ii. interpretation}, Phys. Rev. B 5
  (1972) 306--315 (Jan 1972).
\newblock \href {https://doi.org/10.1103/PhysRevB.5.306}
  {\path{doi:10.1103/PhysRevB.5.306}}.
\newline\urlprefix\url{https://link.aps.org/doi/10.1103/PhysRevB.5.306}

\bibitem{sp_dft_nio}
T.~M. Schuler, D.~L. Ederer, S.~Itza-Ortiz, G.~T. Woods, T.~A. Callcott, J.~C.
  Woicik, \href{https://link.aps.org/doi/10.1103/PhysRevB.71.115113}{Character
  of teh insulating state in $nio$: A mixture of charge-transfer and
  mott-hubbard character}, Phys. Rev. B 71 (2005) 115113 (Mar 2005).
\newblock \href {https://doi.org/10.1103/PhysRevB.71.115113}
  {\path{doi:10.1103/PhysRevB.71.115113}}.
\newline\urlprefix\url{https://link.aps.org/doi/10.1103/PhysRevB.71.115113}

\bibitem{SIC_nio}
A.~Svane, O.~Gunnarsson,
  \href{https://link.aps.org/doi/10.1103/PhysRevLett.65.1148}{Transition-metal
  oxides in the self-interaction--corrected density-functional formalism},
  Phys. Rev. Lett. 65 (1990) 1148--1151 (Aug 1990).
\newblock \href {https://doi.org/10.1103/PhysRevLett.65.1148}
  {\path{doi:10.1103/PhysRevLett.65.1148}}.
\newline\urlprefix\url{https://link.aps.org/doi/10.1103/PhysRevLett.65.1148}

\bibitem{LAD+U_nio}
V.~I. Anisimov, J.~Zaanen, O.~K. Andersen,
  \href{https://link.aps.org/doi/10.1103/PhysRevB.44.943}{Band theory and mott
  insulators: Hubbard u instead of stoner i}, Phys. Rev. B 44 (1991) 943--954
  (Jul 1991).
\newblock \href {https://doi.org/10.1103/PhysRevB.44.943}
  {\path{doi:10.1103/PhysRevB.44.943}}.
\newline\urlprefix\url{https://link.aps.org/doi/10.1103/PhysRevB.44.943}

\bibitem{Gw_nio}
F.~Aryasetiawan, O.~Gunnarsson,
  \href{https://link.aps.org/doi/10.1103/PhysRevLett.74.3221}{Electronic
  structure of $nio$ in the $\mathit{GW}$ approximation}, Phys. Rev. Lett. 74
  (1995) 3221--3224 (Apr 1995).
\newblock \href {https://doi.org/10.1103/PhysRevLett.74.3221}
  {\path{doi:10.1103/PhysRevLett.74.3221}}.
\newline\urlprefix\url{https://link.aps.org/doi/10.1103/PhysRevLett.74.3221}

\bibitem{GW1_niO}
S.~Massidda, A.~Continenza, M.~Posternak, A.~Baldereschi,
  \href{https://link.aps.org/doi/10.1103/PhysRevB.55.13494}{Quasiparticle
  energy bands of transition-metal oxides within a model gw scheme}, Phys. Rev.
  B 55 (1997) 13494--13502 (May 1997).
\newblock \href {https://doi.org/10.1103/PhysRevB.55.13494}
  {\path{doi:10.1103/PhysRevB.55.13494}}.
\newline\urlprefix\url{https://link.aps.org/doi/10.1103/PhysRevB.55.13494}

\bibitem{mark_ScGW_niO}
S.~V. Faleev, M.~van Schilfgaarde, T.~Kotani,
  \href{https://link.aps.org/doi/10.1103/PhysRevLett.93.126406}{All-electron
  self-consistent $gw$ approximation: Application to $si, mno, and nio$}, Phys.
  Rev. Lett. 93 (2004) 126406 (Sep 2004).
\newblock \href {https://doi.org/10.1103/PhysRevLett.93.126406}
  {\path{doi:10.1103/PhysRevLett.93.126406}}.
\newline\urlprefix\url{https://link.aps.org/doi/10.1103/PhysRevLett.93.126406}

\bibitem{GW_nio_Stefen}
J.-L. Li, G.-M. Rignanese, S.~G. Louie,
  \href{https://link.aps.org/doi/10.1103/PhysRevB.71.193102}{Quasiparticle
  energy bands of $nio$ in the $gw$ approximation}, Phys. Rev. B 71 (2005)
  193102 (May 2005).
\newblock \href {https://doi.org/10.1103/PhysRevB.71.193102}
  {\path{doi:10.1103/PhysRevB.71.193102}}.
\newline\urlprefix\url{https://link.aps.org/doi/10.1103/PhysRevB.71.193102}

\bibitem{nio_para_nochange}
B.~Brandow, \href{https://doi.org/10.1080/00018737700101443}{Electronic
  structure of mott insulators}, Advances in Physics 26~(5) (1977) 651--808
  (1977).
\newblock \href
  {http://arxiv.org/abs/https://doi.org/10.1080/00018737700101443}
  {\path{arXiv:https://doi.org/10.1080/00018737700101443}}, \href
  {https://doi.org/10.1080/00018737700101443}
  {\path{doi:10.1080/00018737700101443}}.
\newline\urlprefix\url{https://doi.org/10.1080/00018737700101443}

\bibitem{nio_magnetic_influ}
O.~Tjernberg, S.~S\"oderholm, G.~Chiaia, R.~Girard, U.~O. Karlsson, H.~Nyl\'en,
  I.~Lindau,
  \href{https://link.aps.org/doi/10.1103/PhysRevB.54.10245}{Influence of
  magnetic ordering on the nio valence band}, Phys. Rev. B 54 (1996)
  10245--10248 (Oct 1996).
\newblock \href {https://doi.org/10.1103/PhysRevB.54.10245}
  {\path{doi:10.1103/PhysRevB.54.10245}}.
\newline\urlprefix\url{https://link.aps.org/doi/10.1103/PhysRevB.54.10245}

\bibitem{nio_charge_density}
W.~Jauch, M.~Reehuis,
  \href{https://link.aps.org/doi/10.1103/PhysRevB.70.195121}{Electron density
  distribution in paramagnetic and antiferromagnetic $nio$: A
  $\ensuremath{\gamma}$-ray diffraction study}, Phys. Rev. B 70 (2004) 195121
  (Nov 2004).
\newblock \href {https://doi.org/10.1103/PhysRevB.70.195121}
  {\path{doi:10.1103/PhysRevB.70.195121}}.
\newline\urlprefix\url{https://link.aps.org/doi/10.1103/PhysRevB.70.195121}

\bibitem{Vollhardt_nio}
X.~Ren, I.~Leonov, G.~Keller, M.~Kollar, I.~Nekrasov, D.~Vollhardt,
  \href{https://link.aps.org/doi/10.1103/PhysRevB.74.195114}{$\mathrm{LDA}+\mathrm{DMFT}$
  computation of the electronic spectrum of $nio$}, Phys. Rev. B 74 (2006)
  195114 (Nov 2006).
\newblock \href {https://doi.org/10.1103/PhysRevB.74.195114}
  {\path{doi:10.1103/PhysRevB.74.195114}}.
\newline\urlprefix\url{https://link.aps.org/doi/10.1103/PhysRevB.74.195114}

\bibitem{kang2019nature}
B.~Kang, S.~Choi, The nature of the two-peak structure in $nio$ valence band
  photoemission, arXiv preprint arXiv:1908.05643 (2019).

\bibitem{nio_EXP2}
C.~Y. {Kuo}, T.~{Haupricht}, J.~{Weinen}, H.~{Wu}, K.~D. {Tsuei}, M.~W.
  {Haverkort}, A.~{Tanaka}, L.~H. {Tjeng}, {Challenges from experiment:
  electronic structure of NiO}, European Physical Journal Special Topics
  226~(11) (Jul 2017).
\newblock \href {https://doi.org/10.1140/epjst/e2017-70061-7}
  {\path{doi:10.1140/epjst/e2017-70061-7}}.

\bibitem{Shen}
Z.-X. Shen, R.~S. List, D.~S. Dessau, B.~O. Wells, O.~Jepsen, A.~J. Arko,
  R.~Barttlet, C.~K. Shih, F.~Parmigiani, J.~C. Huang, P.~A.~P. Lindberg,
  \href{https://link.aps.org/doi/10.1103/PhysRevB.44.3604}{Electronic structure
  of $nio$: Correlation and band effects}, Phys. Rev. B 44 (1991) 3604--3626
  (Aug 1991).
\newblock \href {https://doi.org/10.1103/PhysRevB.44.3604}
  {\path{doi:10.1103/PhysRevB.44.3604}}.
\newline\urlprefix\url{https://link.aps.org/doi/10.1103/PhysRevB.44.3604}

\bibitem{PBEsol}
J.~P. Perdew, A.~Ruzsinszky, G.~I. Csonka, O.~A. Vydrov, G.~E. Scuseria, L.~A.
  Constantin, X.~Zhou, K.~Burke,
  \href{https://link.aps.org/doi/10.1103/PhysRevLett.100.136406}{Restoring the
  density-gradient expansion for exchange in solids and surfaces}, Phys. Rev.
  Lett. 100 (2008) 136406 (Apr 2008).
\newblock \href {https://doi.org/10.1103/PhysRevLett.100.136406}
  {\path{doi:10.1103/PhysRevLett.100.136406}}.
\newline\urlprefix\url{https://link.aps.org/doi/10.1103/PhysRevLett.100.136406}

\bibitem{NiO_PBESol}
M.~Råsander, M.~A. Moram, \href{https://doi.org/10.1063/1.4932334}{On the
  accuracy of commonly used density functional approximations in determining
  the elastic constants of insulators and semiconductors}, The Journal of
  Chemical Physics 143~(14) (2015) 144104 (2015).
\newblock \href {http://arxiv.org/abs/https://doi.org/10.1063/1.4932334}
  {\path{arXiv:https://doi.org/10.1063/1.4932334}}, \href
  {https://doi.org/10.1063/1.4932334} {\path{doi:10.1063/1.4932334}}.
\newline\urlprefix\url{https://doi.org/10.1063/1.4932334}

\bibitem{NiO_U1}
V.~I. Anisimov, F.~Aryasetiawan, A.~I. Lichtenstein,
  \href{https://iopscience.iop.org/article/10.1088/0953-8984/9/4/002/meta}{First-principles
  calculations of the electronic structure and spectra of strongly correlated
  systems: {the $LDA$ + $U$ method}}, Journal of Physics: Condensed Matter
  9~(4) (1997) 767--808 (jan 1997).
\newblock \href {https://doi.org/10.1088/0953-8984/9/4/002}
  {\path{doi:10.1088/0953-8984/9/4/002}}.
\newline\urlprefix\url{https://iopscience.iop.org/article/10.1088/0953-8984/9/4/002/meta}

\bibitem{NiO_U2}
S.~Mandal, K.~Haule, K.~M. Rabe, D.~Vanderbilt,
  \href{https://link.aps.org/doi/10.1103/PhysRevB.100.245109}{Influence of
  magnetic ordering on the spectral properties of binary transition metal
  oxides}, Phys. Rev. B 100 (2019) 245109 (Dec 2019).
\newblock \href {https://doi.org/10.1103/PhysRevB.100.245109}
  {\path{doi:10.1103/PhysRevB.100.245109}}.
\newline\urlprefix\url{https://link.aps.org/doi/10.1103/PhysRevB.100.245109}

\bibitem{Jarell_ana_cont}
M.~Jarrell, O.~Biham,
  \href{https://link.aps.org/doi/10.1103/PhysRevLett.63.2504}{Dynamical
  approach to analytic continuation of quantum monte carlo data}, Phys. Rev.
  Lett. 63 (1989) 2504--2507 (Nov 1989).
\newblock \href {https://doi.org/10.1103/PhysRevLett.63.2504}
  {\path{doi:10.1103/PhysRevLett.63.2504}}.
\newline\urlprefix\url{https://link.aps.org/doi/10.1103/PhysRevLett.63.2504}

\end{thebibliography}
\end{document}